\documentclass[11pt]{article} 

\usepackage{amsmath,amsfonts,amsthm,amssymb} 
\usepackage[pdftex]{graphicx}
\usepackage[letterpaper, margin=1in]{geometry} 
\usepackage{setspace} 
\usepackage{mathrsfs} 
\usepackage{color}
\usepackage{float} 
\usepackage{algorithm}
\usepackage{algpseudocode}
\usepackage{amsthm}
\usepackage{subfig}

 \usepackage{multirow}
 \usepackage[table,xcdraw]{xcolor}
 \usepackage{hyperref,xcolor}

\hypersetup{linkcolor=blue}

\usepackage{bm}
\usepackage{url}
\usepackage{enumitem}

\graphicspath{{Figures/}}
\DeclareGraphicsExtensions{.pdf,.jpeg,.png}

\DeclareMathOperator*{\argmax}{argmax} 
\DeclareMathOperator*{\argmin}{argmin}

\usepackage[numbers,sort]{natbib}
\numberwithin{equation}{section} 

\newcommand{\vp}[1]{\textcolor{black}{#1}}
\newcommand{\vpB}[1]{\textcolor{black}{#1}} 

\begin{document}

\title{Deep Reinforcement Learning Algorithm for Dynamic Pricing of Express Lanes with Multiple Access Locations}

\author{Venktesh Pandey, 
	  Evana Wang,  
        and Stephen D. Boyles
\thanks{All authors are affiliated with the Department of Civil, Architectural and Environmental Engineering, The University of Texas at Austin, Austin, TX, 78712 USA.  Corresponding author's e-mail: \texttt{venktesh@utexas.edu}.}}

\markboth{Pandey et al. 2019}%
{Pandey \MakeLowercase{\textit{et al.}}: Deep RL}
\date{}
\maketitle

\begin{abstract}

This article develops a deep reinforcement learning (Deep-RL) framework for dynamic pricing on managed lanes with multiple access locations and heterogeneity in travelers' value of time, origin, and destination. \vp{This framework relaxes assumptions in the literature by considering multiple origins and destinations, multiple access locations to the managed lane, \textit{en route} diversion of travelers, partial observability of the sensor readings, and stochastic demand and observations.} The problem is formulated as a partially observable Markov decision process (POMDP) and policy gradient methods are used to determine tolls as a function of real-time observations. \vp{Tolls are modeled as continuous and stochastic variables, and are determined using a feedforward neural network}. The method is compared against a feedback control method used for dynamic pricing. We show that Deep-RL is effective in learning toll policies for maximizing revenue, minimizing total system travel time, and other joint weighted objectives, when tested on real-world transportation networks. \vpB{The Deep-RL toll policies outperform the feedback control heuristic for the revenue maximization objective by generating revenues up to 9.5\% higher than the heuristic and for the objective minimizing total system travel time (TSTT) by generating TSTT up to 10.4\% lower than the heuristic.} We also propose reward shaping methods for the POMDP to overcome undesired behavior of toll policies, like the \textit{jam-and-harvest} behavior of revenue-maximizing policies. Additionally, we test transferability of the algorithm trained on one set of inputs for new input distributions and offer recommendations on real-time implementations of Deep-RL algorithms. The source code for our experiments is available online at \url{https://github.com/venktesh22/ExpressLanes_Deep-RL}.
\\

\end{abstract}

\newpage
\section{Introduction}
	
	\subsection{Background and Motivation}
	{Priced} managed lanes (MLs), also referred to as express lanes or high-occupancy/toll lanes, are increasingly being used by many cities to mitigate traffic congestion and provide reliable travel time by using the existing capacity of the roadway. As of January 2019, there are 41 managed lane projects across the United States~\cite{trbManagedLaneDatabase}.
	On these lanes, travelers pay a toll which changes with the time of day, or dynamically based on the congestion pattern, to experience less congested travel time from their origin to their destination. In  recent years, managed lane networks have become increasingly complex, spanning longer corridors and having multiple entrance and exit locations. For example, the LBJ TEXpress lanes in Dallas, TX have 17 entrance ramps and 18 exit ramps, and three tolling segments with different time-varying toll values~\cite{lbjtex}.

	Dynamic pricing for express lanes with multiple access points is a complex control problem due to the heterogeneity in lane choice behavior of travelers belonging to different classes. Vehicles differ in their values of time and their destination of travel, both of which impact the pricing structure. Predicting driver behavior is difficult. A recent study showed that a binary logit model, commonly used for modeling lane choice, is inadequate in predicting heterogeneity in lane choice decisions~\cite{burris2018unrevealed}.
	
	Several dynamic pricing algorithms have been explored in the literature that optimize tolls under varying assumptions on driver behavior. These include methods using stochastic dynamic programming~\cite{yang2012distance}, hybrid model predictive control (MPC)~\cite{tan2018hybrid, toledo2015simulation}, reinforcement learning (RL)~\cite{zhu2015reinforcement,pandey2018multiagent}, and approximate dynamic programming~\cite{pandey2018dynamic}. While these algorithms do well against existing heuristics, they make some or all of the following restricting assumptions, which we relax: 
	\begin{enumerate}
		\item Restricted access for travelers: travelers do not exit the managed lane once they enter till their exit is reached~\cite{yang2012distance, zhu2015reinforcement} and that they only consider the first entry point as the decision point for the lane choice decision~\cite{tan2018hybrid}
		\item Fully observable system: toll operators have access to measurements of traffic density throughout the network for optimizing tolls~\cite{yang2012distance, tan2018hybrid, zhu2015reinforcement, pandey2018dynamic, pandey2018multiagent}
		\item Ignored traveler heterogeneity: a single vehicle class is considered with a single origin and destination~\cite{yang2012distance, zhu2015reinforcement, pandey2018dynamic}
		\item Simplified traffic dynamics: for example, the flow dynamics on general-purpose lanes are assumed independent of vehicles using the managed lane~\cite{yang2012distance}; or the proportion of flow split at diverge points is assumed identical for all origins~\cite{tan2018hybrid}
	\end{enumerate} 
	
	In addition, there are relatively few analyses on the conflict between optimization of multiple objectives with realistic constraints. Pandey and Boyles~\cite{pandey2018dynamic} showed that  the revenue-maximizing tolls exhibit a \emph{jam-and-harvest} (JAH) nature where the parallel general purpose lanes (GPLs) are intentionally jammed to congestion earlier in the simulation to harvest more revenue towards the end. Handling such undesirable behavior of optimal policies has not been studied in the literature.
	
	Furthermore, practical applicability of these algorithms in the real world environments is a less-explored question. \vp{Algorithms that optimize prices using a simulation model can be applied in real-time using lookup tables. However, the transferability analysis of such lookup tables to new input distributions is not considered~\cite{yang2012distance,zhu2015reinforcement,
	pandey2018dynamic}. The hybrid MPC algorithm in Tan and Gao~\cite{tan2018hybrid} uses a simulation model to predict boundary traffic as an exogenous input and an optimization model that incorporates real-time measurements of traffic densities and vehicle queue length to optimize tolls over a finite horizon. The computation time for solving the model is in the range of $1.2$--$2.6$ seconds for a 30 seconds optimization horizon, sufficient for a real-time implementation; however, the tests conducted are limited with analysis only on one test network under two scenarios of demand, assuming full observability of the system. Solving an MPC-based model with heterogeneous vehicle classes and partial observability of the system is complex and not fully studied. We thus require scalable algorithms for real-world networks that relax the assumptions on driver behavior and traffic flow, and transfer well from simulation settings to new input distributions.}

	In this article, we use deep reinforcement learning (Deep-RL) algorithms for optimizing tolls while relaxing simplifying assumptions in the earlier literature. In the recent years, Deep-RL algorithms have been successfully used for applications in playing computer games like Atari and planning motion of humanoid robots like \texttt{MuJoCo}~\cite{arulkumaran2017deep}. Similar algorithms have been applied in the areas of traffic signal control~\cite{shabestary2018deep}, active traffic management (like ramp metering)~\cite{belletti2017expert}, and control of autonomous vehicles in mixed autonomy~\cite{wu2017flow}. These traffic control applications indicate the usefulness of Deep-RL algorithms for solving the dynamic pricing problem for managed lanes with complex access structure. 
	
	We formulate and solve the dynamic pricing problem as a Deep-RL problem, and compare its performance against an existing feedback control method. We focus our attention on pricing algorithms that rely on real-time density observations using sensors (such as loop detectors) located only at certain locations around the network without access to any information about the demand distribution or driver characteristics like the value of time (VOT) distribution. Our framework thus relaxes assumptions in the literature by considering multiple origins and destinations, multiple access points to the managed lane facility, \textit{en route} diversion of vehicles at each diverge point, and partial observability of the systems. 
	We investigate the usefulness of Deep-RL as a tool for dynamic pricing, and explain its advantages and limitations by experiments on four different test networks. 
	
	\subsection{Related Work}
	Many control problems have been studied in the area of transportation engineering. These include active traffic management strategies such as ramp metering, variable speed limits, dynamic lane use control, and adaptive traffic signal control (ATSC). Control problems in the area of transportation are broadly solved using three methods: open-loop optimal control methods (that solve the optimal control problem \vp{without incorporating real-time measurements}), closed-loop control methods like MPC (that incorporate the feedback of real-time measurements and optimize over a rolling horizon), and lately RL methods where the optimal control is learnt with an iterative interaction with the environment, possibly in simulated offline settings which can then be translated in real-time settings. A broad overview of all control problems in the transportation domain is out of scope of this article. 
	
	The managed lane pricing problem is also a traffic control problem, where the chosen control directly impacts the driver behavior and thus the congestion pattern. There are three component models to the ML pricing problem~\cite{gardner2013development}: a \emph{lane choice model} that determines how travelers choose a lane given the tolls and travel times, a \emph{traffic flow model} that models the interaction of vehicles in simulated environments, and a \emph{toll pricing model} which determines the toll pricing objectives and how the optimization problem is solved to achieve the best value of the objective. 
	Pandey~\cite{pandey2016optimal} presented a tabular comparison of component models for the existing models in the literature. In this research, we focus on the \emph{toll pricing models}. 
	
	Toll pricing models for MLs with a single access point are commonly studied. Gardner et al.~\cite{gardner2013development} argued that for ML with a single entrance and exit, the tolls minimizing the total system travel time (TSTT) also utilize the managed lanes to full capacity at all times. The authors developed an analytical formulation for tolls minimizing TSTT which send as many vehicles to the ML at each time step as is the capacity of the lane. Lou et al.~\cite{lou2011optimal} used a self-learning approach for optimizing toll prices where the average VOT values were learnt using real-time measurements. Toledo et al.~\cite{toledo2015simulation} used a rolling horizon approach to optimize future tolls with predicted demand from traffic simulation; however, the method of exhaustive search to solve the non-convex control problem does not scale well for large managed lane networks. 
	
	For managed lanes with multiple access points, Tan and Gao~\cite{tan2018hybrid} presented a formulation where the proportion of vehicles entering the managed lane is optimized instead of directly optimizing the toll prices. The authors showed a one-to-one mapping between optimal toll prices and the proportion values, and transformed the control problem into a mixed-integer linear program which can be solved efficiently for networks with multiple access points. Dorogush and Kurzhanskiy~\cite{dorogush2015modeling} used a similar method and optimized split ratios at each diverge, which are then used to determine toll prices; however, their analysis ignored the variation of incoming flow at each diverge. Apart from these optimal control based methods, Zhu and Ukkusuri~\cite{zhu2015reinforcement} and Pandey and Boyles~\cite{pandey2018dynamic} used RL methods, where the control problem is formulated as a Markov decision process (MDP) and the value function (or its equivalent Q-function) is learned by iterative interactions with the environment. However, the tests are conducted for discrete state and action spaces assuming full observability of the system. The present article is guided by advances in RL methods,  and improves these earlier RL-based approaches for dynamic pricing.
	
	Deep-RL improves  traditional RL \vp{by using deep neural networks as function approximators, which has been effective in various control problems}. See Arulkumaran et al. \cite{arulkumaran2017deep} for a survey of Deep-RL applications. Deep-RL works well because it learns the system/environment characteristics by repeated interactions with the environment, without requiring knowledge of the component model. It is a form of end-to-end learning where learning can be done using direct observations, in contrast to the sequential learning methods which use observations to calibrate \vp{component models like the input VOT distribution} and then optimize toll prices. Classical control methods that rely heavily on the model behind the system can be very complex, especially for the dynamic pricing problem \cite{tan2018hybrid}. These methods require simplifying traffic flow and driver behavior assumptions for relaxing the non-convex optimal control problem.  Considering the amount of uncertainty in a dynamic pricing system, Deep-RL based methods can prove effective and we investigate this as a hypothesis in this article. 
	
	Application of Deep-RL algorithms for traffic control problems is not new. Belletti et al.~\cite{belletti2017expert} developed an ``expert-level" control of coordinated ramp metering using Deep-RL methods with multiple agents and achieved precise adaptive metering without requiring model calibration that does better than the traditional benchmark algorithm named ALINEA. 
	Wu et al.~\cite{wu2017flow} used Deep-RL algorithms to solve the control problem of selecting the acceleration and brake of multiple autonomous vehicles (AVs) under conditions of mixed human vehicles and AVs to mitigate traffic congestion. When compared against classical approaches, their approach generated 10-20\% lower TSTT. Other applications of Deep-RL algorithms are in the domain of ATSC including traditional one signal control \cite{genders2016using, shabestary2018deep}, coordinated control of traffic signals \cite{van2016deep}, and large-scale multiagent control using Deep-RL methods \cite{chu2019multi}. See Yan et al.~\cite{yau2017survey} for a review of RL algorithms in the area of ATSC.
		
	Inspired by the open-source benchmark called FLOW~\cite{wu2017flow}, which is a microscopic deep reinforcement learning framework for traffic management, the multiclass mesoscopic traffic flow environment developed in this article is made open-source so future tests on improving the algorithms for dynamic pricing can be benchmarked. 
	
	\subsection{Contributions and Outline}
	The key contributions of this article are:
	\begin{itemize}[noitemsep]
		\item We demonstrate the usefulness of Deep-RL algorithms for solving dynamic pricing control problem under partial observability, and show that it performs well against existing heuristics, without requiring restricting assumptions on driver behavior or traffic dynamics.
		\item We apply multi-objective optimization methods for joint optimization of multiple objectives and overcome undesirable JAH characteristics of revenue-maximizing optimal policies.
		\item We conduct tests to verify the transferability of learned Deep-RL algorithms to new input distributions and make recommendations on real-time implementation of the algorithm.
		\item We develop an open-source framework for dynamic pricing using multiclass cell transmission model available for benchmarking future dynamic pricing experiments.
	\end{itemize}
	
	The rest of the paper is organized as follows. Section \ref{sec:model} introduces the notation and presents the details of the model. Section 3 explains the chosen Deep-RL algorithms and the feedback control heuristic against which the algorithm is compared. Section 4 presents the experimental analysis of Deep-RL algorithms on four test networks and discusses transferability analysis, multi-objective optimization, and comparison of the performance with another heuristic. Section 5 concludes the paper and suggests topics for future work.

\section{Model for Deep Reinforcement Learning}
\label{sec:model}

\subsection{Network Notation}

\begin{figure}[h]%
		\centering
		\includegraphics[width=\columnwidth]{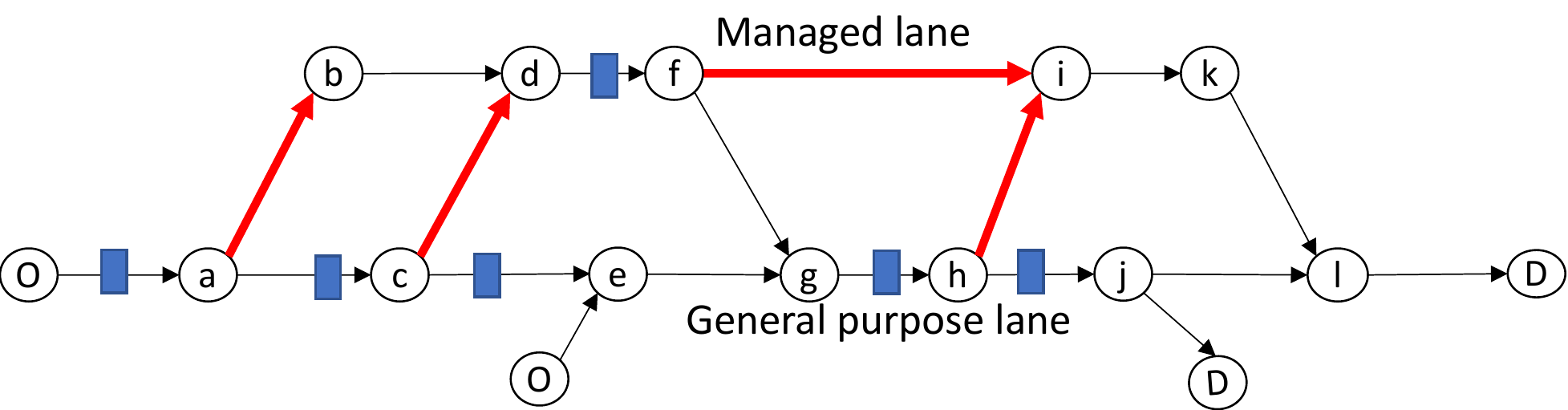}%
		\caption{Managed lane network with multiple entrances and exits where links with higher thickness are tolled, and links with a box are observed by the toll operator }%
		\label{fig:MEME}%
	\end{figure}

Consider the directed network shown in Figure \ref{fig:MEME} which is an abstraction of a managed lane network. The upper set of links form MLs, the lower set of links form GPLs, and the ramps connect the two lanes at various access points. As we describe the network, we label the assumptions made in our model as ``A$\#$". We also label ideas for future work as ``$\text{FW}\#$".

Let $N$ represent the set of all nodes and $A=\{(i,j)~|~i,j\in N \}$ represent the set of all links in the network. 
Let $N_o$ denote the set of all origins and $N_d$ denote the set of all destinations. We assume that origins and destinations connect to the network through nodes on the GPLs (A$\#1)$ and the only way to access the MLs is through on-ramps leading towards the lane. \vpB{This is a reasonable assumption as most current ML installations allow access to MLs only through ramps from the GPL. If there is a direct access to the ML from outside the network, the  current framework can still be used by appropriately adjusting the lane choice model explained in Section \ref{subsec:lanechoice}.}

\begin{figure}[h]%
		\centering
		\includegraphics[scale=0.5]{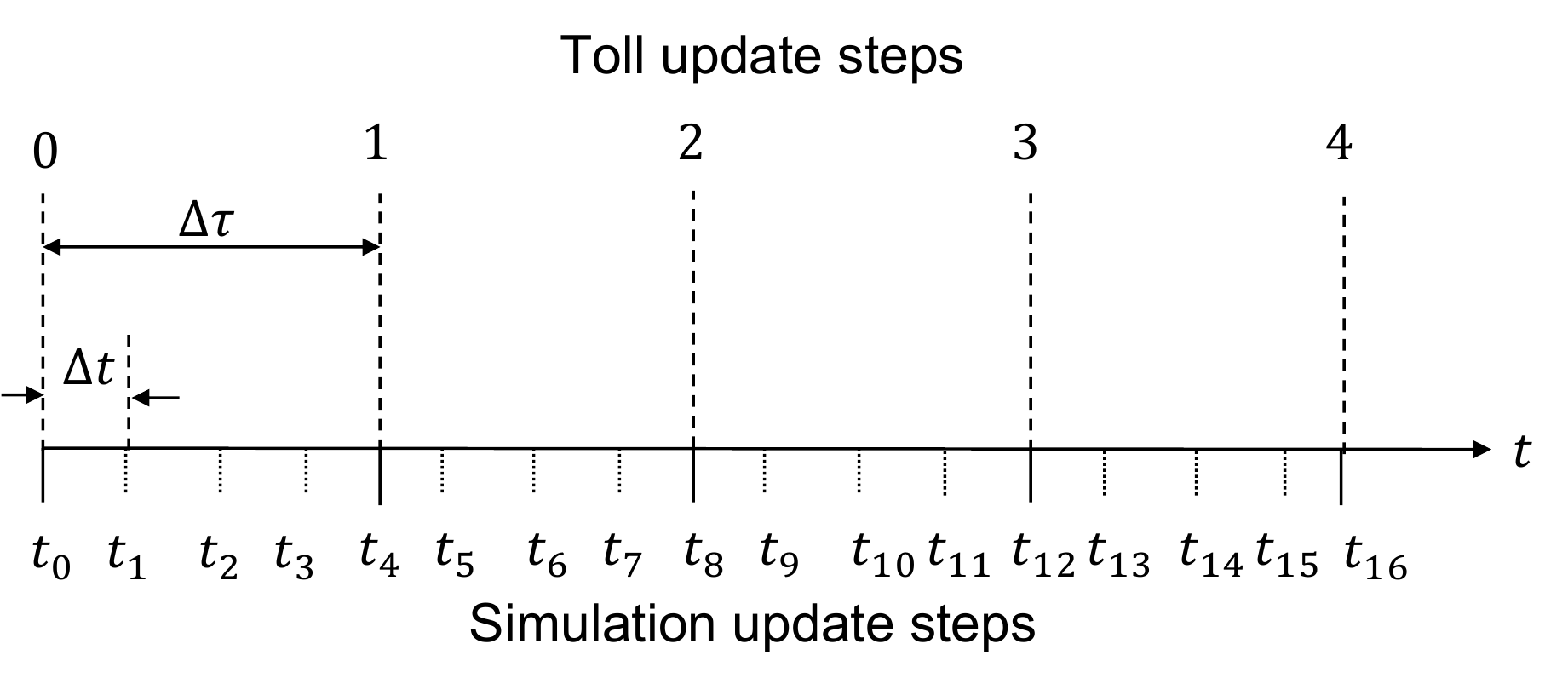}%
		\caption{Representation of a time scale}%
		\label{fig:timeScale}%
	\end{figure}

	The time horizon is divided into equal time steps, each $\Delta t$ units long. The set of all time periods is given by $\mathscr{T}=\{t_0,t_1,t_2,\ldots,t_{T/{\Delta t}} \}$, where $T$, \vp{an integral multiple of $\Delta t$}, is the time horizon. Tolls are updated after every $\Delta \tau = m\Delta t$ time units, where $m$ is a positive integer fixed by the tolling agency.  Define $\mathscr{T}_\tau = \{ k~|~t_{km}\in \mathscr{T}, \text{ where } k \in \{0,1,2,\ldots \} \}$ as the set of time periods where tolls are updated, indexed in increasing order of positive integers. Then, $|\mathscr{T}_\tau|=T/{\Delta \tau}+1$. For example, Figure \ref{fig:timeScale} shows different elements of time where $m=4$ and $T=16\Delta t$.  For the figure, $\mathscr{T}=\{t_0,t_1,t_2,\ldots,t_{16} \}$ and $\mathscr{T}_\tau=\{0,1,2,3,4 \}$.
	
	The demand between an origin and a destination is a random variable. A toll operator does not know the demand distribution, but only relies on the observed realizations of demand. However, for simulation purposes, we model the demand of vehicles from origin $r \in N_o$ to destination $s \in N_d$ at time $t \in \mathscr{T}$ to be a rectified Gaussian random variable with mean $d_{rs}(t)$ and standard deviation $\sigma_d$, \vp{and ignore correlations of demand between different origin-destination (OD) pairs and across time.} The mean demand $d_{rs}(t)$ can be estimated by observing the historical data of the managed lane facility or from the regional model.

	Let $V$ denote the set of all values of VOT (assumed to be a discrete distribution for the population, A$\#2$) and $p_v$ be the proportion of demand with VOT $v$, for any $v \in V$. The $p_v$ values are unknown to a toll operator. For simulation purposes, we choose the VOT distribution ($p_v~|~v\in V$) and $\sigma_d$ to be identical for all origin-destination pairs. Though dynamic traffic assignment models have been used in the literature for optimization of toll prices for express lanes~\cite{zhang2018calibration}, we focus on real-time optimization of toll prices and ignore route-choice equilibration of travelers (A$\#3)$. \vp{We thus assume travelers base their decisions only on real-time information provided at diverge points. The lane choice models are discussed in Section \ref{subsec:lanechoice}.}
	
	\vp{Traffic flow models can either be microscopic or macroscopic. With the exception of Belletti et al.~\cite{belletti2017expert}, all other Deep-RL models in transportation domain use microsimulation to capture the vehicle-to-vehicle interactions. In this article, we use macroscopic models to represent traffic flow for the simplicity they provide.} In contrast to the cell-based representation of managed lane network \vp{in macroscopic traffic models} from the literature, where MLs and GPLs are modeled as part of the same cell~\cite{tan2018hybrid, yang2012distance,dorogush2015modeling}, we divide each link into individual cells, where the links for GPLs are separate from that of MLs. This choice lets us use the cell transmission model (CTM) equations from Daganzo \cite{daganzo1995cell} for modeling traffic flow. Let $\mathscr{C}_{(i,j)}$ represent the set of all cells for link $(i,j)\in A$ and $\mathscr{C} = \bigcup_{(i,j)\in A} \mathscr{C}_{(i,j)}$ denote the set of all cells in the network.
The length  of each cell $c \in \mathscr{C}$, denoted by $l_c$, is determined as usual (the distance traveled at free flow in time $\Delta t$)~\cite{daganzo1995cell}, and is assumed constant for all links in the network (A$\#4)$. We thus require all link lengths to be integral multiples of the cell length. Let $l_{ij}, \nu_{ij}, q_{{\max},ij}, w_{ij},$ and $k_{\text{jam},ij}$ represent the length, free-flow speed, capacity, back-wave speed, and jam density, respectively, for link $(i,j) \in A$ as its fundamental diagram parameters, \vp{which we assume has a trapezoidal shape (A$\#5$)}.

A toll operator is assumed to manage the toll rate at each on-ramp and diverge point beyond a diverge on a ML (A$\#6)$. We assume this toll structure in contrast to the generic structure of separate toll values for each origin-destination (OD) pair, like in Yang et al.~\cite{yang2012distance} and Tan and Gao~\cite{tan2018hybrid}, because it inherently models the constraint that traveling longer distance on the ML levies a higher toll than traveling shorter distance. For a detailed discussion on various options to charge toll on a managed lane network with multiple accesses, see Pandey and Boyles~\cite{pandey2019comparing}. Let $A_{\text{toll}}$ represent the links where tolls are collected. Figure \ref{fig:MEME} highlights these links in bold.  We denote the toll charged on link $(i,j)\in A_{\text{toll}}$ for any $t \in \mathscr{T}$ by $\beta_{ij}(t)$.

\subsection{Lane Choice Model}
\label{subsec:lanechoice}

	Travelers make routing decisions at each diverge locations while traveling towards their destination. Nodes $a,c,f$, and $h$ are the diverge locations for the network in Figure \ref{fig:MEME}. At each diverge node, travelers receive information about the current travel time and toll values. We assume that the information about the current travel time is provided by measuring instantaneous travel time (A$\#7$), and that all travelers make their lane choice decision only using the instantaneous/real-time information and do not rely on historic information (obtained from prior experience) for making lane choices (A$\#8$). Assumptions A$\#7$ and A$\#8$ are only made for simulation purposes, as the Deep-RL model only requires the realization of lane choice by each traveler in form of observed density measurements at detector locations. If we have an estimate of experienced travel time on each route, the simulations can be based on experienced travel time. Assumptions A$\#3$ and A$\#8$ are related: because we assume no prior experience for the drivers, users do not find an equilibrium over route choices. Considering dynamic equilibrium while optimizing a dynamic stochastic control is a complex problem and will be studied as part of the future work $(\text{FW}\#1$).
	
	There are several models proposed in the literature to model lane choice of travelers, including a binary logit model that models stochastic lane choice of travelers over two routes connecting current diverge to the destination, and a decision route model that evaluates deterministic lane choice of multiple vehicle classes comparing utilities over a set of routes connecting current diverge to the merge after the first exit from the ML. For a detailed discussion on the decision route model, refer to Pandey and Boyles~ \cite{pandey2018dynamic}. A recent analysis in Pandey and Boyles  \cite{pandey2019comparing} showed that a decision route model has the least error compared to the optimal route choice model for rational travelers; however, a logit model can capture irrational driver behavior, where a rational traveler is defined as the one who always chooses the route minimizing her utility. Conceptually, the lane choice models can be categorized based on three characteristics: the number of routes over which travelers compare the utility, whether or not the lane choice is stochastic/deterministic, and the heterogeneity in vehicles' value of time (single class vs multiple classes). Table \ref{tab:laneChoice} shows the combinations of categories and models used in the literature. Certain combination have not been used directly, but they could be used. For example, combining decision routes with stochastic lane choice can result in models like multinomial logit or mixed logit, but the assumption that the choices are independent may not hold true (the choices in this setting being the different routes).
	
	\begin{table}[h]
		\centering
		\caption{Categorization of lane choice models for managed lanes with multiple entrances and exits}
		\label{tab:laneChoice}
		\begin{tabular}{|c|c|c|c|}
		\hline
		\textbf{\begin{tabular}[c]{@{}c@{}}Number of \\ VOT classes\end{tabular}} & \textbf{\begin{tabular}[c]{@{}c@{}}Number of routes \\ over which  the utility(s) \\ is (are) compared\end{tabular}} & \textbf{\begin{tabular}[c]{@{}c@{}}Deterministic \\ or Stochastic\end{tabular}} & \textbf{\begin{tabular}[c]{@{}c@{}}Reference(s) in the\\  literature using \\ this lane choice\end{tabular}} \\ \hline
Single & Two & Deterministic &  \cite{gardner2013development}\\ \hline
Single & Two & Stochastic & \cite{tan2018hybrid}\cite{toledo2015simulation}\cite{zhu2015reinforcement}\cite{yang2012distance} \\ \hline
Single & Decision routes & Deterministic &  None \\ \hline
Single & Decision routes & Stochastic & None \\ \hline
Multiple & Two & Deterministic & \cite{gardner2015robust} \\ \hline
Multiple & Two & Stochastic & None \\ \hline
Multiple & Decision routes & Deterministic & \cite{pandey2018dynamic},\cite{pandey2018multiagent} \\ \hline
Multiple & Decision routes & Stochastic & None \\ \hline
		\end{tabular}
\end{table}
	
	The Deep-RL algorithm developed in this article is agnostic to the lane choice model. For simulation purposes, we focus our attention on two models: multiple VOT classes with two routes and stochastic choice (\textit{multiclass binary logit model}) and  multiple VOT classes with decision routes and deterministic choice (\textit{multiclass decision route model}). For simulation purposes, we evaluate the utility of a route as the linear combination of the toll and route's travel time, converted to the same units using the VOT for the class (A$\#9$).

\subsection{Partially Observable Markov Decision Process}
\vp{MDPs are a discrete time stochastic control process that provide a framework for solving problems that involve sequential decision making~\cite{sutton2018reinforcement}. At each time step, the system is in some \textit{state}. The decision maker takes an \textit{action} in that \textit{state}, and the system transitions to the next \textit{state} depending on the transition probabilities, which are only a function of the current \textit{state} and the \textit{action} taken (called the Markov property). Given an \textit{action}, this transition from one \textit{state} to the other generates a reward for each time step and the decision maker seeks to maximize the expected reward across all time steps. Control problems in transportation do not necessarily have the Markov property because of the temporal dependence of congestion pattern. However, by including the simulation time as part of the state, they can be formulated as an MDP.}

\vp{Partially observable Markov decision processes (POMDPs) are MDPs where the state at any time step is not known with certainty, that is, the state is not fully observable. For the dynamic pricing problem where a toll operator does not have access to traffic information throughout the network but only at certain locations, POMDPs are a suitable choice.}  
 We define the control problem for determining the optimal toll as an POMDP with following components:
 \begin{itemize}
 	\item \textbf{Timestep}: Tolls are to be optimized over a finite time horizon for each time $k \in \mathscr{T}_\tau$. A finite horizon can represent a morning or an evening peak period on a corridor, or an entire day.
 	\item \textbf{State}: We first define $x_c^z(t)$ as the number of vehicles in cell $c \in \mathscr{C}$ belonging to class $z \in Z$ at time $t \in \mathscr{T}$, where $Z = \{ (v,d)~|~v \in V, d \in N_d \}$ is the set of all classes, disaggregated by the VOT value and the destination of the vehicle (the origin of a vehicle does not influence lane choice once the vehicle is on the road and is thus ignored). For ML networks where high occupancy vehicles pay a different toll than single/low occupancy vehicles, we can extend $Z$ to include the occupancy level of vehicles, but we leave that analysis for future work ($\text{FW}\#2$). The dimensionality of $Z$ impacts the computational performance of the multiclass cell transmission model. \vp{Similar to the non-atomic flow assumption commonly used in the transportation literature, we consider $x_c^z(t)$ to be a non-negative real number, rather than an integer.} We denote the state of the POMDP by $s$ comprising of the current toll update step $k \in \mathscr{T}_\tau$ and the values $x_c^z(t_{k\Delta\tau})$ for all cells $c \in \mathscr{C}$ and class $z \in Z$. Thus, the state space $S$ can be written as Equation \eqref{eq:stateSpace}. \vpB{Allowing $\Delta \tau$ to be greater than $\Delta t$ ($m>1$) reduces the size of state space compared to choosing $m=1$, which improves the computational efficiency.}
 	\begin{equation}
 		S = \left\lbrace (k, x_c^z(t_{k\Delta\tau}))~|~ k\in \mathscr{T}_\tau , c \in \mathscr{C}, z \in Z \right\rbrace
 		\label{eq:stateSpace}
 	\end{equation}
 	\item \textbf{Observation}: In our model, the observation is done using loop detectors. The detectors measure the total number of vehicles going from one cell to the next and thus cannot distinguish between vehicles belonging to different classes, so the state is not fully observable. This is an advantage of the proposed model, contrasting with the commonly-used full observability assumption~\cite{yin2009dynamic, tan2018hybrid}. The observation space depends on the location of detectors. We conduct sensitivity analyses with respect to changes in the observation space later in the text. Let $\bm{o}(s)$ denote the observation vector for state $s$ and comprise of the measurement of total number of vehicles on each link $(i,j) \in A_{\text{loop}} \subseteq A$ which has a loop detector installed at beginning and end.\footnote{For Figure \ref{fig:MEME}, $A_{\text{loop}} = \{(o,a),(a,c),(c,e),(d,f),(g,h),(h,j)\}$.} That is, $\bm{o}(s) = \{ \sum_{z\in Z} \sum_{c \in \mathscr{C}_\text{(i,j)}}x_c^z(t_{k\Delta \tau}) ~|~ (i,j) \in A_{\text{loop}}\}$. \vp{We assume that we can learn the total number of vehicles on any link by tracking the number of vehicles entering the link (measured at an upstream detector) and the number of vehicles leaving the link (measured at a downstream detector) (A$\#10$).} The actual observation is assumed to be Gaussian random variable with the mean as specified and the standard deviation $\sigma_o$ which models the noise in loop detector measurements. We project negative values of observation, if any, to zero.
 	\item \textbf{Action}: Action $a$ in state $s$ is the toll $\beta_{ij}(t_{k\Delta \tau})$ charged for a toll link $(i,j)\in A_{\text{toll}}$, where $\beta_{ij}(\cdot) \in \left[ \beta_{\min}, \beta_{\max} \right] $. The action is modeled as a continuous variable; the values can be rounded to nearest tenth of a cent or dollar if desired.
 	\item \textbf{Transition function}: The transition of the POMDP from a state $s$ to a new state $s'$ given action $a$, is governed by the traffic flow equations from the CTM model which incorporates the lane choice behavior of travelers. For simulation purposes, we assume that traffic flow throughout the network is deterministic except at diverges where the lane choices of travelers may be stochastic (A$\#11)$. We use a multiclass version of the CTM model similar to the model in Pandey and Boyles~\cite{pandey2018dynamic}.
 	\item \textbf{Reward}: The reward obtained after taking action $a$ in state $s$, denoted by $r(s,a)$, depends on the choice of tolling objective. We consider two objectives, revenue maximization and total system travel time (TSTT) minimization, with following definitions of reward:
 	\begin{itemize}
 		\item Revenue maximization: 
 		\begin{equation}
 			r^{\text{RevMax}}(s,a) = \sum_{x= k\Delta\tau}^{(k+1)\Delta\tau-1} \sum_{(i,j)\in A_{\text{toll}}} \left( \beta_{ij}(t_{k\Delta \tau}) \sum_{(h,i)\in A} y_{hij}(t_x) \right)
 			\label{eq:revMaxObj}
 		\end{equation}
 		where $y_{hij}(t)$ is the total flow moving from link $(h,i)\in A$ to $(i,j)\in A$ from time step $t$ to time step $t+\Delta t$
 		
 		\item Total system travel time minimization: 
 		\begin{equation}
 			r^{\text{TSTTMin}}(s,a) =  - \left( \sum_{x= k\Delta\tau}^{(k+1)\Delta\tau-1} \sum_{c \in \mathscr{C}} \sum_{z\in Z} x_c^{z}(t_x) \right)
 			\label{eq:tsttMinObj}
 		\end{equation}
 		where the negative sign is used to ensure that reward maximization is equivalent to TSTT minimization.
 	\end{itemize}
 \end{itemize}
 	
 	For the dynamic pricing problem, revenue-maximizing tolls often have a JAH nature where the GPLs are jammed to congestion earlier in the simulation to attract more travelers towards the ML later in the simulation generating more revenue~\cite{goccmen2015revenue, pandey2018dynamic}. This undesirable characteristic  of optimal policy is also seen in other applications of RL. For example, for ATSC a simpler definition of reward that maximizes amount of flow during a cycle may lead to ``evil" optimal policies, where the controller agent holds congestion on the mainline and then gains a larger reward by extending the greens for the main approach~\cite{shabestary2018deep}. Similarly, Van der Pol and Oliehoek~\cite{van2016coordinated} show that with inappropriate definitions of reward, the signal control policy may have unusual flips from green to red.
 	
 	To overcome the  undesired JAH nature, \vp{we use reward shaping methods that modify the reward definitions such that the optimal policies have less or no JAH behavior (discussed later in Section~\ref{subsec:multiObjOpt})}. For reward shaping, we quantify the JAH behavior using two statistics defined as a numeric value at the end of simulation. The first statistic, $\text{JAH}_1$, measures the maximum of difference between the number of vehicles in GPLs to the number of vehicles in MLs across all time steps. It is defined as in Equation \eqref{eq:jahStat1}, where $A_{\text{GPL}}(A_{\text{ML}})$ are links on the GPL (ML).
 
 \begin{equation}
 	\text{JAH}_1 = \max_{t \in \mathscr{T}} \left( \sum_{(i,j)\in A_{\text{GPL}}} \sum_{c \in \mathscr{C}_{(i,j)}} \sum_{z\in Z} x_c^z(t) - \sum_{(i,j)\in A_{\text{ML}}} \sum_{c \in \mathscr{C}_{(i,j)}} \sum_{z\in Z} x_c^z(t) \right)
 	\label{eq:jahStat1}
 \end{equation}
 
 The value of $\text{JAH}_1$ is dependent on network properties like number of lanes in GPLs and MLs. We also define an alternate statistic $\text{JAH}_2$ that is network independent. We first define $\zeta(t)$, as in Equation \eqref{eq:zetat}, as the difference between the ratio of current number of vehicles to the maximum number of vehicles allowed in each cell (corresponding to jam density) for all cells on GPLs with that of MLs.
 
 \begin{equation}
 	\zeta(t) = \frac{\sum_{(i,j)\in A_{\text{GPL}}} \sum_{c \in \mathscr{C}_{(i,j)}} \sum_{z\in Z} x_c^z(t)}{\sum_{(i,j)\in A_{\text{GPL}}} \sum_{i \in \mathscr{C}_{(i,j)}} l_{ij} k_{\text{jam},ij}} - \frac{\sum_{(i,j)\in A_{\text{ML}}} \sum_{i \in \mathscr{C}_{(i,j)}} \sum_{z\in Z} x_i^z(t)}{\sum_{(i,j)\in A_{\text{ML}}} \sum_{i \in \mathscr{C}_{(i,j)}} l_{ij} k_{\text{jam},ij}}
 	\label{eq:zetat}
 \end{equation}
 
 $\text{JAH}_2$ can then be defined as a maximum value of $\zeta (t)$ across all time steps, as in Equation  \eqref{eq:jahstat2}. The value of $\text{JAH}_2$ varies between $[-1,1]$ with a high positive value indicating more congestion on GPLs before congestion set in the ML.
 
 \begin{equation}
 	\text{JAH}_2 = \max_{t \in \mathscr{T}} \zeta(t)
 	\label{eq:jahstat2}
 \end{equation}
 
\begin{figure}[h]%
		\centering
		\includegraphics[scale=0.5]{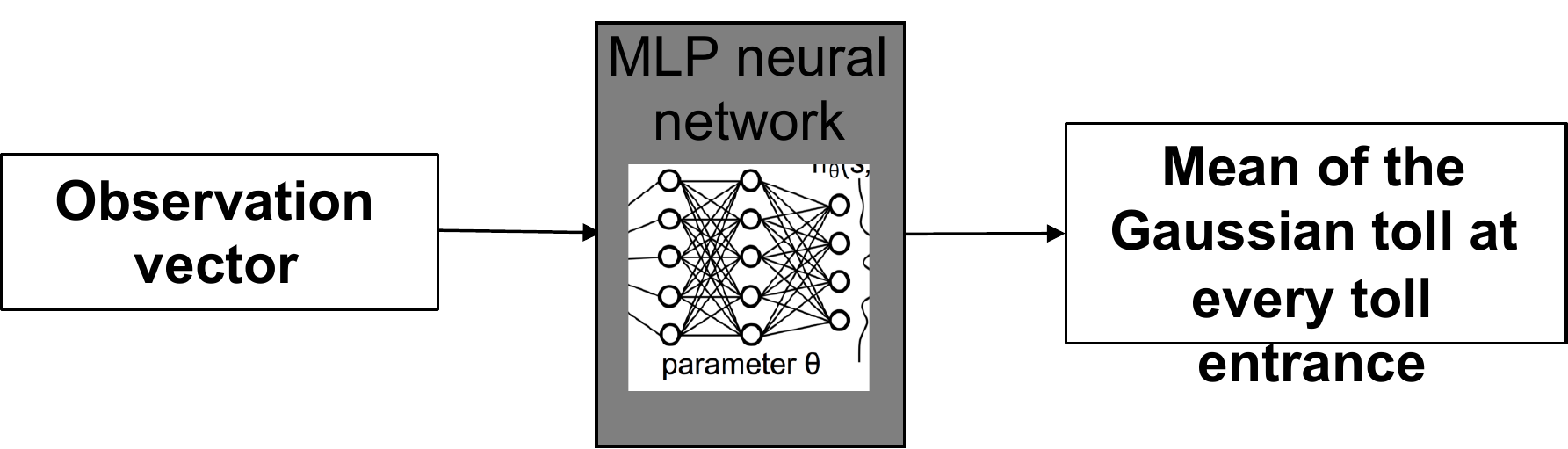}%
		\caption{Abstract representation of the policy}%
		\label{fig:policyNN}%
	\end{figure}

 For the given POMDP, a policy $\pi_{\theta}(a|\bm{o}(s))$ denotes the probability of taking action $a$ given observation $\bm{o}(s)$ in state $s$. We consider stochastic policies parameterized by a vector of real parameters $\theta$. 
 For example, for a policy replaced by a neural network, $\theta$ represents the flattened weights and biases for the nodes in the network. Since the action space for the POMDP is continuous, the neural network outputs the mean of the Gaussian distribution of tolls which is then used to sample continuous actions. For simplicity in Deep-RL training, we assume the covariance of the joint distribution of actions to be a diagonal matrix with constant diagonal terms (A$\#12)$. Figure \ref{fig:policyNN} shows a schematic of the parameterized representation of the policy which takes in the input of observations across the network and returns the mean of the Gaussian toll values for all toll links. MLP stands for multi-layer perceptron which is a feedforward neural network architecture.
 
\subsection{Episodic Reinforcement Learning}
\label{subsec:episodicRL}
In an episodic reinforcement learning problem, an agent's experience is broken into episodes, where an episode is a sequence with a finite number of states, actions, and rewards. Since the POMDP introduced in the previous subsection is finite-horizon, the simulation terminates at time $T/{\Delta t}$. Thus, an episode is formed by a sequence of states, actions, and rewards for each time step $k \in \mathscr{T}_\tau$. 

We first define a trajectory $\aleph$ as a sequence of states and actions visited in an episode, that is $\aleph = (s_0, a_0, s_1, a_1, \cdots, s_{|\mathscr{T}_\tau|-1})$, where $s_k$ is same as the state defined earlier indexed by the time $k$ in that state. Let $r(s_k, a_k)$ be denoted by $r_k$ for all $k \in \mathscr{T}_\tau$.

The goal of the RL problem is to find a policy that maximizes the expected reward over the entire episode.  The optimization problem can then be written as following:

\begin{align}
	\max_{\pi_\theta(\cdot)} & ~J(\pi_\theta) = \mathbb{E}_\aleph [R(\aleph)|\pi]\\
	R(\aleph) &= \sum_{k \in \mathscr{T}_\tau} r_k,
\end{align}

where, $\mathbb{E}_\aleph [R(\aleph)|\pi]=\int R(\aleph)p_\pi(\aleph) d\aleph$ is the expected reward over all possible trajectories obtained after executing policy $\pi$ with $p_\pi(\aleph)$ as the probability distribution of trajectories obtained by executing policy $\pi$.\footnote{Defining an expectation conditioned over a function ($\pi$) instead of a random variable is a slight abuse of notation, but is commonly used in the RL literature.} \vpB{We do not discount  future rewards because tolls are optimized over a short time period (like a day or a morning/evening peak). }

We define a few additional terms used later in the text. Let $V^{\pi}(s_k)= \mathbb{E}_\aleph \sum_{k'=k}^{| \mathscr{T}_\tau|} r_{k'}$ be the value function which evaluates the expected reward obtained from state  $s_k$ till the end of episode following policy $\pi$. Similarly, we define the Q-function, denoted by $Q^{\pi}(s_k, a_k)$, as the expected reward obtained till the end of episode from state  $s_k$ after taking action $a_k$ and following policy $\pi$ thereafter. Last, the advantage function $A^{\pi}(s_k, a_k) = Q^{\pi}(s_k, a_k) - V^{\pi}(s_k)$, defined as the difference between Q-function and value function, determines how much better or worse is an action than other actions on average, given the current policy.
 
The solution of this POMDP is a vector $\theta^*$ that determines the policy which optimizes the objective under certain constraints on the policy space. Commonly considered policy constraints for the dynamic pricing of express lanes include the following:
	\begin{enumerate}
		\item Tolls levied for a longer distance are higher than tolls levied for a shorter distance \vp{from the same entrance}: with the choice of tolling structure (assumption A$\#6$)  where tolls are charged at every diverge, this constraint is already satisfied.
		\item \vp{The ML is always operated at a speed higher than the minimum speed limit (called the speed-limit constraint): in our model, we allow violation of this constraint on the ML. We observe that, given the stochasticity in lane choice of travelers and demand, bottlenecks can occur at merges and diverges which can result in an inevitable spillover on managed lanes during congested cases. Thus, a hard constraint keeping the ML congestion free throughout the learning period is not useful. We instead quantify the violation of the speed-limit constraint using the time-space diagram of the cells on the ML. We define \texttt{\%-violation} as the proportion of cell-timestep pairs on the time-space diagram where the speed limit constraint is violated, expressed as percentage. Mathematically,} 
		\begin{align}
			\texttt{\%-violation} &= \frac{\sum_{(i,j)\in A_{\text{ML}}} \sum_{c \in \mathscr{C}_{(i,j)}} \sum_{t\in \mathscr{T}} I_c^t}{|\mathscr{T}|\sum_{(i,j)\in A_{\text{ML}}} |\mathscr{C}_{(i,j)}|} \times 100
		\end{align}
		\vp{where, $I_c^t$ is an indicator variable which is 1 if the number of vehicles in the cell $c$ in time step $t$ is higher than the desired number of vehicles in the cell and 0 otherwise. The desired number of vehicles in each cell is determined from the density corresponding to the minimum speed limit on the fundamental diagram.}
		\vp{As discussed in Section~\ref{sec:results}, allowing the speed-limit constraint to be violated in our model is not restrictive as the best-found policies for each objective have \texttt{\%-violation} values of less than 2\% for all networks tested.}
		\item Toll variation from one time step to the next is restricted: we do not explicitly model this constraint. If the tolling horizon is ``sufficiently" large (say 5 minutes), a large change in tolls from one toll update to the next can be less of a problem. In our experiments, the optimal tolls are structured and do not oscillate significantly.
		\item Tolls are upper and lower bounded by a value: we model this by clipping the toll output by the function approximator within the desired range $\left[ \beta_{\min}, \beta_{\max} \right]$.
	\end{enumerate}
Next, we discuss the solution methods used to solve the POMDP using Deep-RL methods and other heuristics.

\section{Solution Methods}

\subsection{Deep Reinforcement Learning Algorithms}
Deep reinforcement learning algorithms can be broadly categorized into value-based methods and policy-based methods. The former methods try to learn the value functions and \vp{use approaches based on dynamic programming} to solve the problem, while the latter methods try to learn the policy directly based on the observations. Policy gradient methods work well with continuous state and action spaces, making it a preferred choice for the toll optimization problem.

Derivative-free optimization and gradient-based optimization are two types of policy-based methods. \vp{We focus on the methods relying on derivatives as they are considered to be data efficient~\cite{schulman2016optimizing}.} Providing an overview of the state-of-the-art of policy gradient methods to solve reinforcement learning problems is out of the scope of this work. We refer the reader to Schulman \cite{schulman2016optimizing} for additional details. In this article, we choose two of the commonly used algorithms for solving the problem: the vanilla policy gradient (VPG) algorithm and the proximal policy optimization (PPO) method from Schulman et al.~\cite{schulman2017proximal}.

The algorithms use the derivative of the objective function with respect to  the policy parameters to improve them using stochastic gradient descent. The methods differ in calculation of the derivatives and the update of parameter $\theta$. \vpB{We can express the derivative of $J(\pi_\theta)$ with respect to $\theta$ as:}

\begin{subequations}
	\begin{align}
		\nabla_\theta J(\pi_\theta) &= \nabla_\theta \mathbb{E}_\aleph [R(\aleph)|\pi] & \\
		& = \nabla_\theta \int_{\aleph} P(\aleph|\theta)R(\aleph) d\aleph & \\
		& = \int_{\aleph} \nabla_\theta P(\aleph|\theta) R(\aleph) d\aleph & \\
		& = \int_\aleph P(\aleph|\theta) \nabla_\theta \log P(\aleph|\theta) R(\aleph) d\aleph & \left( \text{since } \nabla_\theta \log P(\aleph|\theta) = \frac{1}{P(\aleph|\theta)} \nabla_\theta P(\aleph|\theta) \right) \label{eq:subeqLogDeri}\\
		& = \mathbb{E}_{\aleph} \left[ \nabla_\theta \log P(\aleph|\theta)R(\aleph) \right] & \\
		& = \mathbb{E}_{\aleph} \left[ \sum_{k=0}^{|\mathscr{T}_\tau|} \nabla_\theta \log (\pi_\theta(a_k|s_k)) R(\aleph)\right] & \label{eq:subeqFinalExp}
	\end{align}
	\label{eq:derivativeGradExp}
	\end{subequations}
\vpB{where we first convert the probability of a trajectory into a product of the probabilities of taking certain actions in each state, and  then convert this product into a sum. As a result, the derivative in the RHS of Equation \eqref{eq:subeqFinalExp} can be easily obtained by performing back propagation on the policy neural network.}

 \vpB{The expectation in Equation \eqref{eq:subeqFinalExp}} can be approximated by averaging over a finite number of trajectories. Let $\mathscr{N}=\{\aleph_i~|~ i\in{1,2,...}\}$ be the set of trajectories obtained using policy $\pi_\theta(\cdot)$. Then, we can write:
 \begin{equation}
		\nabla_\theta J(\pi_\theta) \approx \frac{1}{|\mathscr{N}|} \sum_{\aleph \in \mathscr{N}} \left[ \sum_{k=0}^{|\mathscr{T}_\tau|} \nabla_\theta \log (\pi_\theta(a_k|s_k)) R(\aleph)\right]
		\label{eq:gradProbExp}
	\end{equation}
 
 \vp{In the above formulation the likelihood of actions taken along the trajectory is affected by reward over entire trajectory. However, it is more intuitive for an action to influence the reward obtained only after the time step when it was implemented. It can be shown that the right hand side of the expression in Equation \eqref{eq:gradProbExp} is equivalent to the following expression:}
 \begin{equation}
		\frac{1}{|\mathscr{N}|} \sum_{\aleph \in \mathscr{N}} \left[ \sum_{k=0}^{|\mathscr{T}_{\tau}|} \nabla_\theta \log (\pi_\theta(a_k|s_k)) \hat{R}(k)\right]
	\end{equation}
	
	\vp{where $\hat{R}(k)$ is the reward-to-go function at time $k$, given by $\hat{R}(k) = \sum_{k'=k}^{|\mathscr{T}_\tau|} r_{k'}$. This new expression for the gradient of the objective requires sampling of fewer trajectories and generates a low-variance sample estimate of the gradient.}
	
	Additionally, the variance can be further reduced by using the advantage function estimates instead of reward-to-go function~\cite{schulman2015high}. VPG uses the following form for approximating the derivative:

\begin{equation}
	\nabla_\theta J(\pi(\theta)) \approx \frac{1}{|\mathscr{N}|} \sum_{\aleph \in \mathscr{N}}  \left[ \sum_{k=0}^{|\mathscr{T}_{\tau}|} \nabla_\theta \log(\pi_\theta(a_k | s_k)) \hat{A}_k \right]
	\label{eq:polcygrad}
\end{equation}

where $\hat{A}_k$ is the estimate of advantage function, $A^{\pi_\theta}(s_k,a_k)$, from current time $k$ till the end of episode, following the policy from which the given trajectory is sampled. We use the generalized advantage estimation (GAE) technique to estimate $\hat{A}_k$ \vpB{which requires an estimate of the value function~\cite{schulman2015high}. We use value function approximation to estimate of $V^{\pi}(s_k)$ using a neural network as the functional approximator. Let $\hat{V}_\phi(s_k)$ denote the estimate of $V^{\pi}(s_k)$, parameterized by a real vector of parameters $\phi$. The algorithm starts with an estimate of $\phi$ ($\phi_0$) and iteratively improves it by minimizing the squared difference with the reward-to-go value from the trajectory. The update in $\phi$ parameters are evaluated using Equation \eqref{eq:VFA}:}

\begin{equation}
	\phi_{n+1} = \argmin_{\phi} \frac{1}{|\mathscr{N}||\mathscr{T}_\tau|} \sum_{\aleph \in \mathscr{N}} \sum_{k=0}^{|\mathscr{T}_\tau|} \left( V_{\phi}(s_k) - \hat{R}(k) \right) ^2
	\label{eq:VFA}
\end{equation}
More details on GAE are provided in Schulman et al.~\cite{schulman2015high}.

VPG updates the value of $\theta$ parameter from iteration $n$ to $n+1$ using the standard gradient ascent formula:

\begin{equation}
	\theta_{n+1} = \theta_n + \alpha \nabla_\theta J(\pi(\theta_n))
	\label{eq:vpgUpdate}
\end{equation}

\vp{In Equation \eqref{eq:vpgUpdate}, an inappropriate choice of the learning rate $\alpha$ can lead to large policy updates from one iteration to the next which can cause the objective values to fluctuate.} The PPO algorithm modifies the policy update to take the biggest possible improvement using the data generated from current policy while ensuring improvement in the objective. It performs specialized clipping to discourage large changes in the policy. The policy update for PPO is given by:

\begin{equation}
	\theta_{n+1} = \argmax_{\theta} \frac{1}{|\mathscr{N}|} \sum_{\aleph \in \mathscr{N}} \left[ \sum_{k=0}^{|\mathscr{T}_\tau|} \min \left( r_k(\theta)\hat{A}^{\pi_{\theta_n}}(s_k,a_k), \texttt{clip}(r_k(\theta), 1-\epsilon, 1+\epsilon)\hat{A}^{\pi_{\theta_n}}(s_k,a_k) \right) \right]
	\label{eq:polcygrad2}
\end{equation}

where $r_k(\theta)$ is the ratio of probabilities following a policy and the policy in the current iteration ($\theta_{n}$) given by Equation \eqref{eq:policygrad3}, and the \texttt{clip}($\cdot$) function, given by Equation \eqref{eq:clipFunction}, restricts the value of first argument between the next two arguments. 

\begin{equation}
	r_k(\theta) =  \frac{\pi_\theta(a_k | s_k)}{\pi_{\theta_{n}}(a_k| s_k)} 
	\label{eq:policygrad3}
\end{equation}

\begin{equation}
  \texttt{clip}(r,1-\epsilon, 1+\epsilon)=
  \begin{cases}
    1-\epsilon, & \text{if $r \leq 1-\epsilon$}\\
    r, & \text{if $1-\epsilon < r < 1+\epsilon$} \\
    1+\epsilon, & \text{if $r \geq 1+\epsilon$}.
  \end{cases}
  \label{eq:clipFunction}
\end{equation}

\vpB{The clipping operation selects the policy parameters in the next iteration such that the ratio of action probabilities in iteration $n+1$ to iteration $n$ are between $\left[ 1-\epsilon, 1+\epsilon \right]$, where $\epsilon$ is a small parameter, typically 0.01. Policy updates for PPO can be solved using the Adam gradient ascent algorithm, a variant of stochastic gradient ascent with adaptive learning rates for different parameters~\cite{kingma2014adam,schulman2017proximal}.}

The structure for both algorithms is presented in Algorithm \ref{cp2:algo:disAlgo}. \vp{For the experiments, we develop a new RL environment for macroscopic simulation of traffic similar to the current RL benchmarks (called ``gym" environments) and customize the open-source implementation of both algorithms provided by OpenAI Spinningup \cite{spinningUp2019} to work with our new environment.} 

\begin{algorithm}[H]
			\caption{Policy gradient algorithm for dynamic pricing~\cite{spinningUp2019}}
			\label{cp2:algo:disAlgo}
			\begin{algorithmic}[]
			\State Input: initialize policy parameters $\theta_0$ and value function parameters $\phi_0$
			\For $~n=0,1,2,\cdots$
			\State Collect set of trajectories $\mathscr{N}_n=\{ \aleph_n\}$ by running policy $\pi_n = \pi_{\theta_n}$ in the environment
			\State Compute rewards to go $\hat{R}_k$
			\State Compute advantage estimates using rewards-to-go and generalized advantage estimation
			\State Update policy parameters:
				\begin{itemize}[noitemsep]
					\item \textbf{VPG}: Estimate policy gradients using Equation \eqref{eq:polcygrad} and update policy parameters using Equation \eqref{eq:vpgUpdate}
					\item \textbf{PPO}: Update policy parameters by solving Equation \eqref{eq:polcygrad2} using Adam gradient ascent algorithm
				\end{itemize}
			
			\State Update value function approximation parameter (used for advantage estimation) in Equation \eqref{eq:VFA} using Adam gradient descent			
		\EndFor
			
			\end{algorithmic}
		\end{algorithm}

\subsection{Feedback Control Heuristic}
We compare the performance of Deep-RL algorithms against a feedback control heuristic based on the measurement of total number of vehicles in the links on ML. We customize the \texttt{Density} heuristic in Pandey and Boyles~\cite{pandey2018dynamic} to charge varying tolls for different toll links.

Define $\text{ML}(i,j)$ as the set of links on the ML used by a traveler upon first entering the ML using the toll link $(i,j)\in A_{\text{toll}}$ \vpB{until the next merge or diverge}. For the network in Figure \ref{fig:MEME}, $\text{ML}(a,b)=\{ (b,d)\}$, $\text{ML}(c,d)=\{ (d,f)\}$, $\text{ML}(f,i)=\{ (f,i)\}$, and $\text{ML}(h,i)=\{ (i,k)\}$. \vpB{This definition allows the sets $\text{ML}(i,j)$ to be mutually exclusive and exhaustive in the space of all links on the ML. That is,}
\begin{align*}
	\text{ML}(i,j) \cap \text{ML}(k,l) &= \Phi & \forall~(i,j)\in A_{\text{toll}}, (k,l)\in A_{\text{toll}}, (i,j)\neq (k,l) \\
	\bigcup_{(i,j)\in A_{\text{toll}}} \text{ML}(i,j) &= A_{\text{ML}} &
\end{align*} 

\vpB{We assume that the feedback control heuristic updates the tolls for each toll link $(i,j)\in A_{\text{toll}}$ based on the density observations on links in $\text{ML}(i,j)$, that is, detectors are installed on each link in the ML and only those detectors are used to update the toll (A$\#13$).} The toll value for an update time $(k+1) \in \mathscr{T}_\tau$ is based on the toll value in the previous update step adjusted by the difference between the desired and current numbers of vehicles. The toll update is given by  Equation \eqref{eq:densityHeuristic}. 

\begin{equation}
				\beta_{ij}(t_{(k+1)\Delta \tau})  = \beta_{ij}(t_{k\Delta \tau})+P \times \left( X_{\text{ML}(i,j)}(k)-X_{\text{ML}(i,j)}^{\text{desired}} \right)
				\label{eq:densityHeuristic}
\end{equation}

where $X_{\text{ML}(i,j)}(k)$ is the total number of vehicles on links in $\text{ML}(i,j)$ before updating tolls at time $k+1$ and $X_{\text{ML}(i,j)}^{\text{desired}}$ be the desired value of the number of vehicles on the links in $\text{ML}(i,j)$. $P$ is the regulator parameter, with units $\$/$veh, controlling the influence of difference between the desired and current number of vehicles on the toll update. A typical desired value is the number of vehicles corresponding to the critical density on the ML link. We generalize the desired number of vehicles by defining $X_{\text{ML}(i,j)}^{\text{desired}}$ as:

\begin{equation}
	X_{\text{ML}(i,j)}^{\text{desired}} = \sum_{(g,h)\in \text{ML}(i,j)} \eta k_{\text{critical},(g,h)}l_{gh}
\end{equation}

where, $k_{\text{critical},(g,h)}$ is the critical density for link $(g,h)\in A$ and $\eta$ is the scaling parameter varying between $(0,1]$ that sets the desired number of vehicles to a proportion value of the number of vehicles at critical density. We calibrate the feedback control heuristic for different values of desired density and regulator parameter. \vp{In principle, both $\eta$ and $P$ can vary with time and the toll location; however, determining the ``optimal" variability in these parameters is a control problem in itself, exploring which is left as part of the future work ($\text{FW}\#3$).}

We do not include other algorithms for comparison because of lack of compatibility \vp{due to the full-observability assumption}. The algorithms in Zhu and Ukkusuri~\cite{zhu2015reinforcement} and Pandey and Boyles~\cite{pandey2018dynamic} do not scale for continuous action space and tolls. Comparing the performance of Deep-RL methods against the hybrid MPC method in Tan and Gao~\cite{tan2018hybrid} requires extensive analysis and will be a part of the future work ($\text{FW}\#4$).

\section{Experimental Analysis}
\label{sec:results}

\subsection{Preliminaries}
We conduct our analysis on four different networks. The first is a network with single entrance and single exit (SESE) commonly used in the managed lane pricing literature. The next two are the double entrance single exit (DESE) network and the network for toll segment 2 of the LBJ TEXpress lanes in Dallas, TX (LBJ). The DESE network includes two toll locations for modeling \textit{en route} lane changes. The LBJ network has four toll locations. Last is the network of the northbound Loop 1 (MoPac) Express lanes in Austin, TX. The MoPac network has three entry locations to the express lanes and two exit locations. 

\begin{figure}[h]
	\includegraphics[scale=0.45]{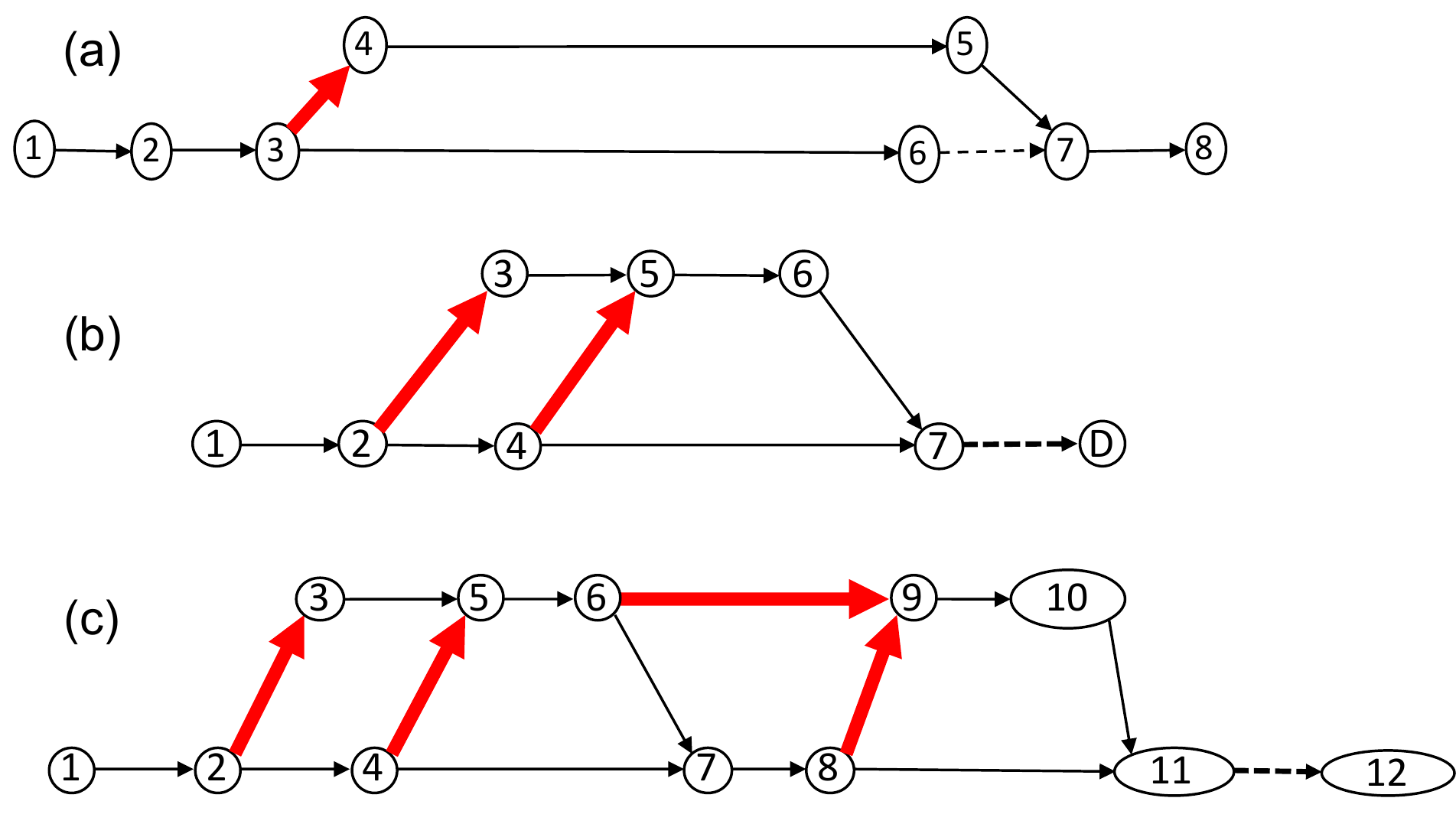}
	\includegraphics[scale=0.45]{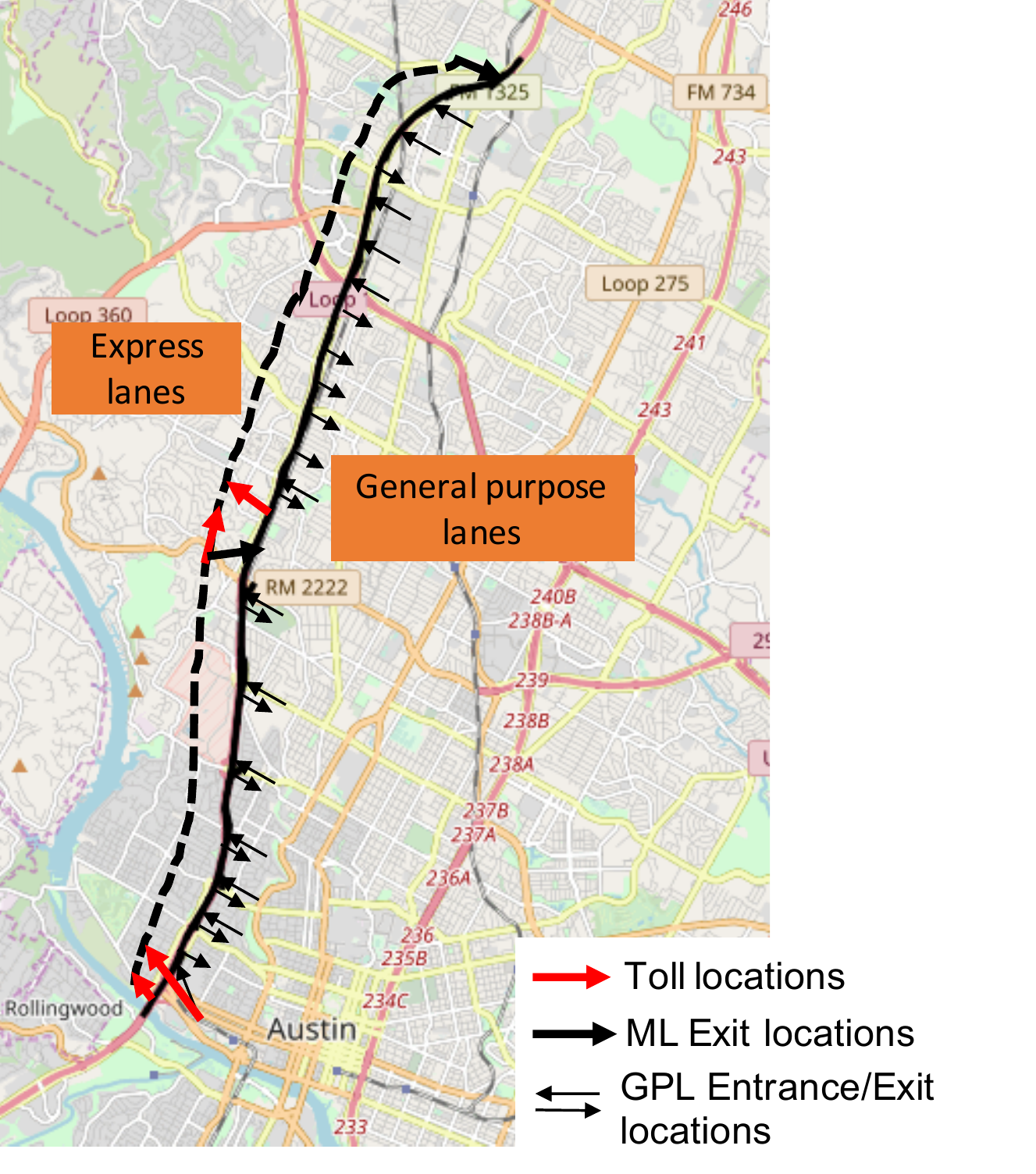}
	\centering
	\caption{Abstract representation of (a) single entrance single exit (SESE) network, (b) double entrance single exit (DESE) network, (c) LBJ network, and (d) Northbound MoPac express lane network (latitude-longitude locations of express lanes are shifted to the left to show the locations of toll points and exits from the managed lane). The tolls are collected on the links with higher thickness.}
	\label{fig:allNetworks}
\end{figure}

\begin{figure}[h]
\centering
\subfloat[\label{fig:demandDist}]{\includegraphics[width=0.45\textwidth]{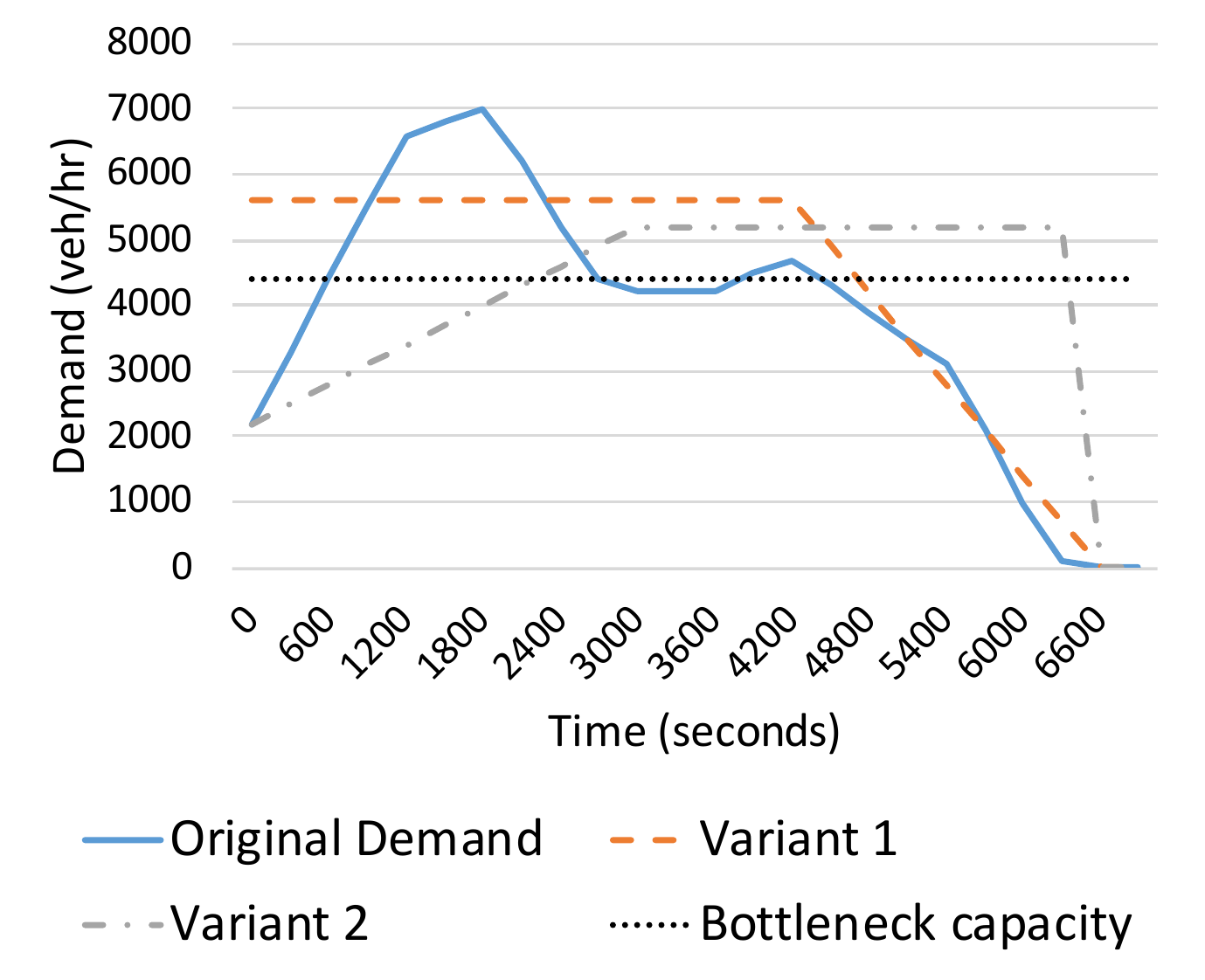}}\hfill
\subfloat[\label{fig:votDist}] {\includegraphics[width=0.55\textwidth]{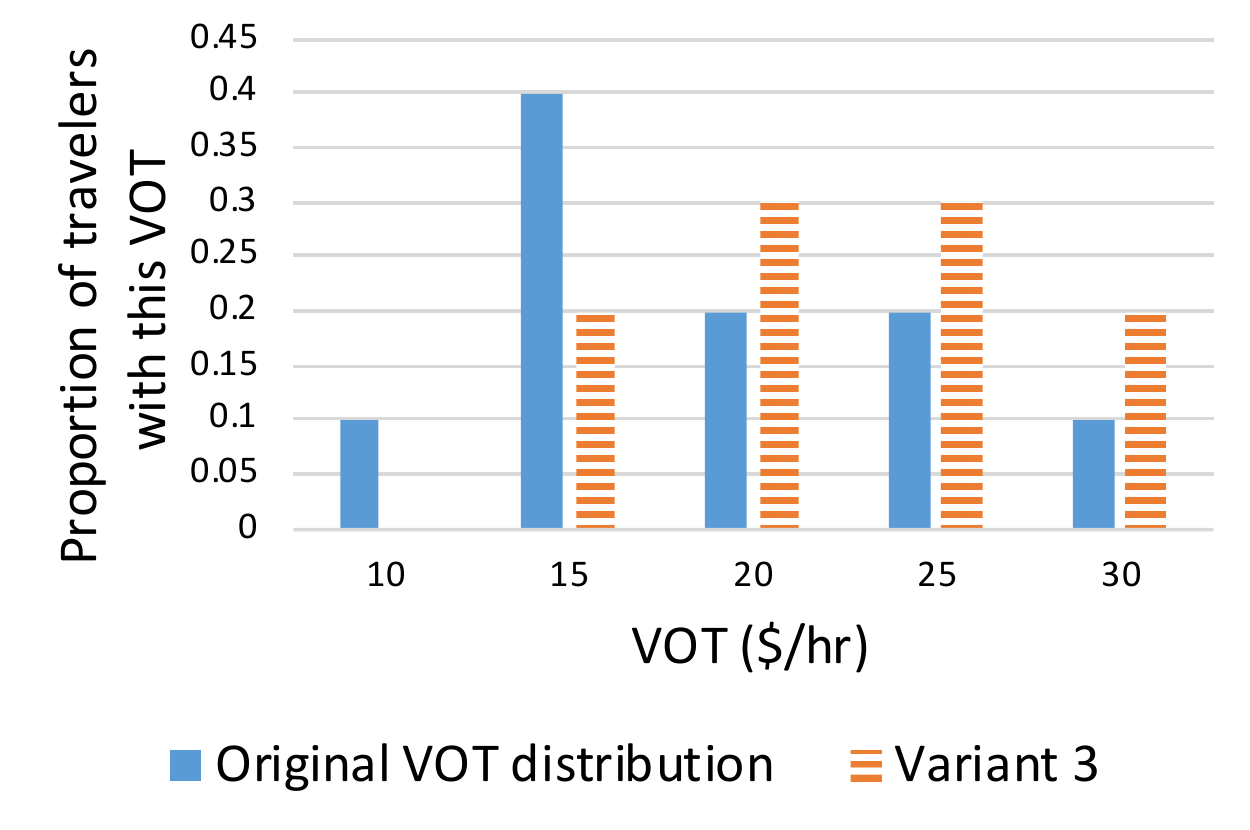}}\hfill
\caption{(a) Demand distributions used for the SESE, DESE and LBJ networks and its variants, and (b) VOT distribution and its variant} \label{fig:variantsTransferability}
\end{figure}

Figure \ref{fig:allNetworks} shows the networks, where the thick lines denote the links where tolls are collected. The demand distribution for the first three networks is artificially generated and follows a two-peak pattern (refer to the original demand curve in Figure \ref{fig:demandDist}), while the demand for the MoPac network is derived from a dynamic traffic assignment model of the Travis County region. There are a total of $105$ origin-destination pairs in the MoPac network with a total demand of $49{,}273$ vehicles using the network in three hours of the evening peak.

Table \ref{tab:parameters} shows the values of parameters used for different networks. Five VOT classes were selected for each network and the same VOT distribution was used. Figure \ref{fig:votDist} shows this VOT distribution (labelled ``original"; in some experiments we vary this distribution.)

\begin{table}[H]
\centering
\caption{Values of parameters used in the simulation}
\label{tab:parameters}
\begin{tabular}{|c|c|c|c|c|
>{\columncolor[HTML]{C0C0C0}}c |c|c|}
\hline
 & SESE & DESE & LBJ & MoPac & \cellcolor[HTML]{C0C0C0} & Parameter & Value \\ \cline{1-5} \cline{7-8} 
Corridor length (miles) & 7.3 & 1.59 & 2.91 & 11.1 & \cellcolor[HTML]{C0C0C0} & $\beta_\text{min}$ & \$0.1 \\ \cline{1-5} \cline{7-8} 
Simulation duration (hour) & 2 & 2 & 2 & 3 & \cellcolor[HTML]{C0C0C0} & $\beta_\text{max}$ & \$4.0 \\ \cline{1-5} \cline{7-8} 
$\Delta \tau$ (seconds) & 60 & 300 & 300 & 300 & \cellcolor[HTML]{C0C0C0} & $q_{ij}$ (vphpl) & 2200 \\ \cline{1-5} \cline{7-8} 
$\nu_{ij}$ (mph) & 55 & 55 & 55 & 65 & \cellcolor[HTML]{C0C0C0} & $k_{\text{jam},ij}$ (veh/mile) & 265 \\ \cline{1-5} \cline{7-8}
$\sigma_o$ (veh/hr) & 50 & 50 & 50 & 50 & \cellcolor[HTML]{C0C0C0} & $\nu_{ij}/w_{ij}$ & 3 \\ \cline{1-5} \cline{7-8}  
$\sigma_d$ (veh/hr) & 10 & 0 & 0 & 100 & \multirow{-7}{*}{\cellcolor[HTML]{C0C0C0}} & $\Delta t$ (seconds) & 6 \\ \hline
\end{tabular}
\end{table}

A feedforward multilayer perceptron was selected as the neural network. Hyperparameter tuning was conducted, and the architecture with two hidden layers and 64 nodes in each layer was selected. For the MoPac network, three hidden layers with 128 nodes each were selected. The values of other hyperparameters for Deep-RL training are as follows: learning rate for policy update equals $10^{-4}$, learning rate for value function updates is $10^{-3}$, number of iterations for value function updates is $80$, and the $\gamma^{\text{GAE}}$ and $\lambda^{\text{GAE}}$ values for the GAE method are $0.99$ and $0.97$, respectively. Each network was simulated for a number of iterations ranging between $100$ and $200$ where the average in each iteration was reported over $10$ episodes.

\subsection{Validating JAH Statistics}
In this subsection, we discuss how the JAH statistics  defined in Equations \ref{eq:jahStat1} and \ref{eq:jahstat2} are meaningful in capturing the jam-and-harvest nature of the revenue maximizing profiles. We simulate random toll profiles on the LBJ network and record the congestion profiles for three values of $\text{JAH}_2$: $0.22, 0.33,$ and $0.49$.\footnote{The $\text{JAH}_2$ values varied between $0.2$ and $0.5$ for this network as shown in Figure \ref{fig:allParetoPlots}}

\begin{figure}[H]
\centering
\subfloat[\label{fig:point22_ML_TSD}]{\includegraphics[width=0.34\textwidth]{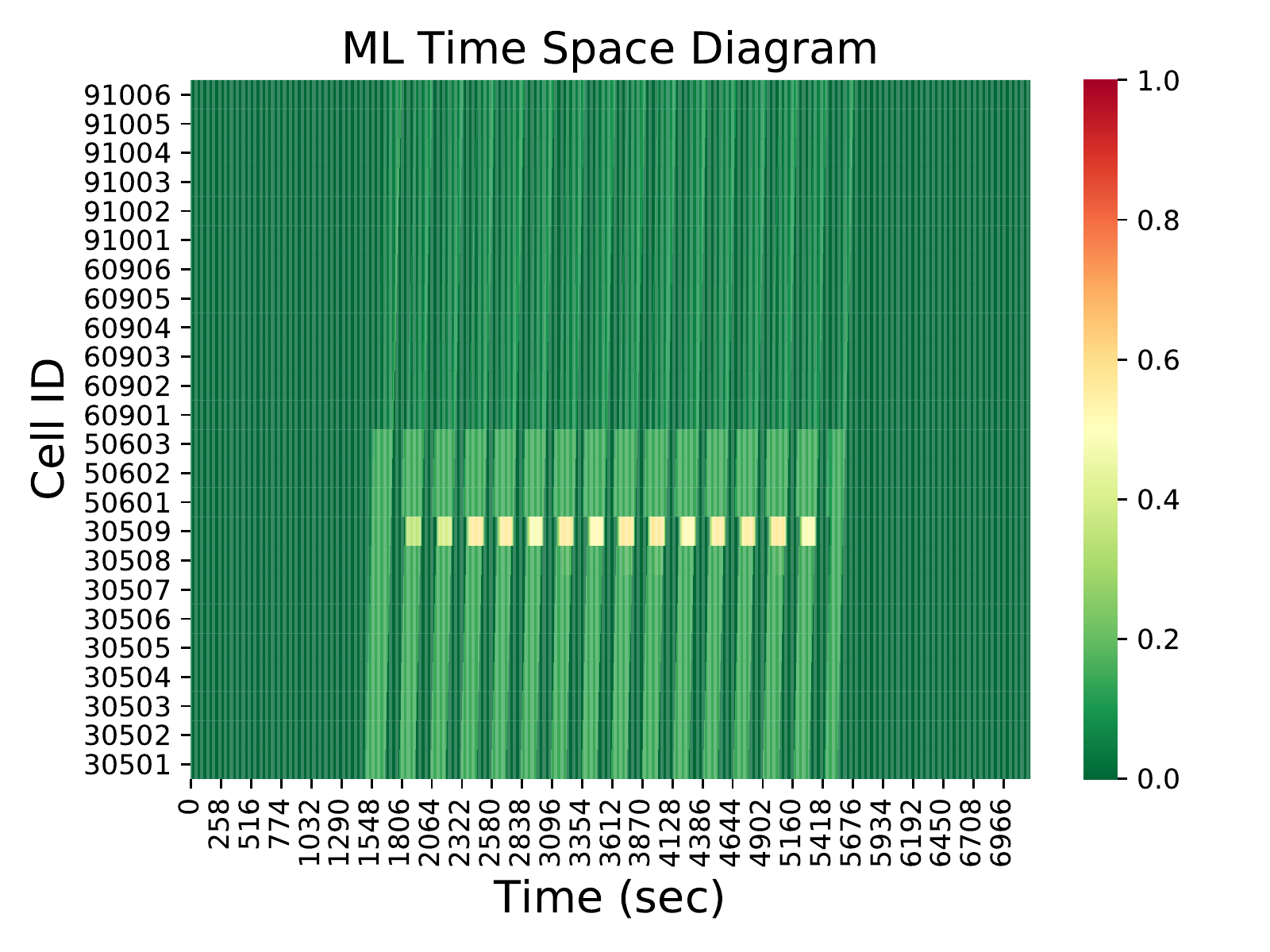}}\hfill
\subfloat[\label{fig:point22_GPL_TSD}] {\includegraphics[width=0.34\textwidth]{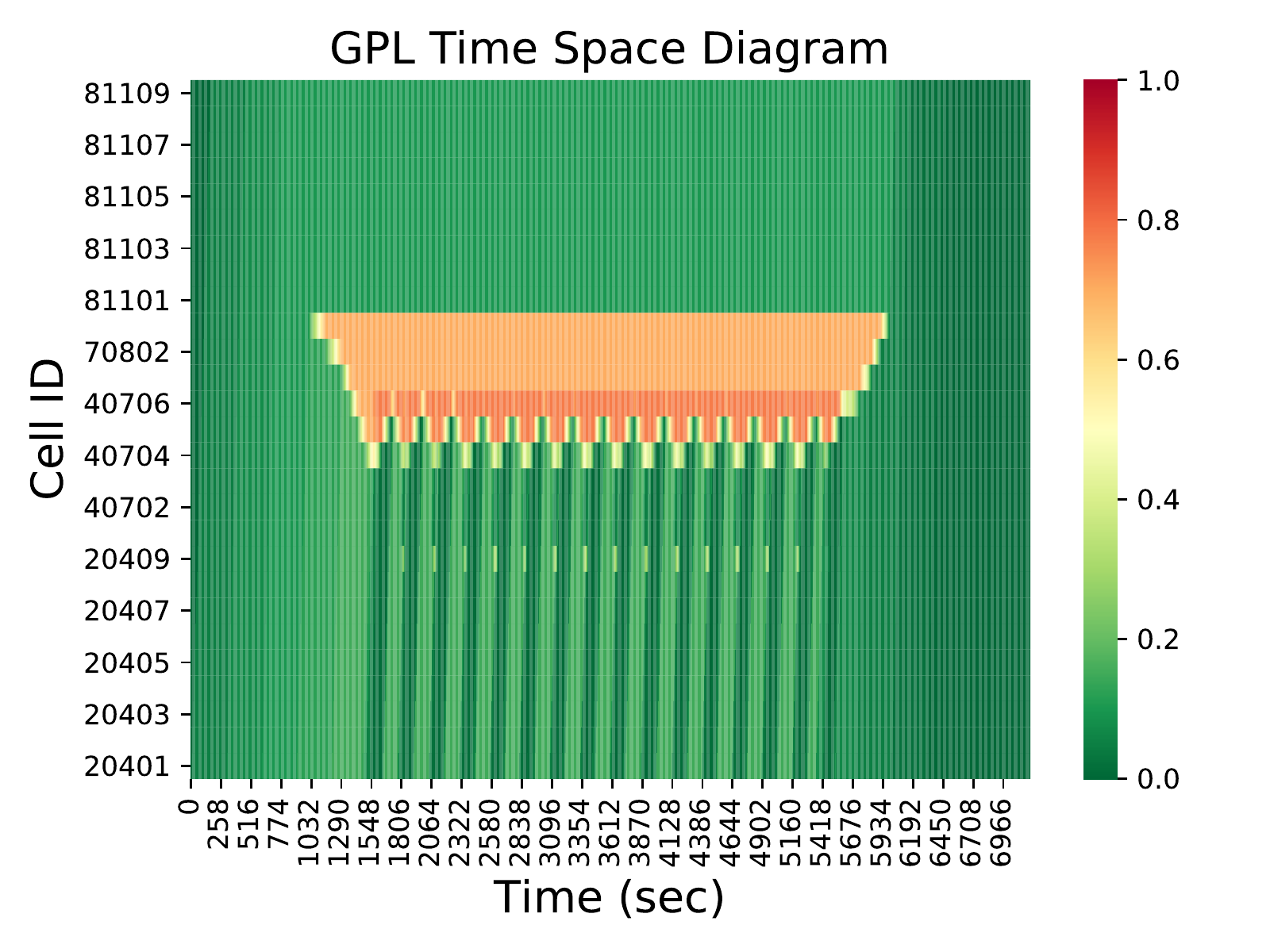}}\hfill
\subfloat[\label{fig:point22_zeta}] {\includegraphics[width=0.3\textwidth]{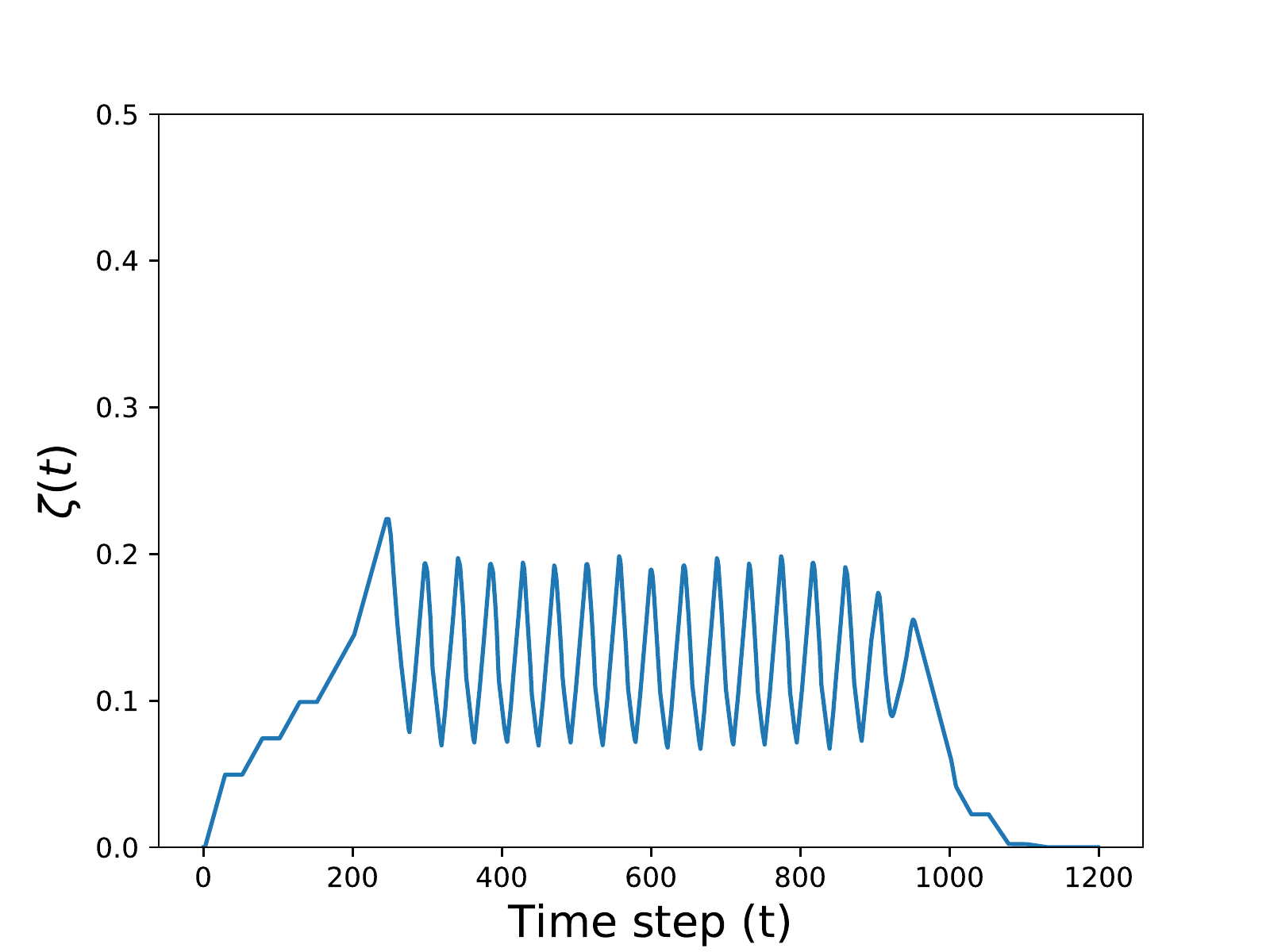}}\hfill
\caption{Plots for $\text{JAH}_2=0.22$} 
\label{fig:jahValidationPoint22}
\end{figure}

\begin{figure}[H]
\centering
\subfloat[\label{fig:point33_ML_TSD}]{\includegraphics[width=0.34\textwidth]{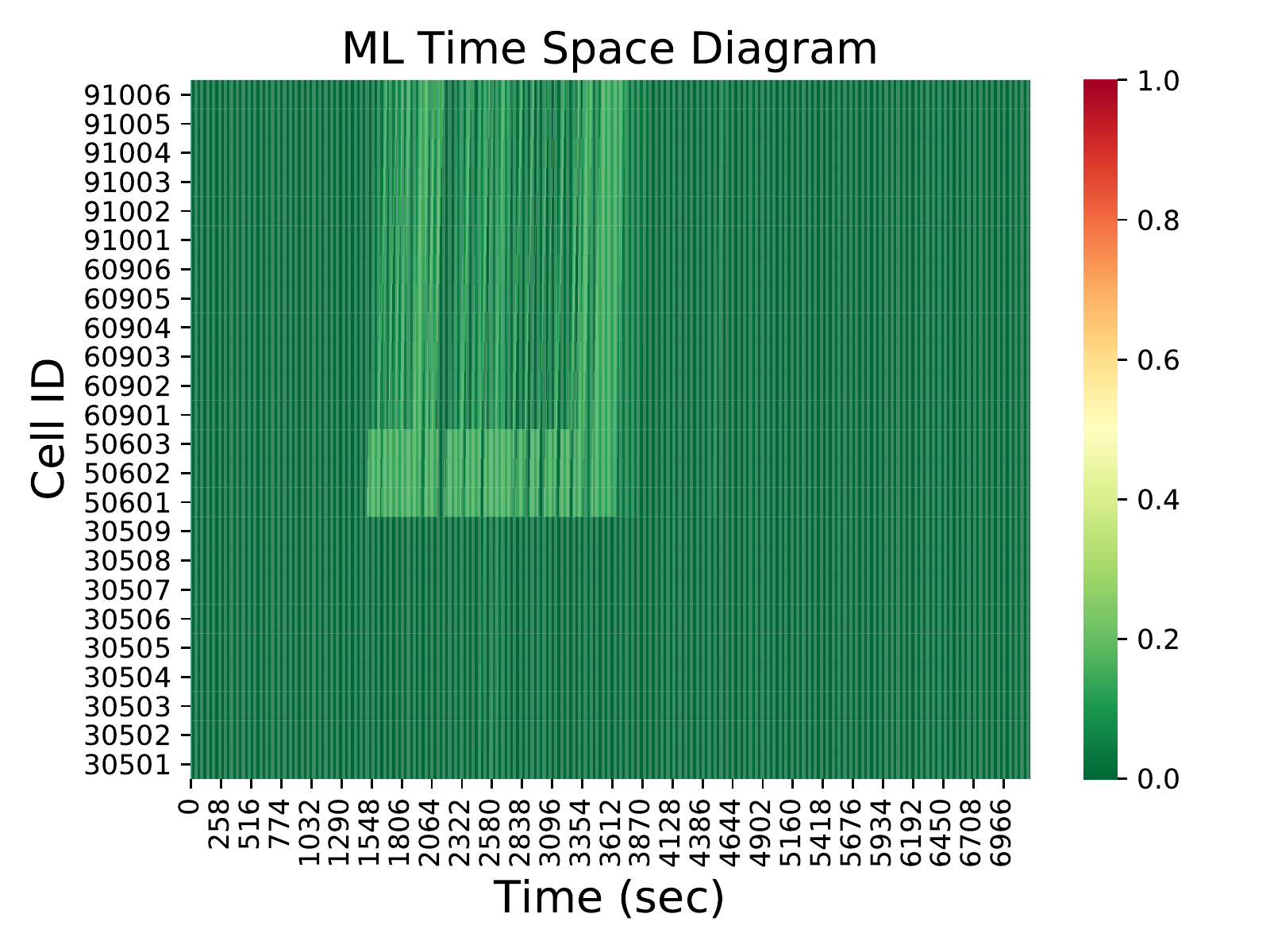}}\hfill
\subfloat[\label{fig:point33_GPL_TSD}] {\includegraphics[width=0.34\textwidth]{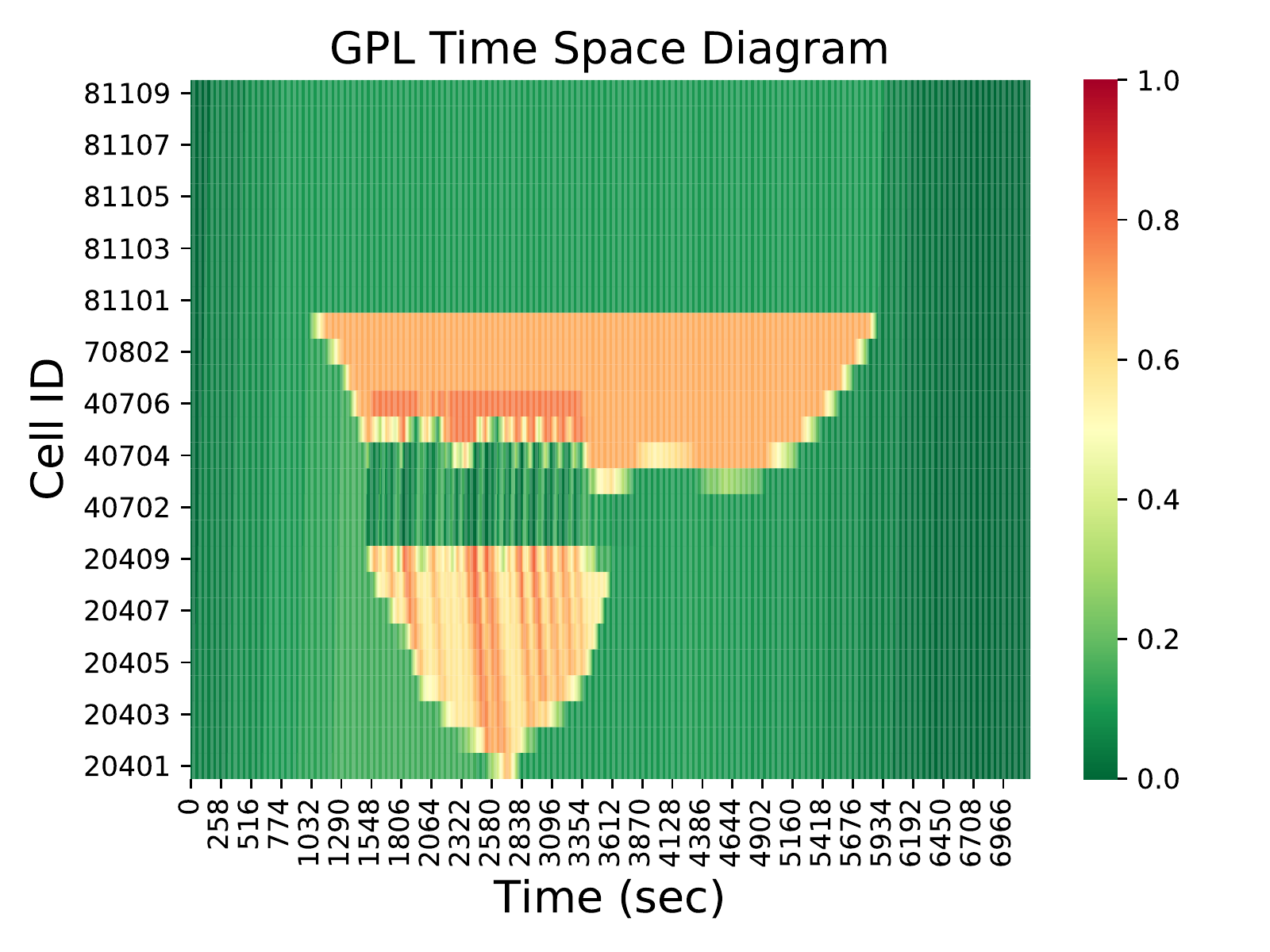}}\hfill
\subfloat[\label{fig:point33_zeta}] {\includegraphics[width=0.3\textwidth]{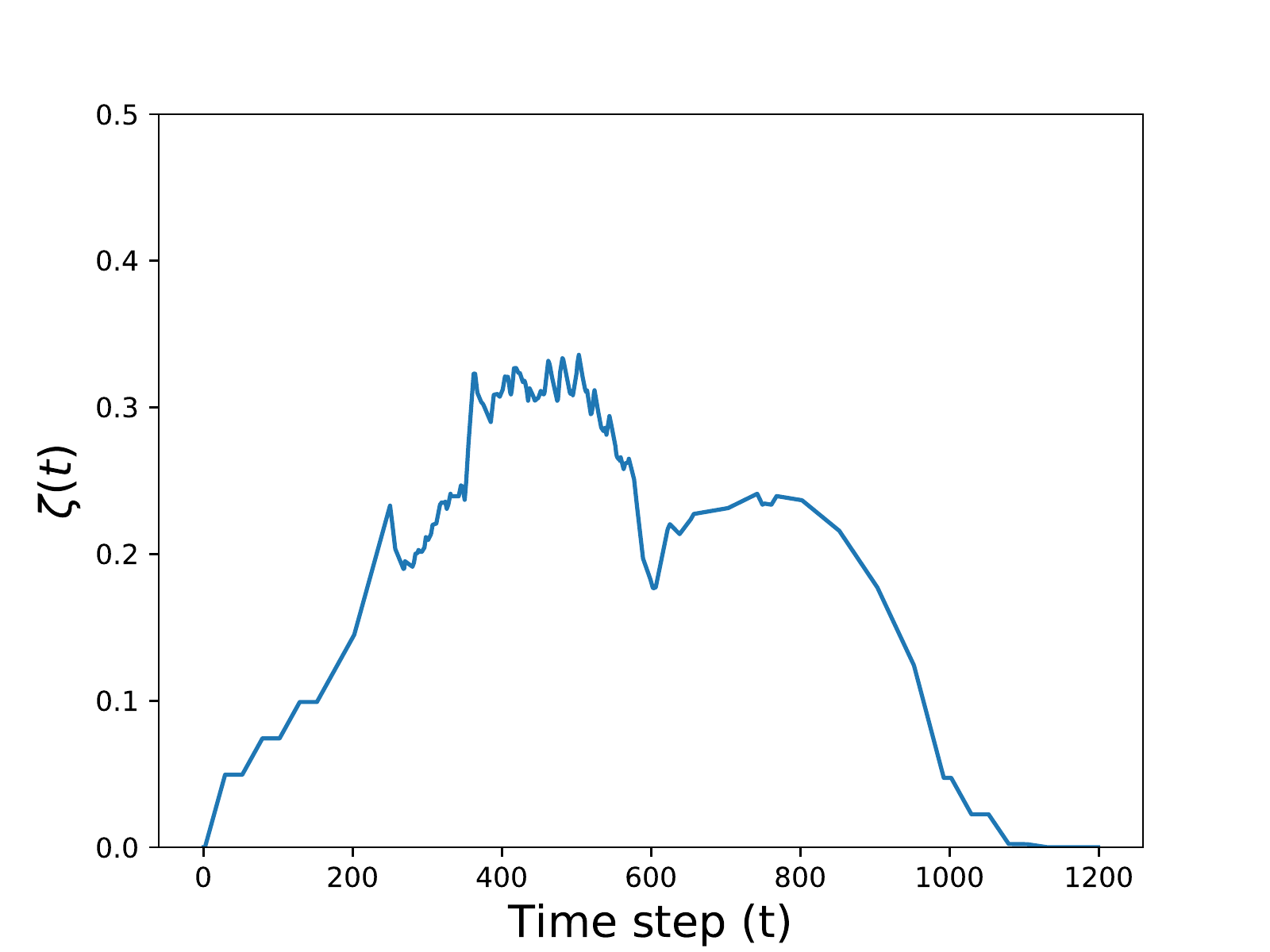}}\hfill
\caption{Plots for $\text{JAH}_2=0.33$} 
\label{fig:jahValidationPoint33}
\end{figure}

\begin{figure}[H]
\centering
\subfloat[\label{fig:point49_ML_TSD}]{\includegraphics[width=0.34\textwidth]{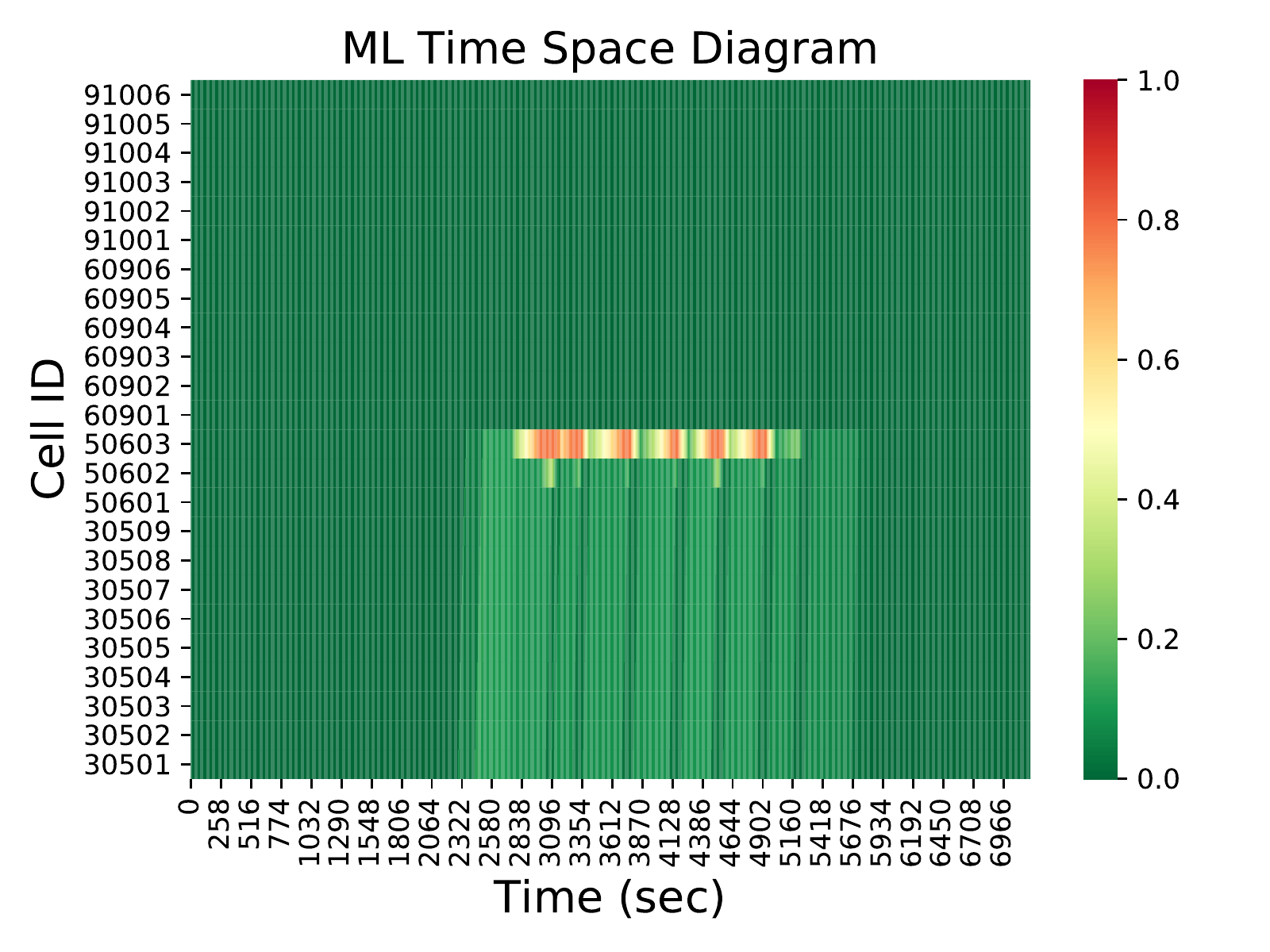}}\hfill
\subfloat[\label{fig:point49_GPL_TSD}] {\includegraphics[width=0.34\textwidth]{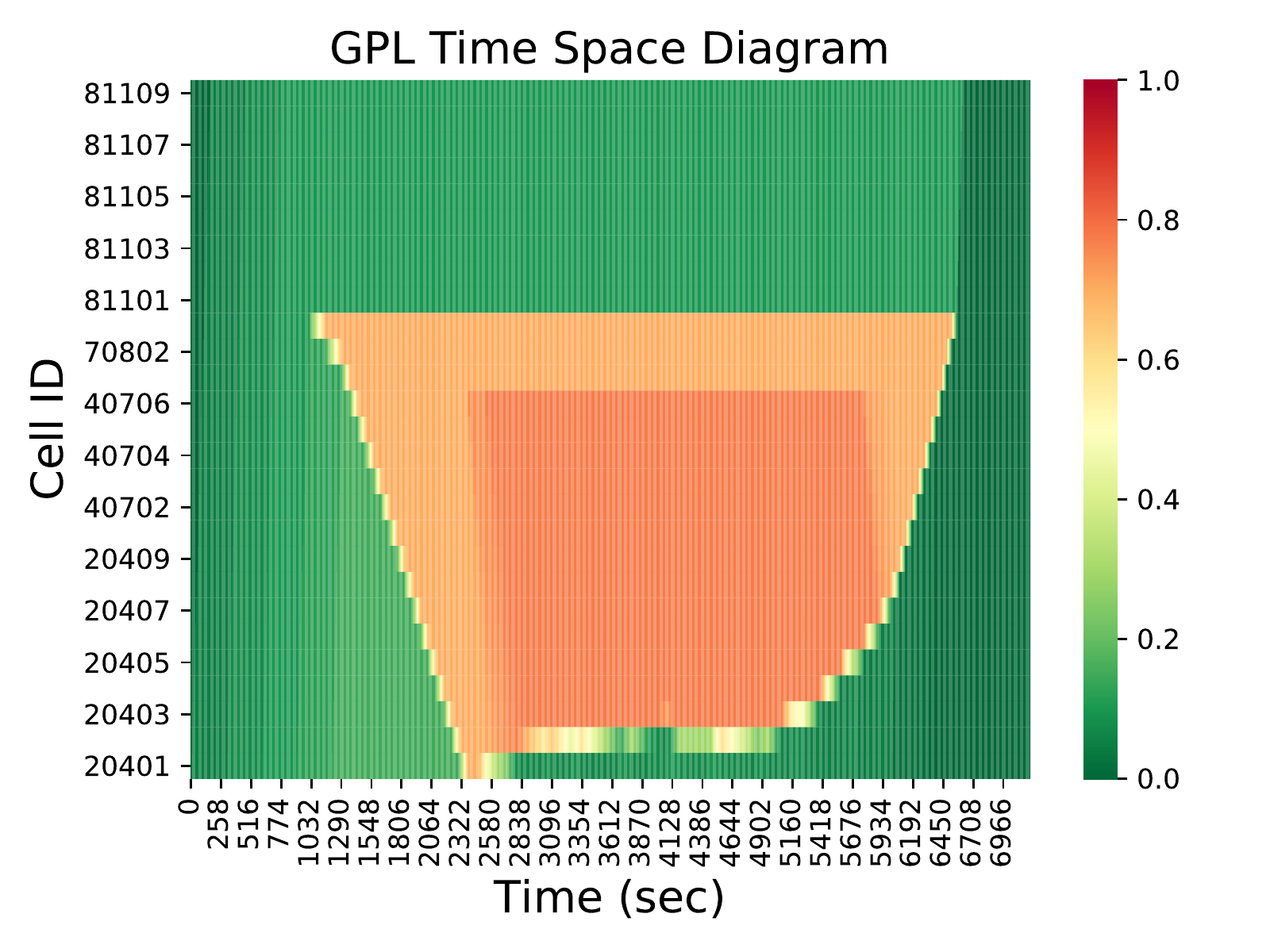}}\hfill
\subfloat[\label{fig:point49_zeta}] {\includegraphics[width=0.3\textwidth]{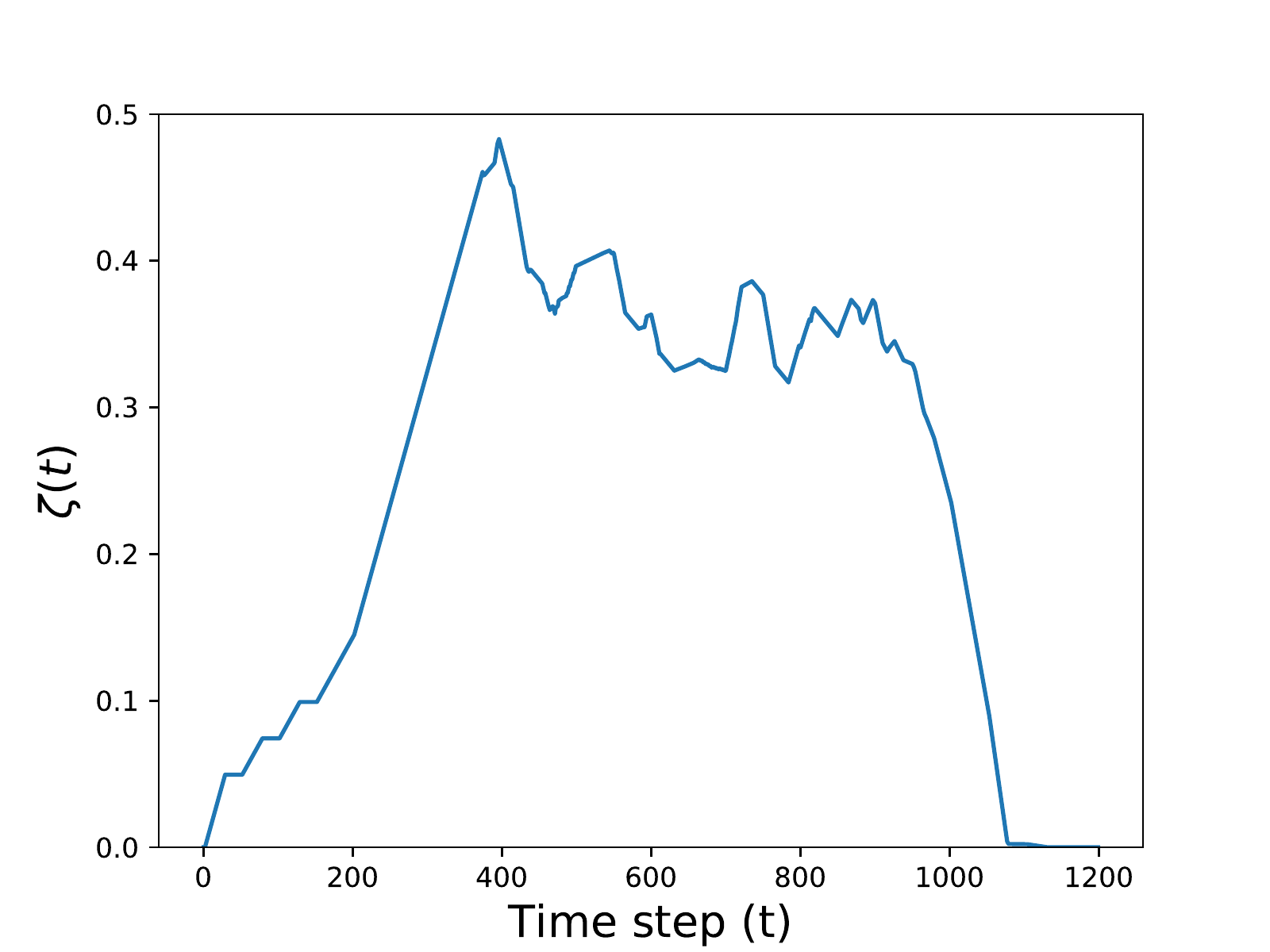}}\hfill
\caption{Plots for $\text{JAH}_2=0.49$} 
\label{fig:jahValidationPoint49}
\end{figure}

 Figures \ref{fig:jahValidationPoint22}, \ref{fig:jahValidationPoint33}, and \ref{fig:jahValidationPoint49} show the plots for the time space diagram on managed lane and general purpose lane, and the variation of $\zeta(t)$ for three different toll profiles leading to $\text{JAH}_2$ values of $0.22, 0.33,$ and $0.49$,  respectively. \vp{The scale on the time-space diagrams varies from $0$, representing no vehicles, to $1$, representing jam density.} The cell id value on the y-axis is a six-digit number where the first two digits are the tail node of the link, the second two digits are the head node of the link, and the last two digits are the index of the cell number on the link starting from index $1$ for the first cell near the tail node. Thus, the increasing value of cell IDs on the y-axis indicates the downstream direction. 

As observed, higher value of JAH statistics results in higher congestion on the GPL relative to the ML. When $\text{JAH}_2=0.22$, vehicles use the ML starting from $1500$ seconds into the simulation. Whereas, when $\text{JAH}_2=0.49$, vehicles do not enter the managed lane until approximately $2300$ seconds into the simulation, by which the GPLs are heavily congested, indicating more jam-and-harvest behavior .

Table \ref{tab:jahStatsTabular} shows values of revenue, TSTT, and $\text{JAH}_1$ for the three toll profiles simulated. We see that the $\text{JAH}_1$ statistic is also high when the $\text{JAH}_2$ statistic is high. The highest revenue is obtained for the highest value of $\text{JAH}_2$ value; however, it is not necessary that a toll profile with high $\text{JAH}_2$ produces more revenue. TSTT values follow the reverse trend as the revenue; high JAH statistic leads to low TSTT except for the case of Figure \ref{fig:jahValidationPoint33}.

\begin{table}[h]
		\centering
		\caption{Value of different statistics for different cases}
		\label{tab:jahStatsTabular}
		\begin{tabular}{|c|c|c|c|c|}
		\hline
		Figure & Revenue (\$) & TSTT (hr) & $\text{JAH}_1$ (vehicles) & $\text{JAH}_2$ \\ \hline
		Figure \ref{fig:jahValidationPoint22} & $1203.68$ & $1018.7$ & $451.73$ & $0.22$ \\ \hline
		Figure \ref{fig:jahValidationPoint33} & $957.12$ & $823.27$ & $721.20$ & $0.33$ \\ \hline
		Figure \ref{fig:jahValidationPoint49} & $4106.03$ & $1421.05$ & $997.23$ & $0.49$ \\ \hline
		\end{tabular}
\end{table}

These experiments help quantify the abstract ``jam-and-harvest" nature used in the literature and will later be used to generate toll profiles with low $\text{JAH}_i$ values ($i=\{1,2\}$). We discuss more about the variation of multiple objectives for different toll profiles in Section \ref{subsec:multiObjOpt}.

\subsection{Learning Performance of Deep-RL}

\subsubsection{Learning for different objectives}
\begin{figure}
\centering
\subfloat[SESE Revenue Maximization.\label{fig:sese_revMax}]{\includegraphics[width=0.5\textwidth]{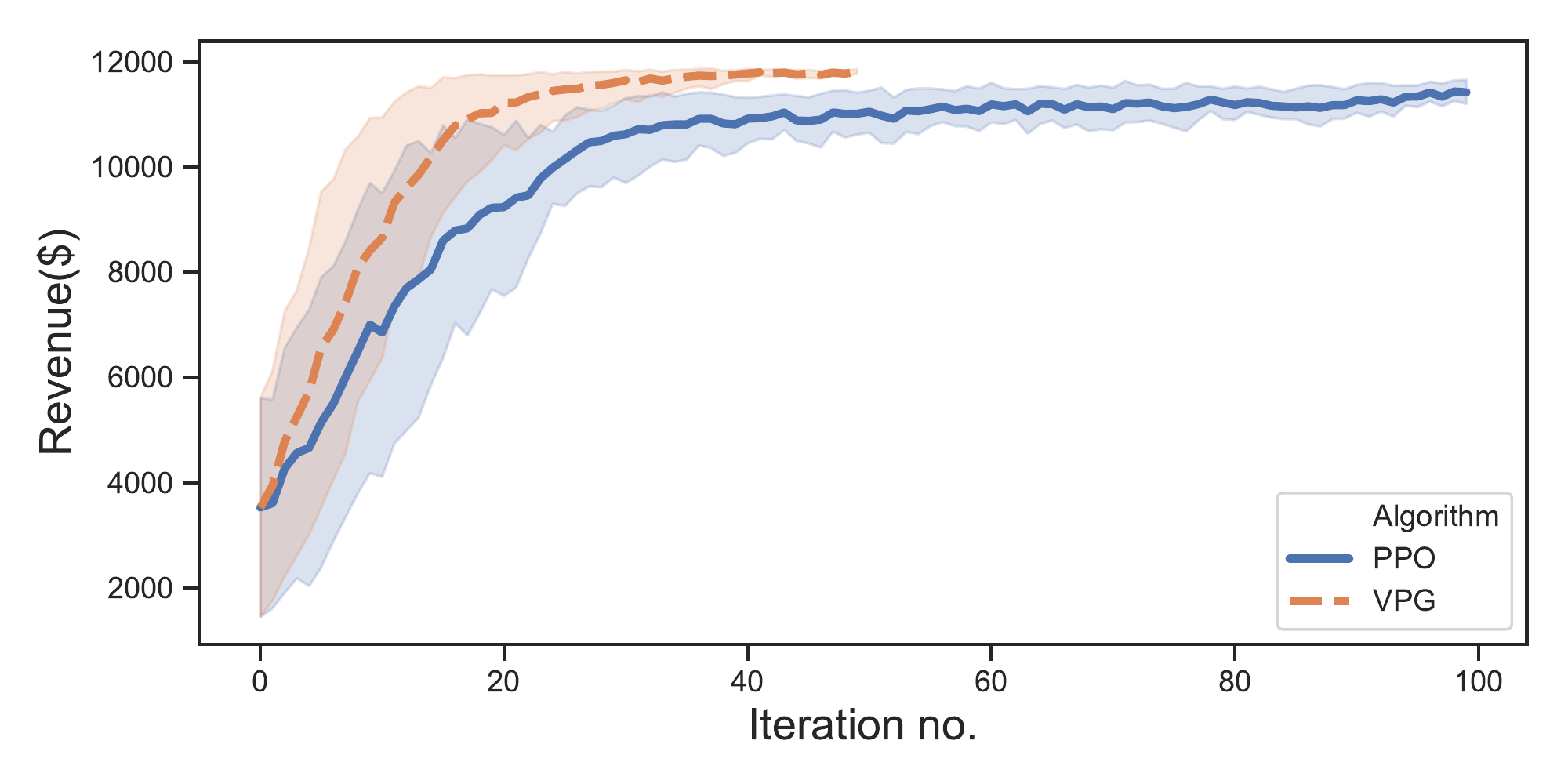}}\hfill
\subfloat[SESE TSTT Minimization.\label{fig:sese_tsttMin}] {\includegraphics[width=0.5\textwidth]{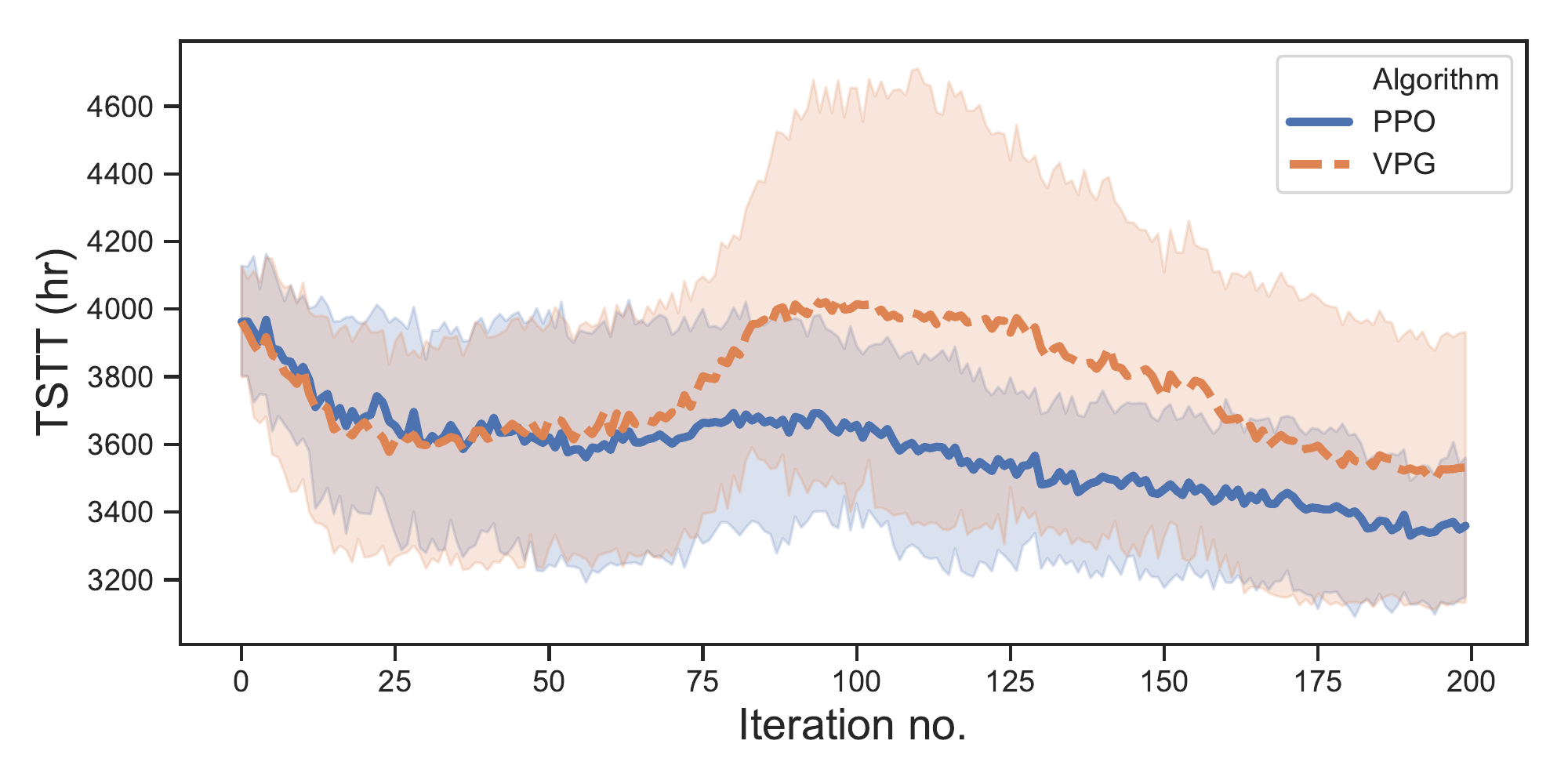}}\hfill
\subfloat[DESE Revenue Maximization.\label{fig:dese_revMax}]{\includegraphics[width=0.5\textwidth]{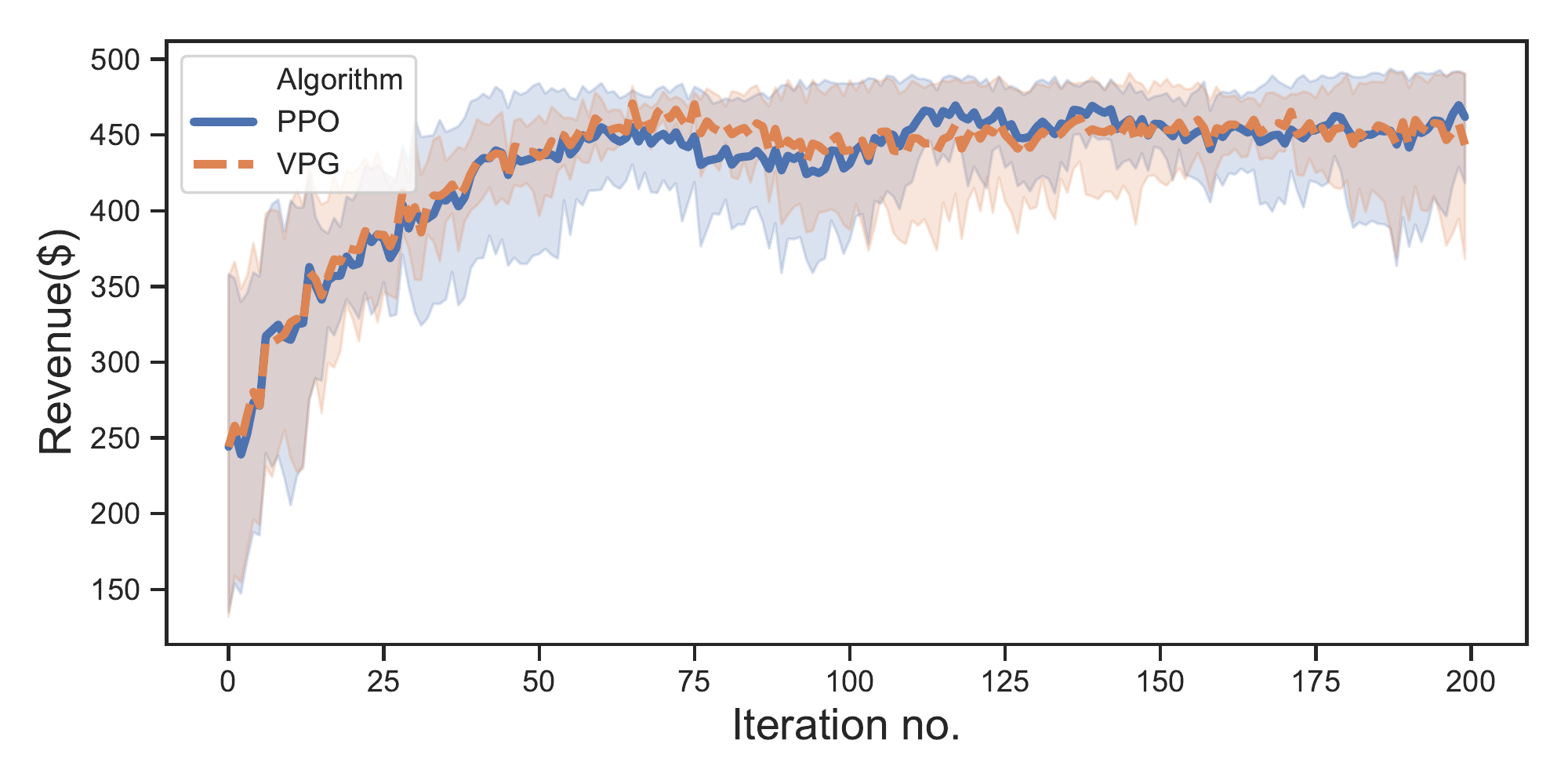}}\hfill
\subfloat[DESE TSTT Minimization.\label{fig:dese_tsttMin}] {\includegraphics[width=0.5\textwidth]{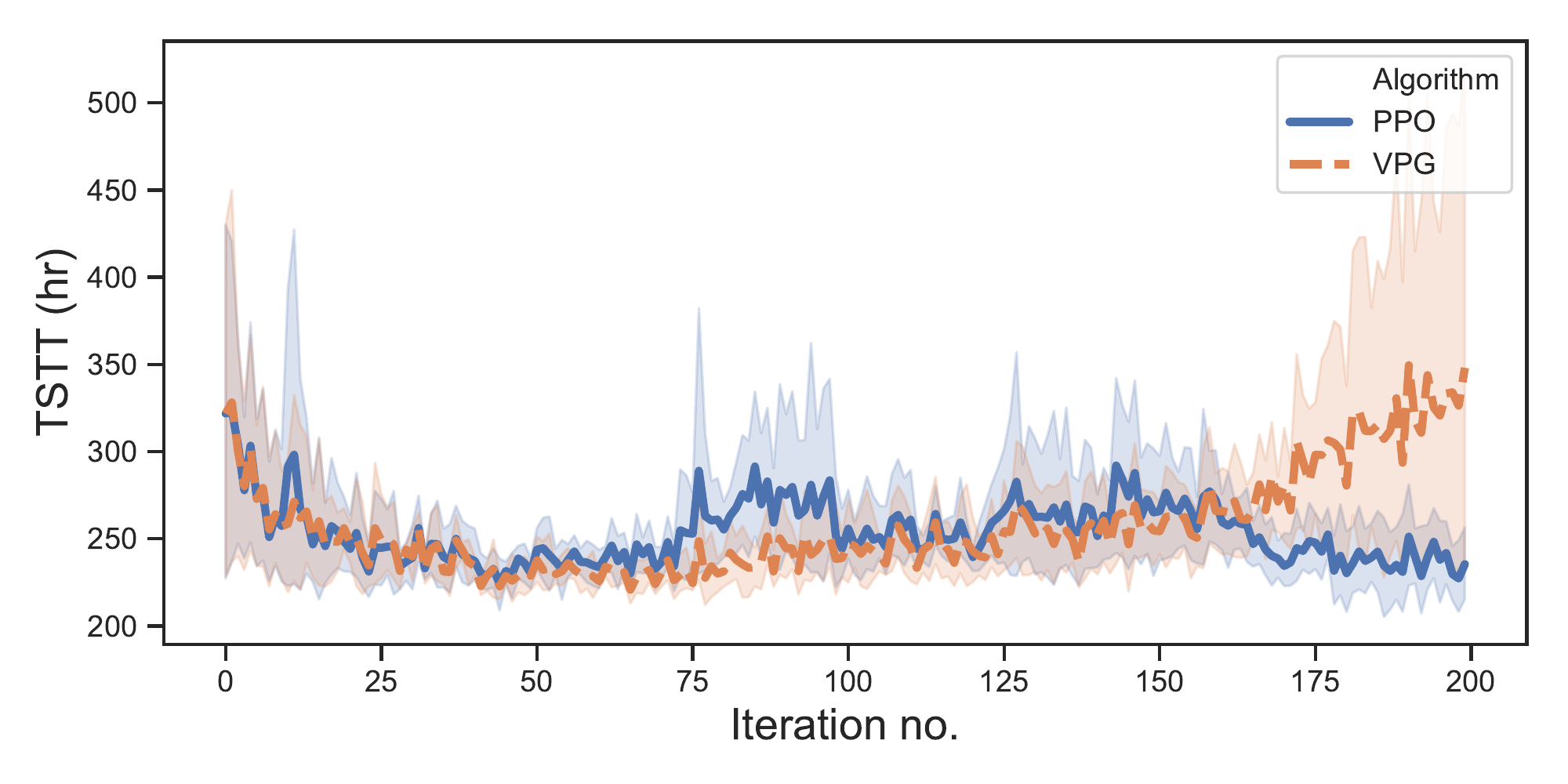}}\hfill
\subfloat[LBJ Revenue Maximization.\label{fig:lbj_revMax}]{\includegraphics[width=0.5\textwidth]{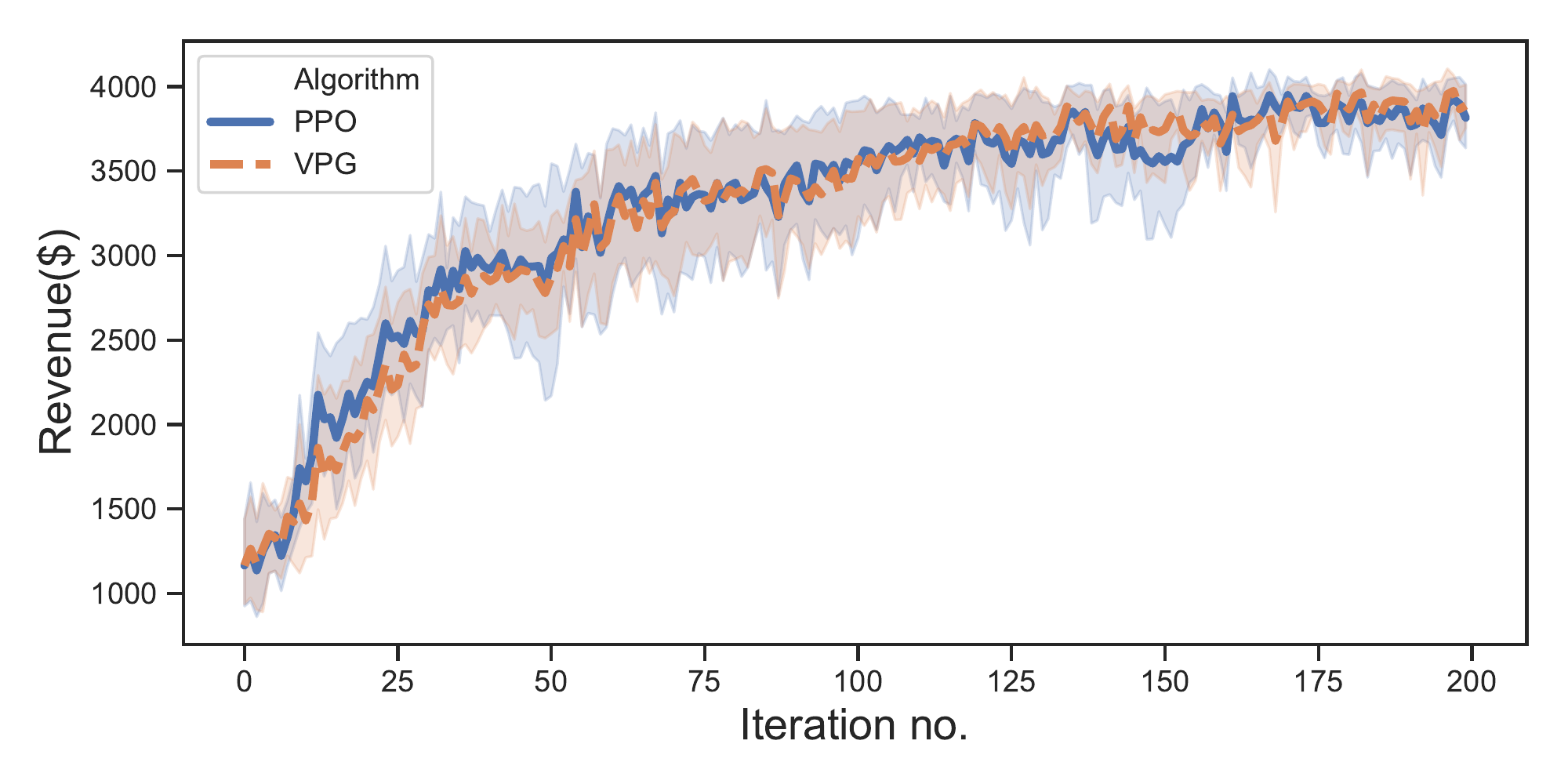}}\hfill
\subfloat[LBJ TSTT Minimization.\label{fig:lbj_tsttMin}] {\includegraphics[width=0.5\textwidth]{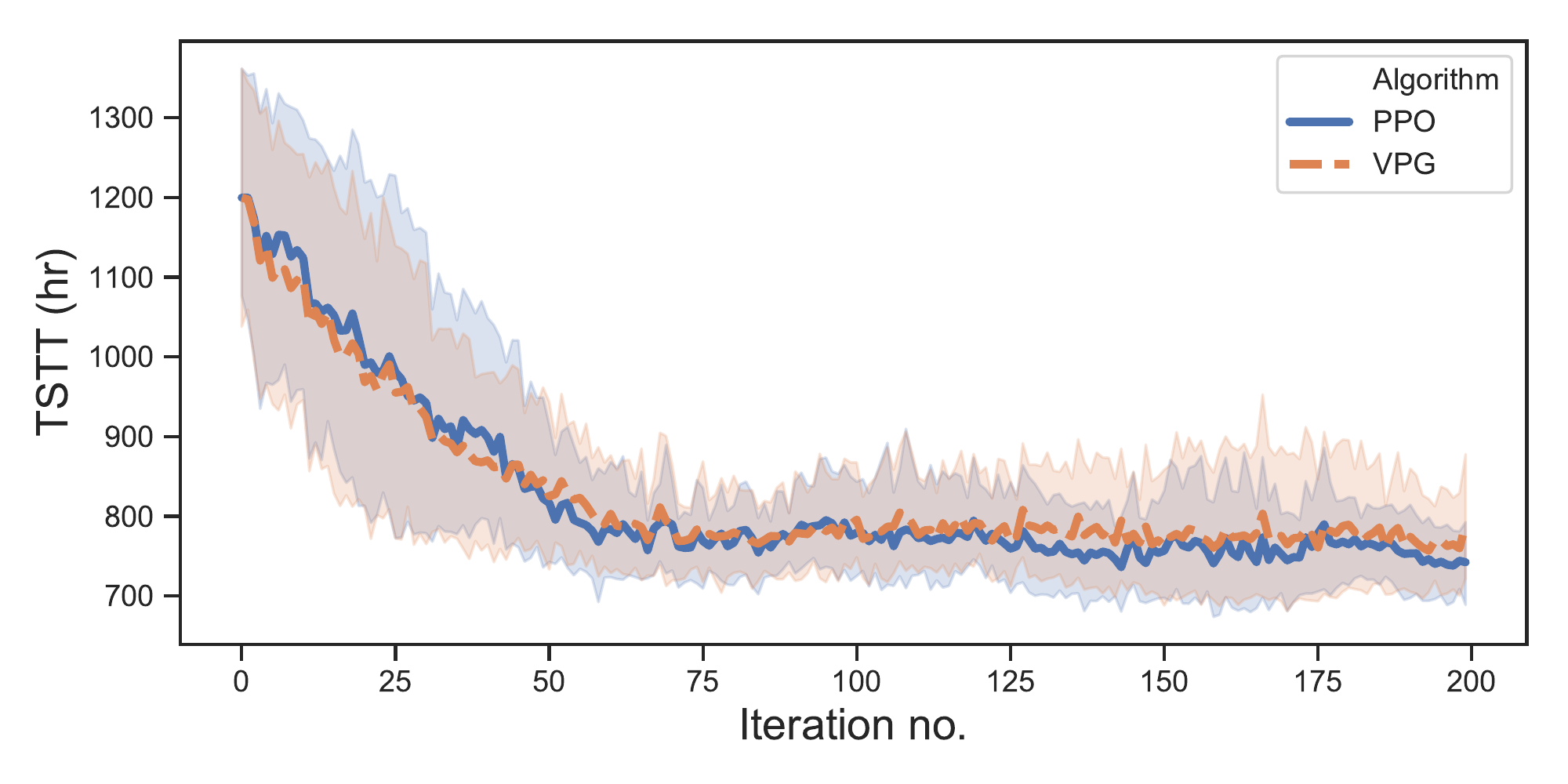}}\hfill
\subfloat[MoPac Revenue Maximization.\label{fig:mopac_revMax}]{\includegraphics[width=0.5\textwidth]{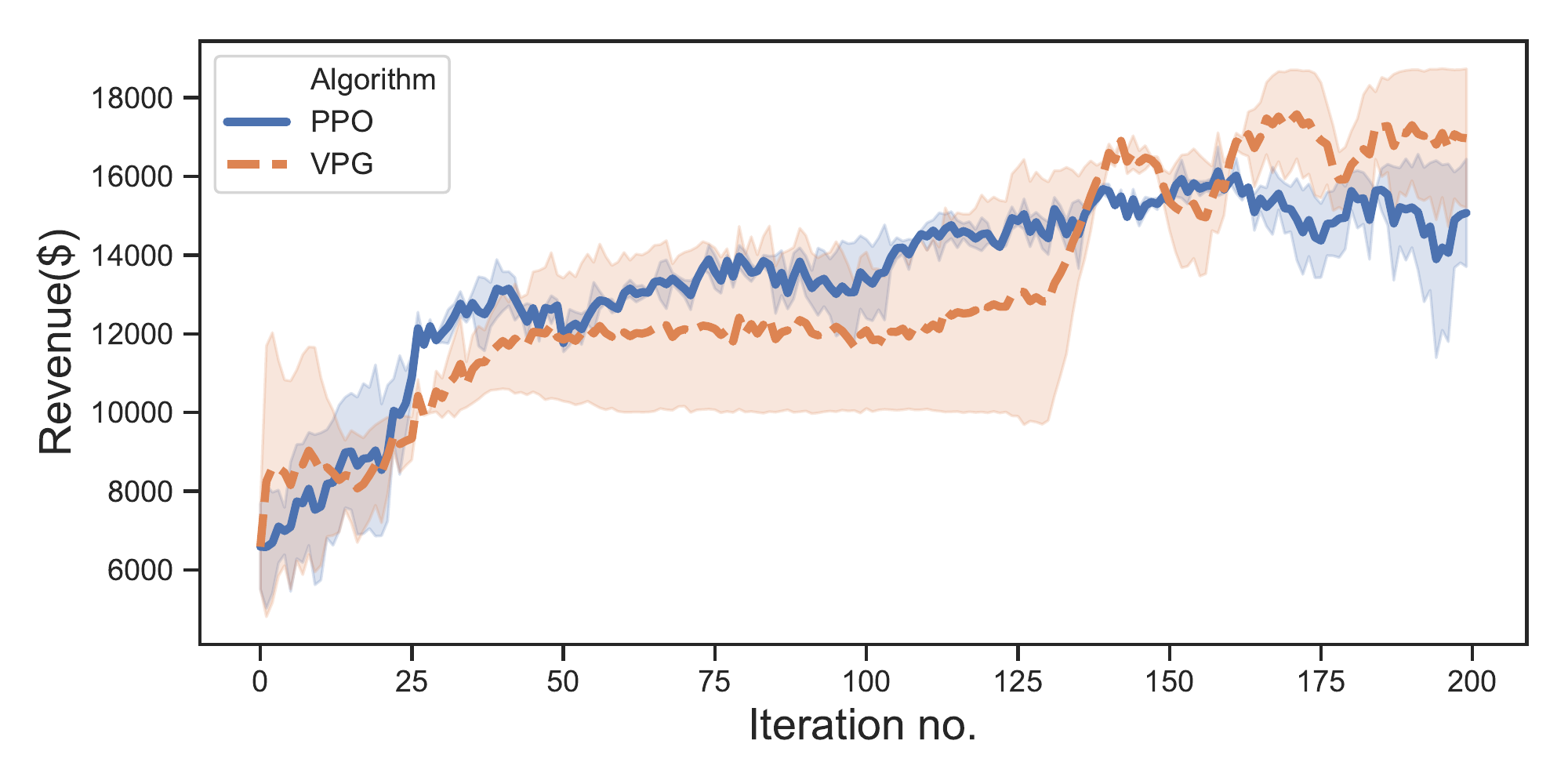}}\hfill
\subfloat[MoPac TSTT Minimization.\label{fig:mopac_tsttMin}] {\includegraphics[width=0.5\textwidth]{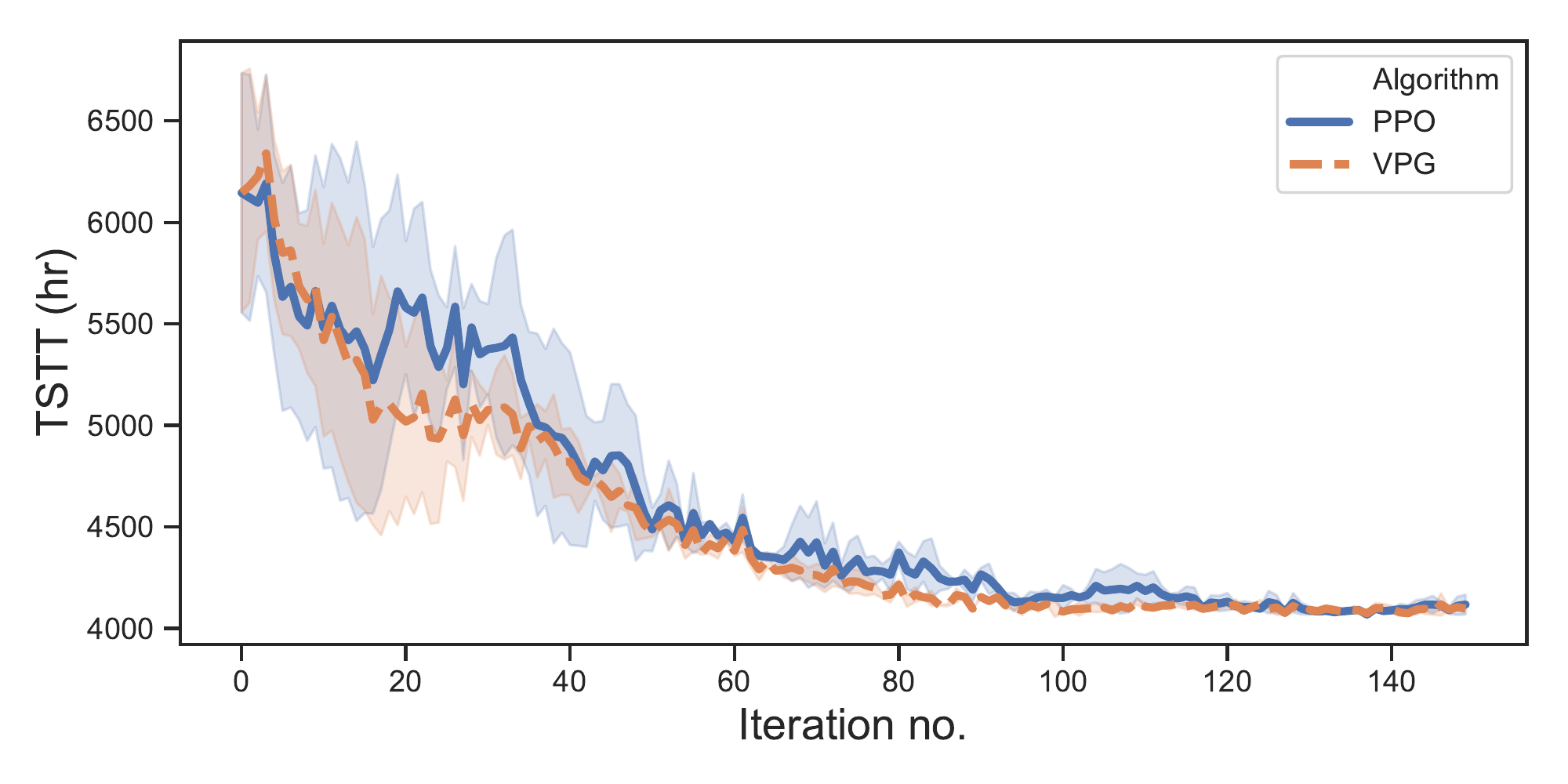}}\hfill
\caption{Plot of average objective value and the confidence interval with iteration over 10 random seeds for the four networks} \label{fig:allRevMaxTSTTMinPlots}
\end{figure}

We next compare the learning performance of the VPG and PPO Deep-RL algorithms for both revenue maximization and TSTT minimization objectives. Figure \ref{fig:allRevMaxTSTTMinPlots} show the plots of variation of learning for two objectives for all four networks over 200 iterations. The average in each iteration is reported over 10 random seeds, and \vp{for each random seed 10 trajectories are simulated to perform policy updates in Equations \eqref{eq:vpgUpdate} and \eqref{eq:polcygrad2}.} 

We make the following observations. First, both Deep-RL algorithms are able to learn ``good" objective values within 200 iterations, evident in the increasing trend of the average revenue for the revenue maximization objective and a decreasing trend of the average TSTT for the TSTT minimization objective. \vp{Contrasting with the learning curves in Pandey and Boyles \cite{pandey2018dynamic}, which used the value function approximation technique to learn value functions with iterations, we observe that policy gradient methods are more efficient in learning than value-based methods.} For the revenue maximization objective, the average revenue values converge to a high value for all networks. For the TSTT minimization objective, the average TSTT values for SESE (Figure \ref{fig:sese_tsttMin}) and DESE (Figure \ref{fig:dese_tsttMin}) networks do not converge; however a decreasing trend is evident. The VPG algorithm for the DESE network in Figure \ref{fig:dese_tsttMin} shows divergence towards the end. \vpB{This behavior can be attributed to the lack of convergence guarantees for gradient-based algorithms in stochastic settings, where the algorithms may converge to a local optimum or may diverge. Therefore, we recommend tracking the ``best" policy parameters over iterations.}

We argue that learning for the revenue maximization objective is easier than learning for the TSTT minimization objective. This is because the reward definition for revenue maximization in Equation \eqref{eq:revMaxObj} involves the action values (in terms of $\beta_{ij}(\cdot)$) and thus incorporates a direct feedback on the efficiency of current toll. This allows the gradient descent algorithm to learn the right tolls quickly. On the other hand, the feedback of whether the toll is ``good enough" is less clear for the TSTT minimization objective. Equation \eqref{eq:tsttMinObj} does not incorporate the toll values directly and the only way to learn whether a set of tolls taken were right is at the end of simulation when the TSTT value is generated. This is known as the \textit{credit assignment problem} in the RL literature, where it is unclear which actions over the entire episode were helpful. The credit assignment problem can potentially be addressed by reframing the reward definition for the TSTT minimization objective, but this analysis is left as part of the future work ($\text{FW}\#5$).

Second, we observe that there is no evident difference in the performance results of VPG and PPO algorithms. For the revenue maximization objectives, the algorithms perform ``almost identically" with values of average revenue of PPO within $\sim 5$\% of the average revenue values of VPG algorithm at any iteration. For the TSTT minimization objective, we observe that PPO prevents high variation in average TSTT values from one iteration to the next, whereas the VPG algorithm shows higher oscillations (evident in Figures \ref{fig:sese_tsttMin} and \ref{fig:dese_tsttMin}). The variance in the average TSTT values is also higher for the VPG algorithm for the TSTT minimization objective.

Last, in contrast to our expectation that a larger network with high dimensional action space might require large number of iterations to converge, we observe that for both LBJ and MoPac networks, the average objectives converge within 200 iterations, which is equivalent to simulating $2000$ episodes with $2000\times 2 \text{ hours}/5 \text{ minutes}=48000$ action interactions with the environment. Both networks mimic the real-world implementations of express lanes, and thus we argue that learning is possible within a reasonable number of interactions with the environment even for real-world networks. The amount of data required for training Deep-RL models is often considered its major limitation~\cite{arulkumaran2017deep}; however, for the dynamic pricing problem it is not a constraining factor.

\vp{Next, we report the computation time needed for training the networks in Table \ref{tab:compTime}. The run times are reported on a Unix machine with 8 GB RAM and are computed starting when the algorithms begin execution till the end of desired number of iterations.} As observed, both Deep-RL algorithms show minor to no difference. The total computation time for training of algorithm for an objective is less than half a hour for the first three networks. For the MoPac network, the computation time is around 23 hours. The bottleneck in the simulation is the traffic flow simulation using multiclass cell transmission model. For the MoPac network $|Z|=65$ and $|\mathscr{C}|=258$, and thus updating $65\times258=16{,}770$ flow variables for every time step is time consuming. Efficient implementation of CTM model with parallel computations can help improve the efficiency of training. We note that the $23.39$ hours spent for training are conducted offline on a simulation model. Once the model is trained, it can be transferred with less effort to real-world settings. We conduct tests on transferability of learned algorithms to new domains in Section \ref{subsec:transferability}.

\begin{table}[H]
\centering
		\caption{Computation time for Deep-RL training}
		\label{tab:compTime}
\begin{tabular}{|c|c|c|c|}
\hline
\multirow{2}{*}{\textbf{Network}} & \multicolumn{2}{c|}{\textbf{\begin{tabular}[c]{@{}c@{}}Computation time per iteration \\ for simulating 10 episodes (seconds)\end{tabular}}} & \multirow{2}{*}{\textbf{\begin{tabular}[c]{@{}c@{}}Total average computation \\ time for training (hours)\end{tabular}}} \\ \cline{2-3}
 & VPG & PPO &  \\ \hline
\textbf{SESE} & 7.00 & 6.99 & 0.39 \\ \hline
\textbf{DESE} & 3.59 & 3.57 & 0.20 \\ \hline
\textbf{LBJ} & 7.51 & 7.49 & 0.42 \\ \hline
\textbf{MoPac} & 420.99 & 419.2 & 23.39 \\ \hline
\end{tabular}
\end{table}

\subsubsection{Impact of observation space}
\begin{figure}[H]
\centering
\subfloat[\label{fig:LBJ_observabilityTestA}]{\includegraphics[width=0.5\textwidth]{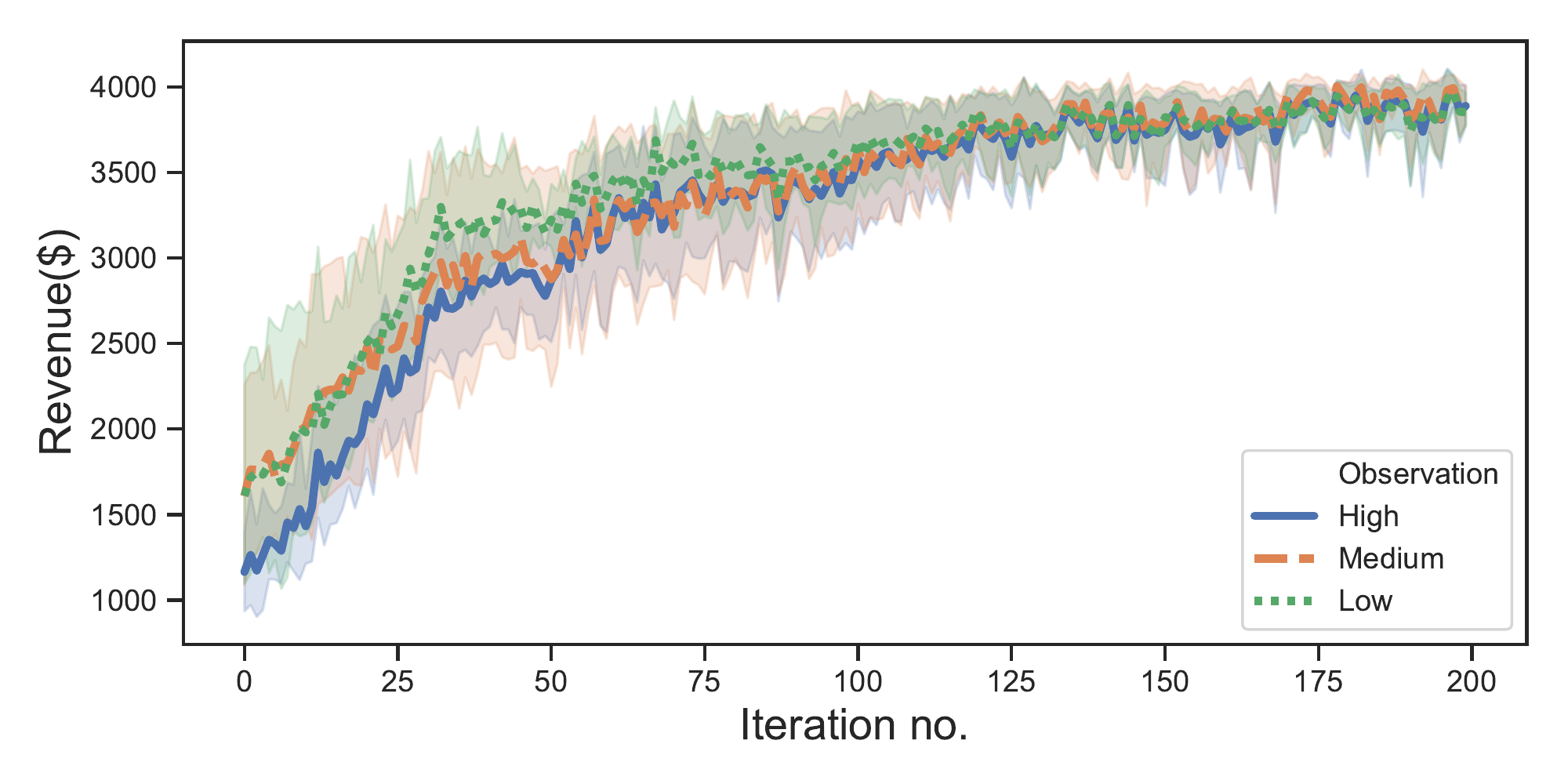}}\hfill
\subfloat[\label{fig:LBJ_observabilityTestB}] {\includegraphics[width=0.5\textwidth]{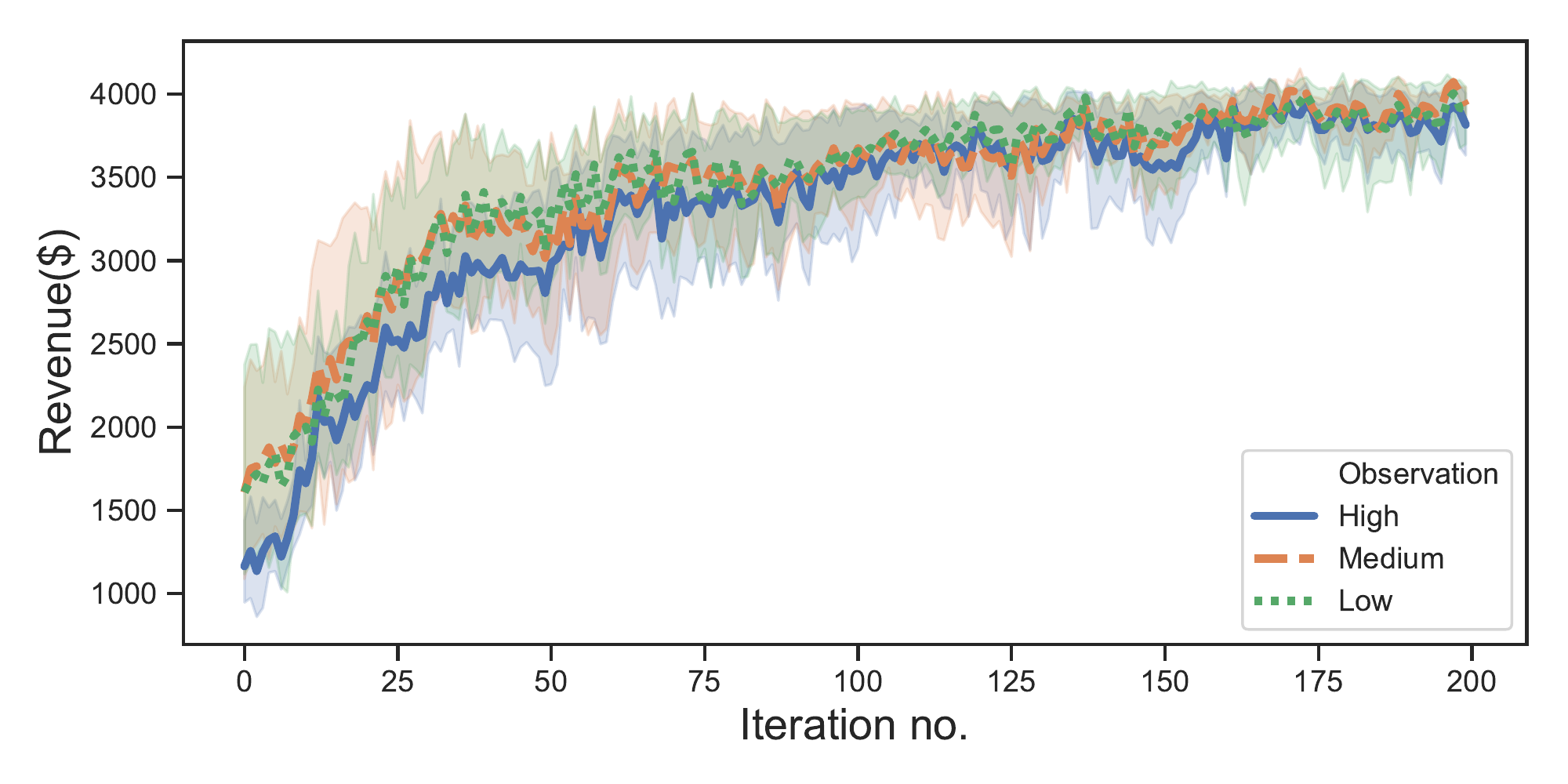}}\hfill
\caption{Plot of the average revenue with iteration over 5 random seeds for the three levels of observation for (a) VPG algorithm, and (b) PPO algorithm for the LBJ network} \label{fig:LBJ_observabilityTest}
\end{figure}

\vpB{We also test the impact of observation space on the learning of Deep-RL algorithms. For the LBJ network, the results in Figures \ref{fig:lbj_revMax} and \ref{fig:lbj_tsttMin} assumed that flows are observed on all links (which we term \texttt{High} observation). We consider two additional observation cases: (a) observing links $(3,5),(4,7),(6,9),$ and $(8,11)$ (\texttt{Medium} observation), and (b) only observing link $(6,9)$ in the network (\texttt{Low} observation). Figure \ref{fig:LBJ_observabilityTest} shows the learning results for revenue maximization objectives for the two algorithms for three levels of observation space.} 

\vpB{We observe that  changing the observation space has a minor impact on learning rate. This result was unexpected, and suggests that good performance can be obtained with relatively few sensors. We \vp{speculate} that this happens due to the spatial correlation of the congestion pattern on a corridor (where observing additional links does not add a new information for setting the tolls). \vp{The computation time differences on using different observation spaces were also not significant.}}

\vpB{These findings indicate that a toll operator can learn toll profiles optimizing an objective without placing sensors on all links, which is a lower cost alternative than observing all links. Future work will be devoted to the cost-benefit analysis of different sensor-location combinations assuming variability in sensing errors across different sensors. (FW$\#6$)}

\subsubsection{Learning for varied inputs and transferability analysis}
\label{subsec:transferability}
\vp{In this section, we consider how Deep-RL algorithms perform for varied set of inputs and how the policies trained on one set of inputs perform when transferred to new inputs without retraining for the new inputs.} This analysis is useful for a toll operator who trains the algorithm in a simulation environment for certain assumptions of input. \vp{For the policy to transfer, the observation space in the new setting must be identical to the setting where the transferred policy is trained}. We only consider cases for changes in input demand distribution, VOT distribution, and lane choice model. Transferability of Deep-RL algorithms trained on one network to other networks or the same network with new origins and destinations requires extensive investigation and is a topic for future research ($\text{FW}\#7$).

We consider the revenue-maximizing policy for the LBJ network and \vp{consider four different input cases}. The first two cases consider new demand distributions (Variant 1 and Variant 2) shown in Figure \ref{fig:demandDist}.  The third case considers a new VOT distribution (Variant 3) shown in Figure \ref{fig:votDist}. And, the last case uses a  multiclass binary logit model with scaling parameter $6$ for modeling driver lane choice~\cite{pandey2019comparing}. \vp{For each case, we also directly apply the policy obtained at the final iteration of training on the LBJ network for the revenue-maximization objective with the original demand, VOT distribution, and lane choice model (Figure \ref{fig:lbj_revMax}).} 

\begin{figure}[h]
\centering
\subfloat[Demand Variant 1. \label{fig:transfer_variant1}]{\includegraphics[width=0.5\textwidth]{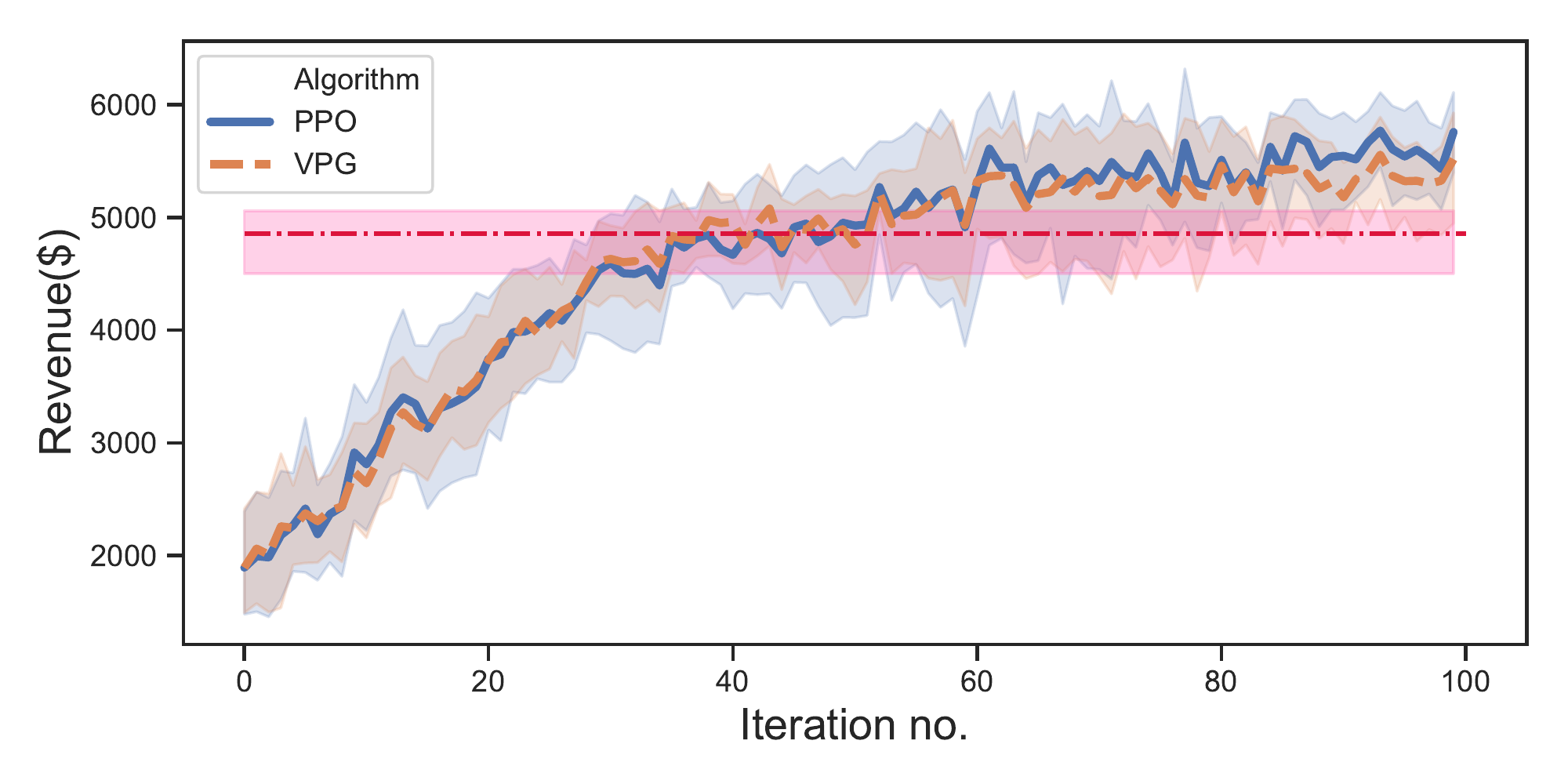}}\hfill
\subfloat[Demand Variant 2.\label{fig:transfer_variant2}] {\includegraphics[width=0.5\textwidth]{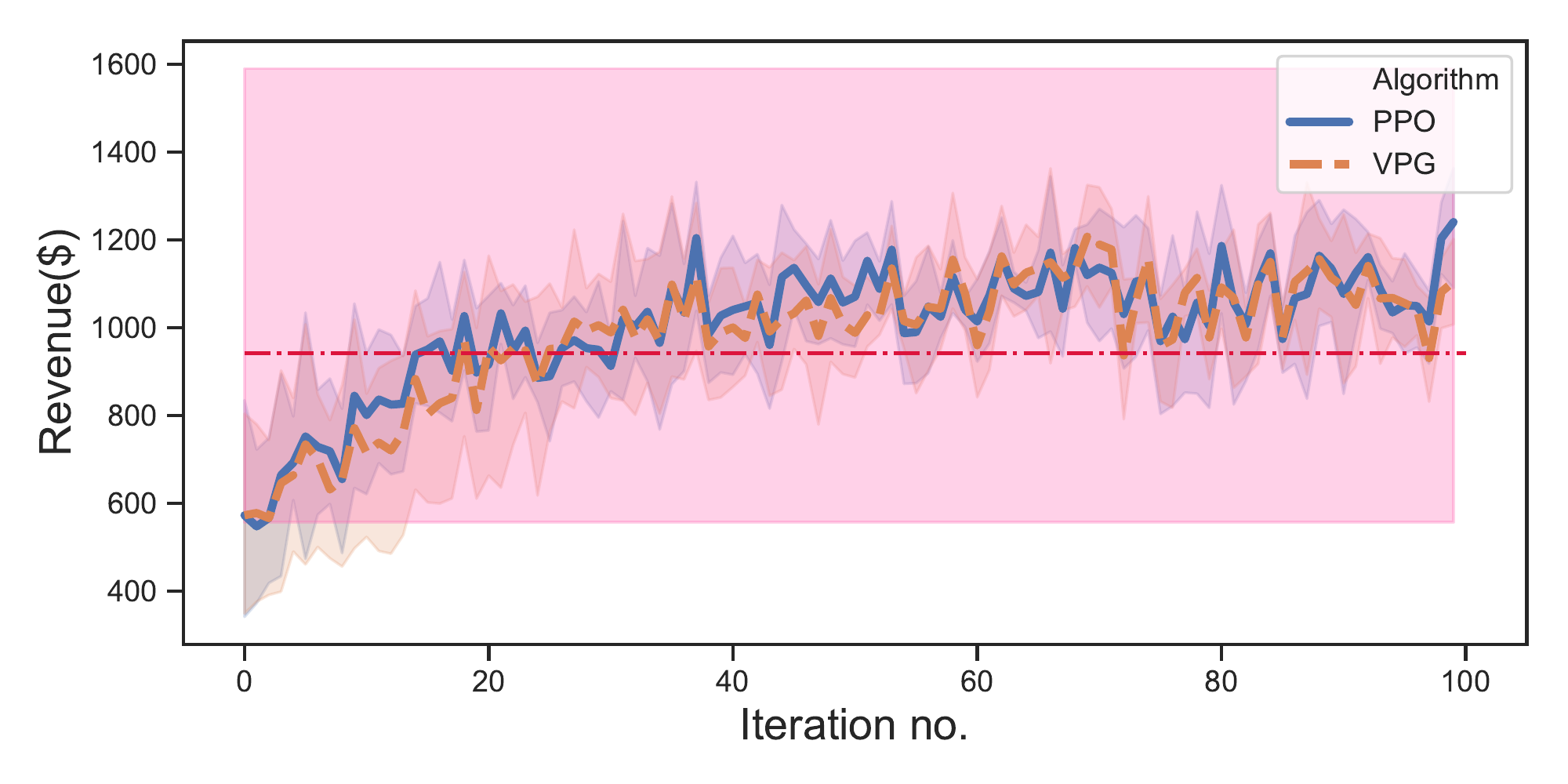}}\hfill
\subfloat[VOT Variant 3.\label{fig:transfer_variant3}] {\includegraphics[width=0.5\textwidth]{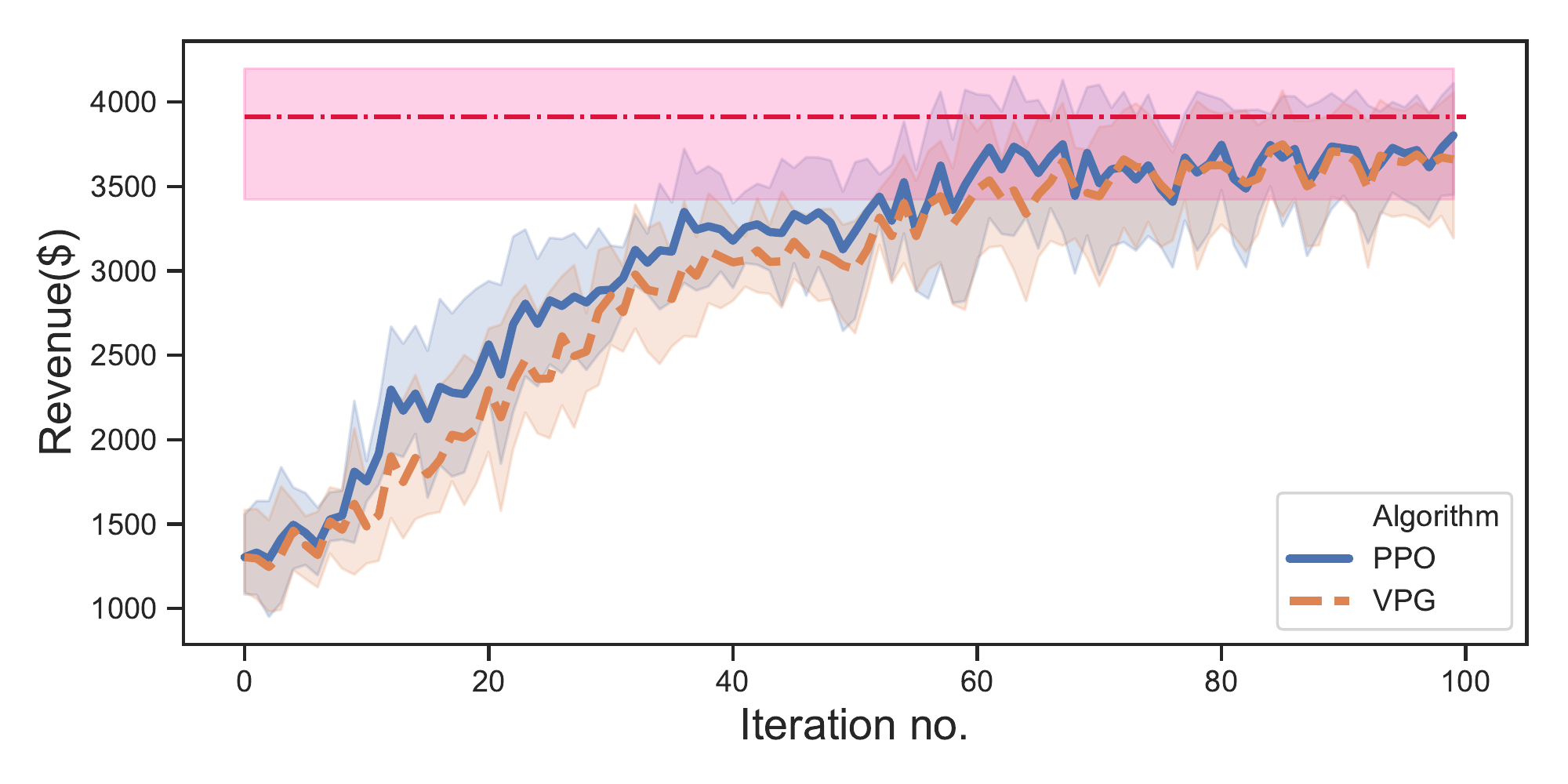}}\hfill
\subfloat[Stochastic lane choice.\label{fig:transfer_variant4}] {\includegraphics[width=0.5\textwidth]{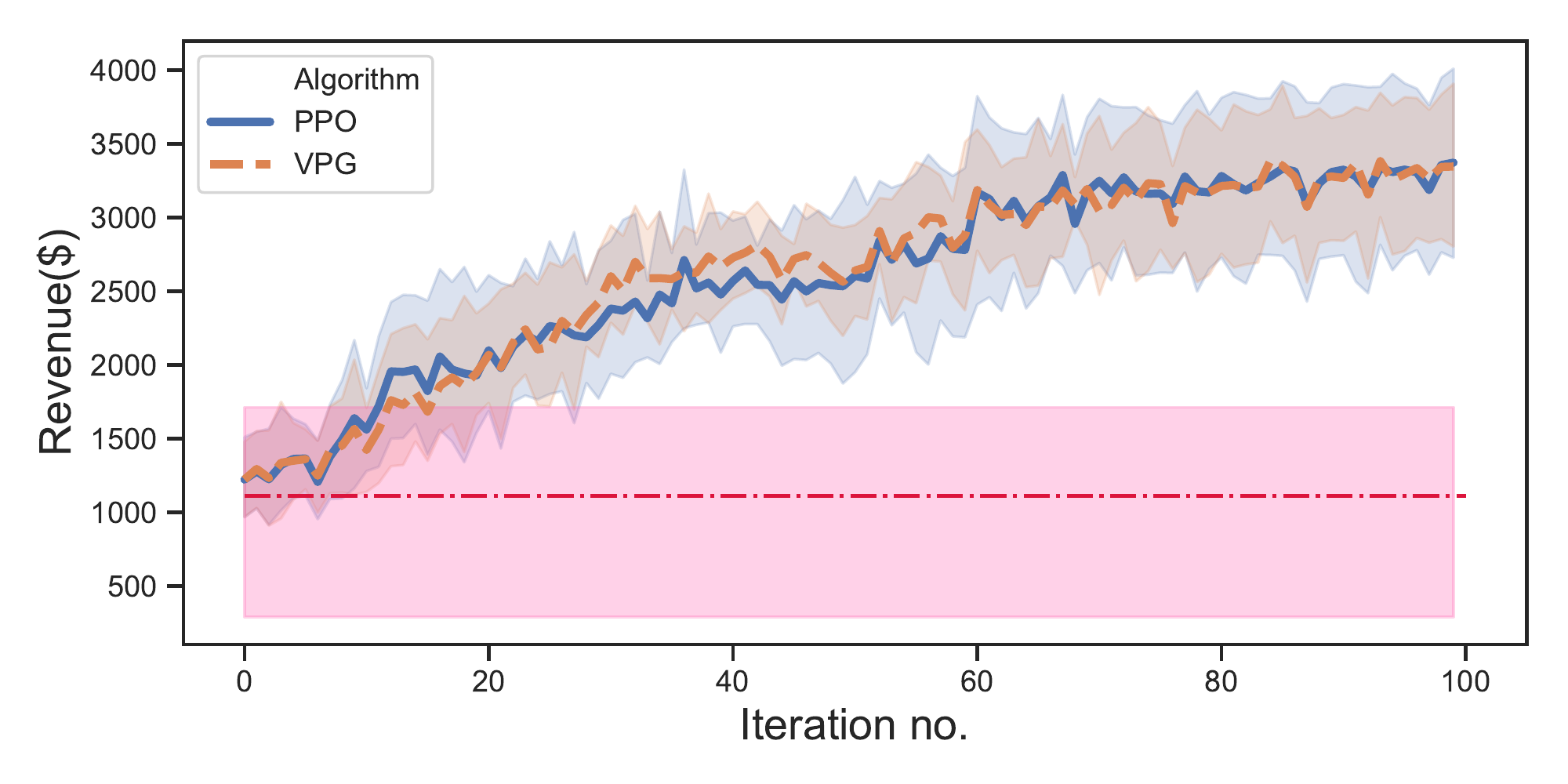}}\hfill
\caption{\vp{Comparing learning-from-scratch performance of the VPG and PPO algorithms on different input distributions with the policy transferred after learning on the original distribution (shown as a horizontal line-dot pattern) for the LBJ network}} \label{fig:variantsTransferabilityResults}
\end{figure}

Figure \ref{fig:variantsTransferabilityResults} show the plots of variation of revenue with iterations while learning from scratch \vp{for both VPG and PPO algorithms} and the average revenue \vp{(and its full range of variation)} obtained from the transferred policy for the new inputs. The average is reported over 100 runs of the transferred policy for new inputs without retraining.

First, we observe that learning for the new input configurations ``converges" within 100 iterations for all four cases.  This observation indicates the Deep-RL algorithms \vp{can iteratively learn ``good" toll profiles regardless of the input distribution}. This is a significant advantage over the MPC-based algorithms in the literature that require assumptions on driver behavior and inputs to solve the optimization problem at each time step. \vp{Similar to the previous cases, both VPG and PPO algorithms perform almost identically with less than $\sim 10\%$ difference in the objective values at any iteration for the four cases. This is in contrast to the other environments used for testing Deep-RL algorithms like Atari games and \texttt{MuJoCo} where the PPO algorithm is significantly better than the VPG algorithm~\cite{schulman2017proximal}. This is because the state update in the ML pricing problem is not drastically influenced by the toll actions, unlike the high uncertainty in the state transition in the Atari and \texttt{MuJoCo} environments. Thus, the VPG algorithm does not produce large-policy updates and has no relative disadvantage over the PPO algorithm, explaining their almost-identical performance.}

Second, the average revenue of the transferred policy is within $5-12\%$ of the average revenue at termination while learning from scratch. For case 3 with VOT variant, the transferred policy does even better than the policy learned from scratch after 100 iterations of training. The observations from the first three cases suggest that even though the Deep-RL algorithms were not trained for the new inputs, they are able to learn characteristics of the congestion in the network and perform well (on an average) on the new inputs. However, for case 2, the transferred policy has a lot of variance in the generated revenue; this is because small changes in input tolls have higher impact on generated revenue for demand Variant 2. 

Third, contrary to the first three cases, the transfer of policy in case 4 did not work well: the average revenue of transferred policy is 40\% of the maximum revenue obtained. This is because the multiclass logit model predicts significantly different proportion of splits of travelers at a diverge and thus have a significant impact on the evolution of congestion. \vp{Both cases 3 and 4 impact the split of travelers at the diverge, yet the performance of transferred policy is very different for both cases. This finding suggests that the driver lane choice model should be carefully selected and calibrated for Deep-RL training for reliable transfer to the real-world environments, whereas the demand and VOT distributions are less important.}

\subsection{Multi-objective Optimization}
\label{subsec:multiObjOpt}
\vp{We next focus our attention on multiple optimization objectives together. In the literature, revenue maximization and TSTT minimization objectives are shown to be conflicting~\cite{pandey2018dynamic}, that is toll policies generating high revenue have a high value of TSTT. Finding toll profiles that satisfy both objectives to a degree is the focus of this section.}

\begin{figure}
	\centering
	\includegraphics[width=\columnwidth]{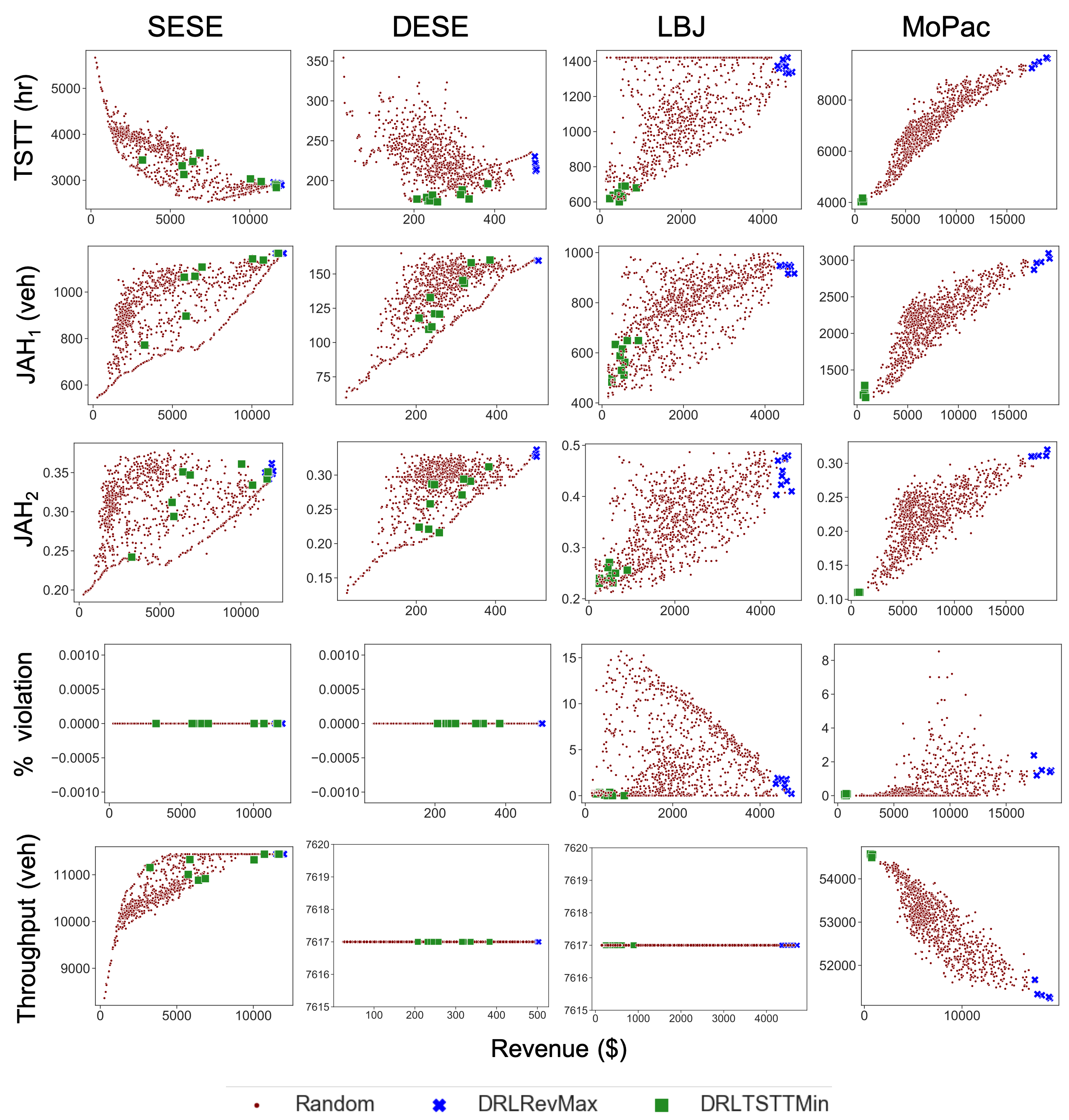}
	\caption{Plot of various objectives against the revenue for 1000 randomly generated toll profiles (Random) and the profiles generated from Deep-RL for revenue maximization (DRLRevMax) and TSTT minimization (DRLTSTTMin) objectives}
	\label{fig:allParetoPlots}
\end{figure}

\vp{We consider how different objectives vary with respect to each other for 1000 randomized toll profiles simulated for all four networks. Figure \ref{fig:allParetoPlots} shows the plots of variation of TSTT, $\text{JAH}_1$, $\text{JAH}_2$, \vp{\texttt{\%-violation}}, and the total number of vehicles exiting the system (throughput) against the revenue obtained from the toll policies. The figure also shows the values of objectives from the toll profiles generated by Deep-RL algorithms where ``DRLRevMax" indicates toll profiles from the revenue maximization objectives and ``DRLTSTTMin" indicates toll profiles from the TSTT minimization objective.}

\vp{We make following observations. First, we observe that the best toll profiles generated from Deep-RL algorithm are the best found among the  other randomly generated profiles for the respective objectives. For the revenue maximization objective, toll profiles generated from Deep-RL algorithms have the highest revenue for all networks. For the TSTT minimization objective, toll profiles from Deep-RL algorithm have the lowest TSTT, except for the SESE network where the Deep-RL algorithm had not converged after 200 iterations (shown in Figure \ref{fig:sese_tsttMin}).}

\vp{Second, similar to the trends in the literature, toll profiles generating high revenue also generate high values of TSTT for the LBJ and MoPac networks. 
However, for the SESE and DESE networks, the trend does not hold as toll profiles generating high revenue also have low values of TSTT. This behavior, where revenue-maximizing tolls do not differ significantly from the TSTT-minimizing tolls is possible for networks where GPLs are jammed quickly enough. Once the GPL is jammed, revenue is maximized by charging the highest possible toll while sending maximum number of vehicles towards the ML. Such tolls will also generate low values of TSTT as they utilize the ML to its full capacity from that time step onwards. This finding indicates that, depending on the network properties and the inputs, the two objectives may not always be in conflict with each other. We leave a detailed analysis of how different network characteristics impact the similarity and differences between revenue-maximizing and TSTT-minimizing tolls for future work (FW$\#8$).}

\vp{Third, we see that tolls generating high revenue also have high values of JAH$_1$ and JAH$_2$ statistics. The tolls generating low TSTT, however, do not have a fixed trend and the behavior depends on networks. For example, for the MoPac networks, tolls generating low TSTT have lower revenue and thus have lower values of JAH statistics; however, for the other networks, JAH statistics are also relatively high for the tolls minimizing TSTT compared to the least JAH statistic value obtained. This finding shows that tolls minimizing TSTT may also exhibit JAH behavior, though the extent of JAH for TSTT-minimizing profiles is always lower than the revenue-maximizing profiles.}

\vp{Fourth, for the LBJ and MoPac networks with multiple access points to the ML, we observe that several toll profiles can cause violation of the speed limit constraint. 
However, the toll profiles optimizing the revenue or TSTT generate \texttt{\%-violation} less than 2\% for both MoPac and LBJ networks. This is intuitive for the revenue maximization objective: a higher revenue is  obtained only when more travelers use ML and the lane is kept congestion free. Similarly, for TSTT minimization objective, low TSTT occurs when travelers spend less time in the network and exit the system sooner which is achieved when ML is ensured to be flowing at its capacity and does not become congested.}

\vp{Last, the trends in throughput depend on the congestion level; if all vehicles clear at the end of simulation, throughput is a constant value equal to the number of vehicles using the system. However, for SESE and MoPac networks congestion persists till the end of simulation. For the MoPac network, tolls generating high revenue have less throughput and the tolls generating low TSTT have a higher throughput. Whereas for the SESE network, for the reasons explained earlier, throughput is high for both TSTT-minimizing and revenue-maximizing profiles.}

Next, we seek toll profiles that optimize two objectives. Multi-objective reinforcement learning is an area that focuses on the problem of optimizing multiple objectives~\cite{liu2014multiobjective}. There are two broad approaches for solving this problem: single-policy approach and multi-policy approach. Single-policy approaches convert the multi-objective problem into a single objective by defining certain preferences among different objectives like defining a weighted combination of multiple objectives. Multi-policy approaches seek to find the policies on the Pareto frontier of multi objective. In this article, we focus on the single-policy approach due to its simplicity. \vp{We consider the weighted-sum and threshold-penalization approaches explained next.}

\vp{First, we apply the weighted-sum approach for finding a single policy that jointly optimizes TSTT and revenue. We define a new joint reward function $r^{\text{joint}}(s,a)$ as a linear combination of two rewards.} 
\begin{equation}
	r^{\text{joint}}(s,a) = \lambda~ r^{\text{RevMax}}(s,a) + r^{\text{TSTTMin}}(s,a)
	\label{eq:jointReward}
\end{equation}

The value of $\lambda$ is the relative weight of revenue (\$) with respect to TSTT (hrs) and has units hr$/\$$. Geometrically, $\lambda$ represents the slope of a line on the TSTT-Revenue plot.

\begin{figure}[H]
\centering
\subfloat[TSTT vs Revenue $\lambda_1$.\label{fig:lbj_multiobj1}]{\includegraphics[width=0.5\textwidth]{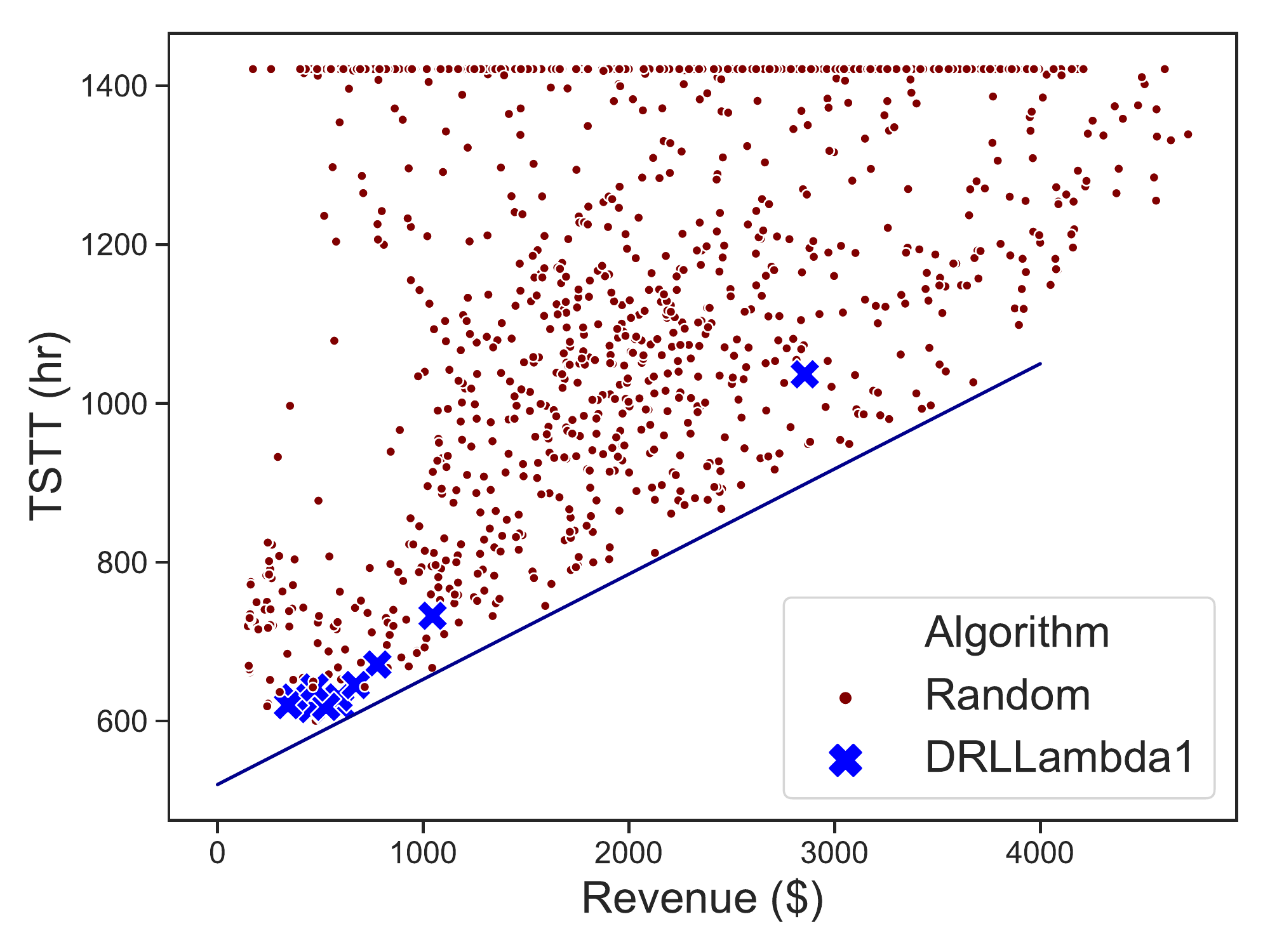}}\hfill
\subfloat[TSTT vs Revenue $\lambda_2$.\label{fig:lbj_multiobj2}] {\includegraphics[width=0.5\textwidth]{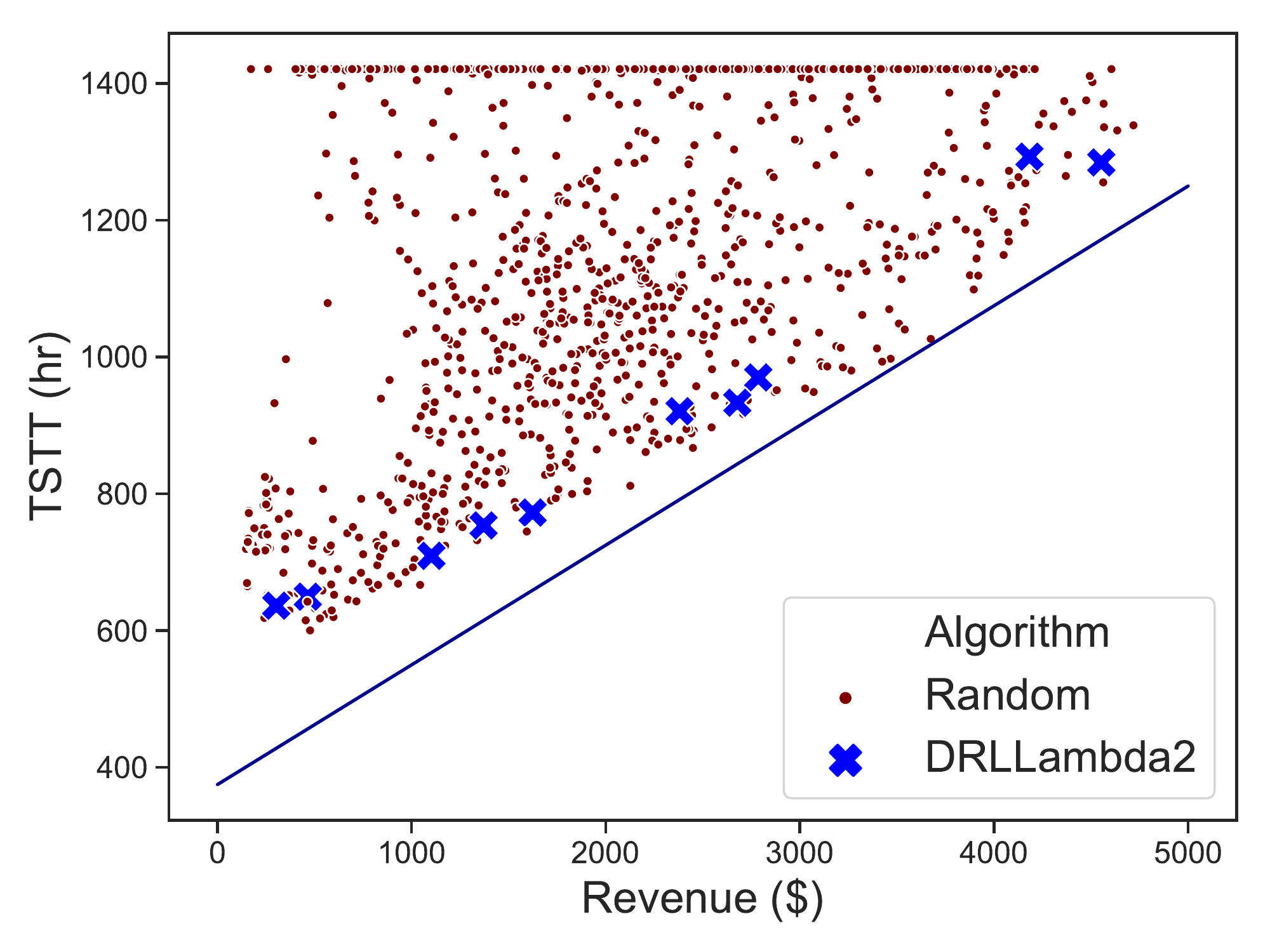}}\hfill
\caption{Plot of TSTT vs revenue for the LBJ network for toll profiles generated randomly and toll profiles generated after optimizing the joint reward for two different values of $\lambda$} \label{fig:lbj_TSTT_vs_Rev_Slope}
\end{figure}

We run VPG and PPO algorithms for the new reward \vpB{on the LBJ network} with two different values of $\lambda$: $\lambda_1=0.1325$ hr/\$ and $\lambda_2=0.175$ hr/\$ (\vpB{the values are chosen so that toll profiles in the mid-region of the TSTT-revenue plot are potentially optimal}). Figure \ref{fig:lbj_TSTT_vs_Rev_Slope} shows the plot of optimal toll profiles obtained from Deep-RL algorithms on the TSTT-Revenue space. The slopes of the lines, equal to the $\lambda$ values, are also shown, and the lines are positioned by moving them from the bottom to the top till they touch the first point among the generated space of points (that is, the line is approximately a tangent to the Pareto frontier).

As observed, Deep-RL algorithms are able to learn toll profiles that maximize the joint reward. For the $\lambda_1$ case, toll profiles are generated very close to the Pareto frontier; however, they are concentrated in the region where both TSTT and revenue are lower indicating the presence of local minima in the region. For the $\lambda_2$ case, the toll profiles are more spread out in terms of their values of TSTT and revenue; however, there are still a few toll profiles that are closer to the Pareto frontier tangent line which the Deep-RL algorithms did not find. This can again be explained by the behavior of policy gradient algorithms which are prone to converge to local optimum because they follow a gradient-descent approach. 

Optimizing using a joint reward definition as Equation \eqref{eq:jointReward} can also be interpreted as following: that a toll operator is willing to sacrifice $\$1$ revenue for a $1/\lambda$ hours decrease in TSTT value. For the two values of $\lambda$, $\lambda_1$ and $\lambda_2$, this is equivalent to sacrificing $\$1$ revenue for a $7.55$ hours and $5.72$ hours decrease in total delay for the system, respectively. If they trade off these objective outside this range, the optimal policy will be the same as solely maximizing revenue or minimizing TSTT.

The second approach for solving multi-objective optimization problem  is the \vp{threshold approach} where we find toll policies that maximum revenue (minimize TSTT) such that TSTT (revenue) is less (higher) than a certain threshold. However, such threshold constraints are hard to model in policy gradient methods working with continuous actions as that requires defining the constraints on the space of actions and projecting the tolled policy after every update onto the feasible action space. One such method is the constrained policy optimization that ensures that a policy satisfies the constraint throughout the training phase~\cite{achiam2017constrained}. However, such methods are complex to model and will be a part of future studies ($\text{FW}\#9$).

In this article, we apply the \vp{threshold-penalization method} to model threshold constraints. This method simulates a policy and if at the end of an episode the constraint is violated, a high negative value is added to the reward to penalize such update. We test this technique  to find tolls that maximize revenue such that $\text{JAH}_1$ statistic is less than a threshold value. We use $\text{JAH}_1$ statistic because it has a physical interpretation and, unlike $\text{JAH}_2$, is not unitless.

\begin{figure}[H]
\centering
\subfloat[\label{fig:lbj_modifiedreward}] {\includegraphics[width=0.5\textwidth]{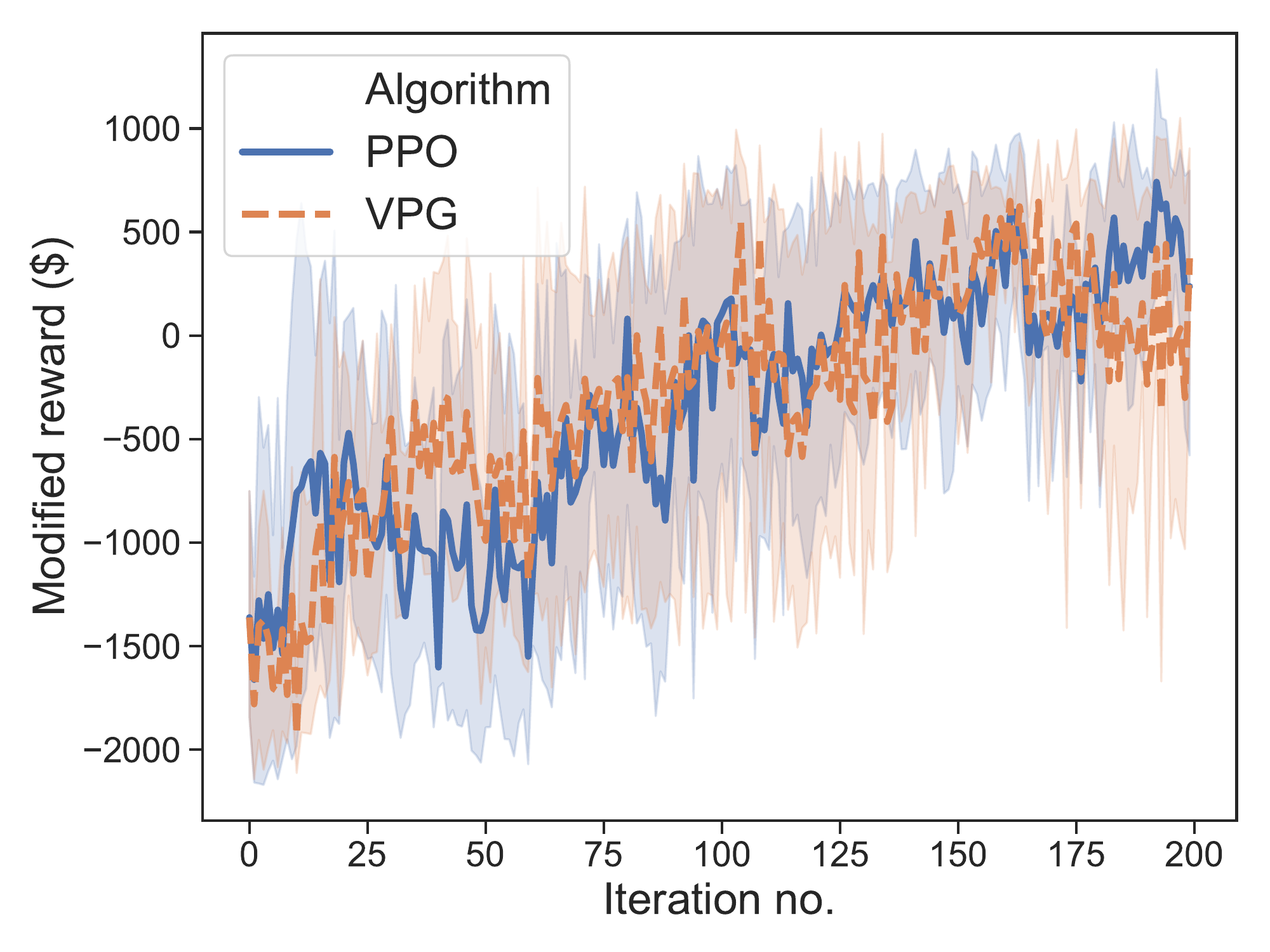}}\hfill
\subfloat[\label{fig:lbj_jahThreshold}]{\includegraphics[width=0.5\textwidth]{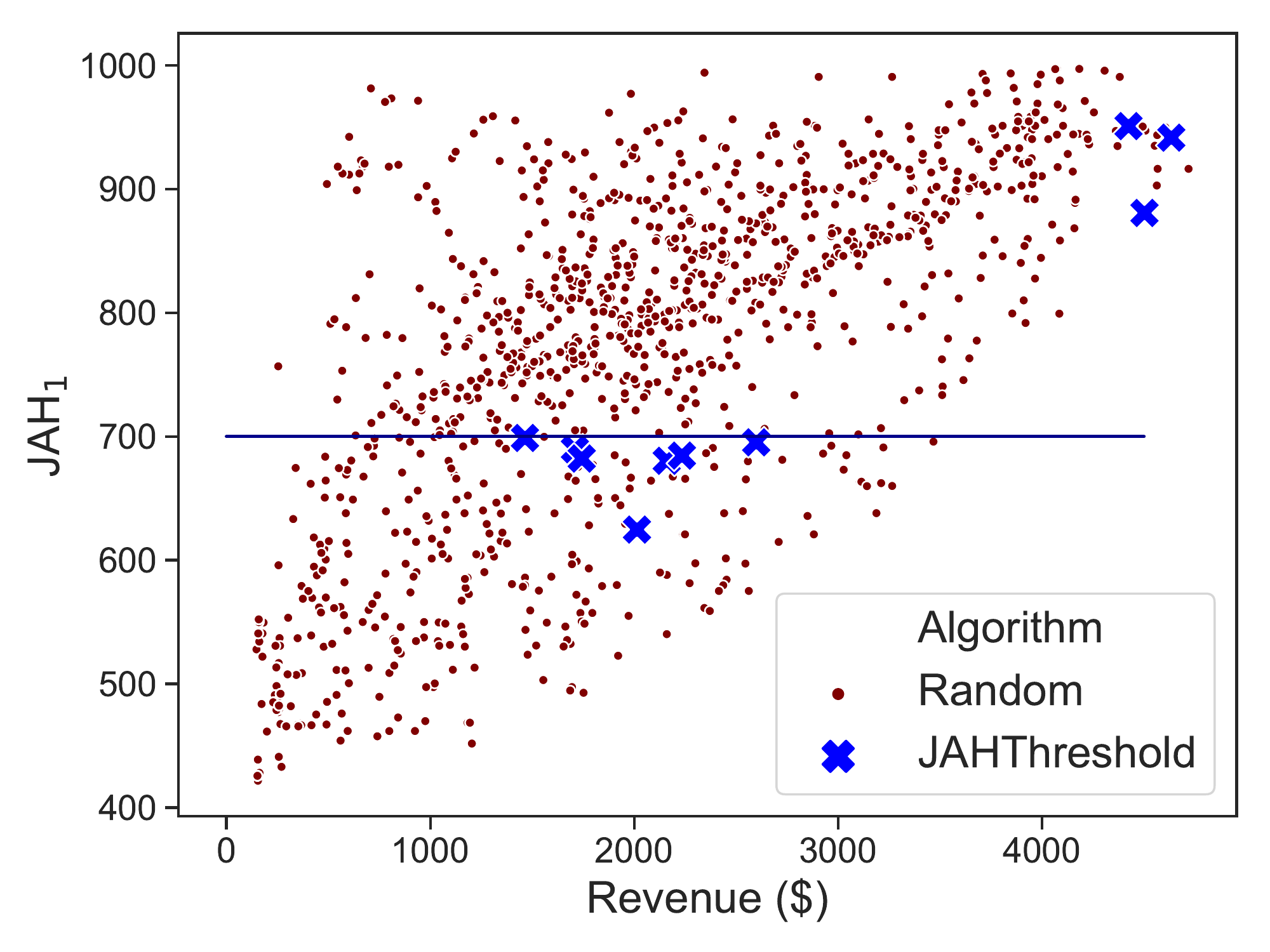}}\hfill
\caption{(a) Plot of average modified reward with iteration while maximizing revenue with a reward penalty of $-\$3000$ if the JAH$_1$ statistic is more than 700 vehicles, and (b) the plot of JAH$_1$ vs revenue for the best-found toll profiles from the threshold-penalization method, along with toll profiles generated randomly} \label{fig:lbj_JAH_vs_rev}
\end{figure}

\vp{We conduct tests for the threshold-penalization technique on the LBJ network with a threshold JAH$_1$ of 700 vehicles and add a reward value of $-\$3000$ to the final reward if at the end of simulation the JAH$_1$ statistic is higher than the threshold. Figure \ref{fig:lbj_modifiedreward} shows the learning curve plotting the variation of modified reward with iterations. We observe that both VPG and PPO algorithms improve the modified reward with iterations, though it is hard to argue that they have converged. Learning is difficult in this case due to the same \emph{credit assignment problem} where it is unclear will toll over an episode resulted in the constraint violation. Figure \ref{fig:lbj_jahThreshold} shows the plot for tolls obtained from threshold-penalization technique on the $\text{JAH}_1$-Revenue space.} 

\vp{As observed, the threshold-penalization method is able to learn toll profiles with desired JAH value for 7 out of 10 random seeds. However, the learned toll profile is not the best found (that is, there are toll profiles with JAH less than 700 but generating revenue higher than $\$2800$, which is the best found revenue). This is because the modified reward did not converge (yet) after 200 iterations. Despite the lack of convergence, we conclude that the penalization method is a useful tool to model constraints on toll profiles. The success of threshold-penalization method depends on the random seed, as that determines which local minimum the algorithm will converge to}. 

\subsection{Comparison with Feedback Control Heuristic}
\begin{figure}[H]
\centering
\subfloat[SESE \label{fig:density_revmax_sese}]{\includegraphics[width=0.3\textwidth]{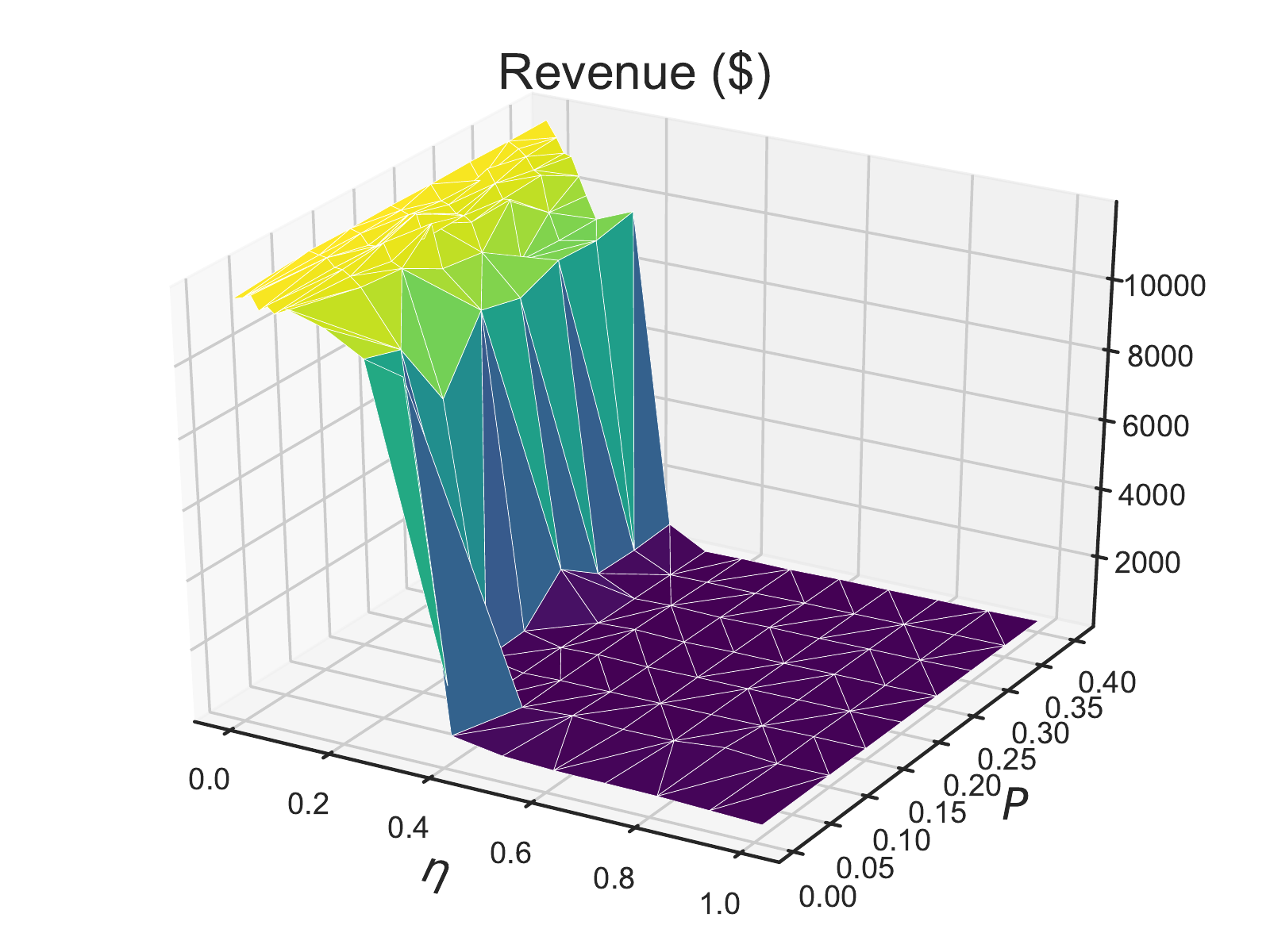}}\hfill
\subfloat[LBJ \label{fig:density_revmax_lbj}]{\includegraphics[width=0.3\textwidth]{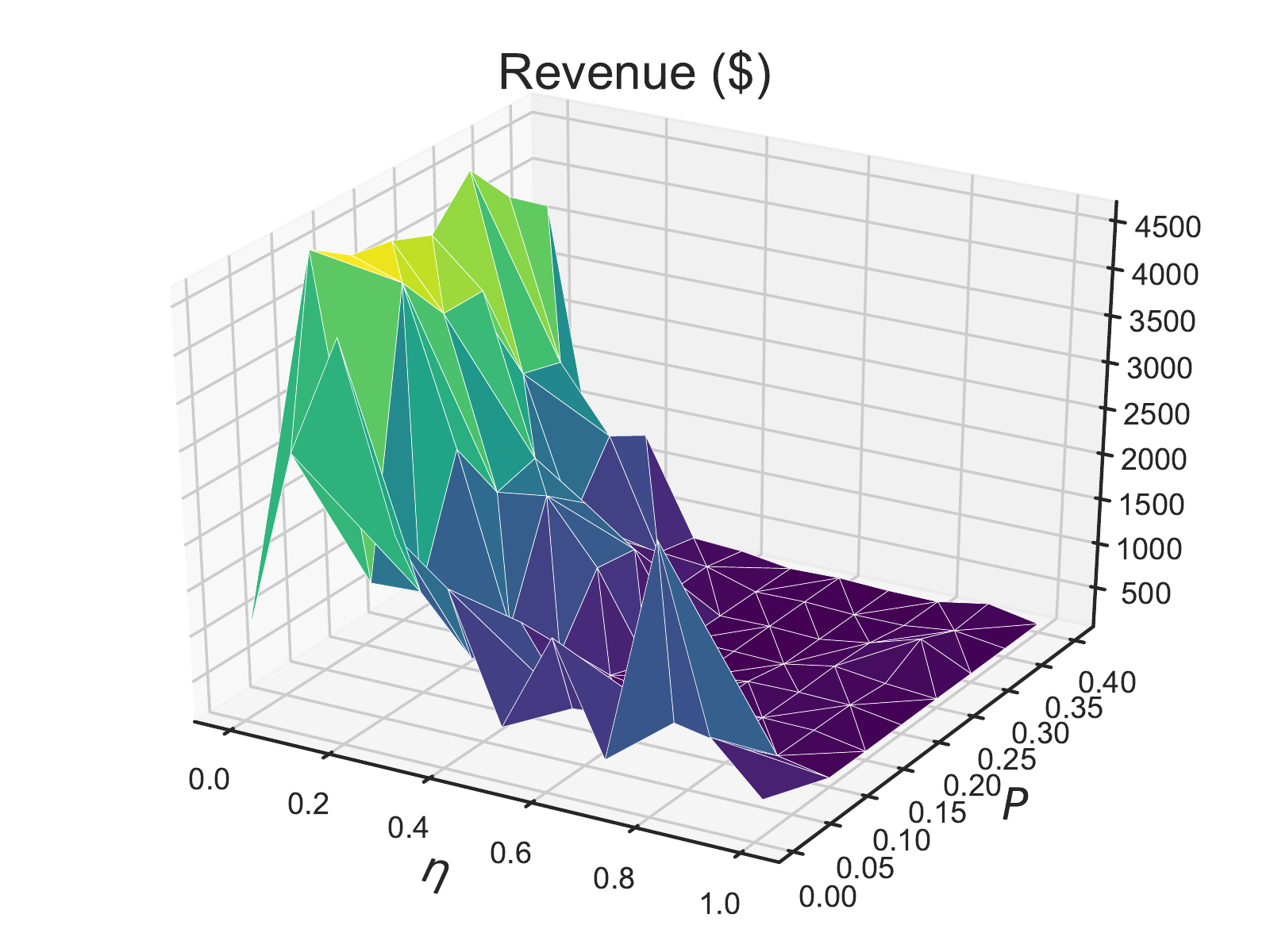}}\hfill
\subfloat[MoPac \label{fig:density_revmax_mopac}]{\includegraphics[width=0.3\textwidth]{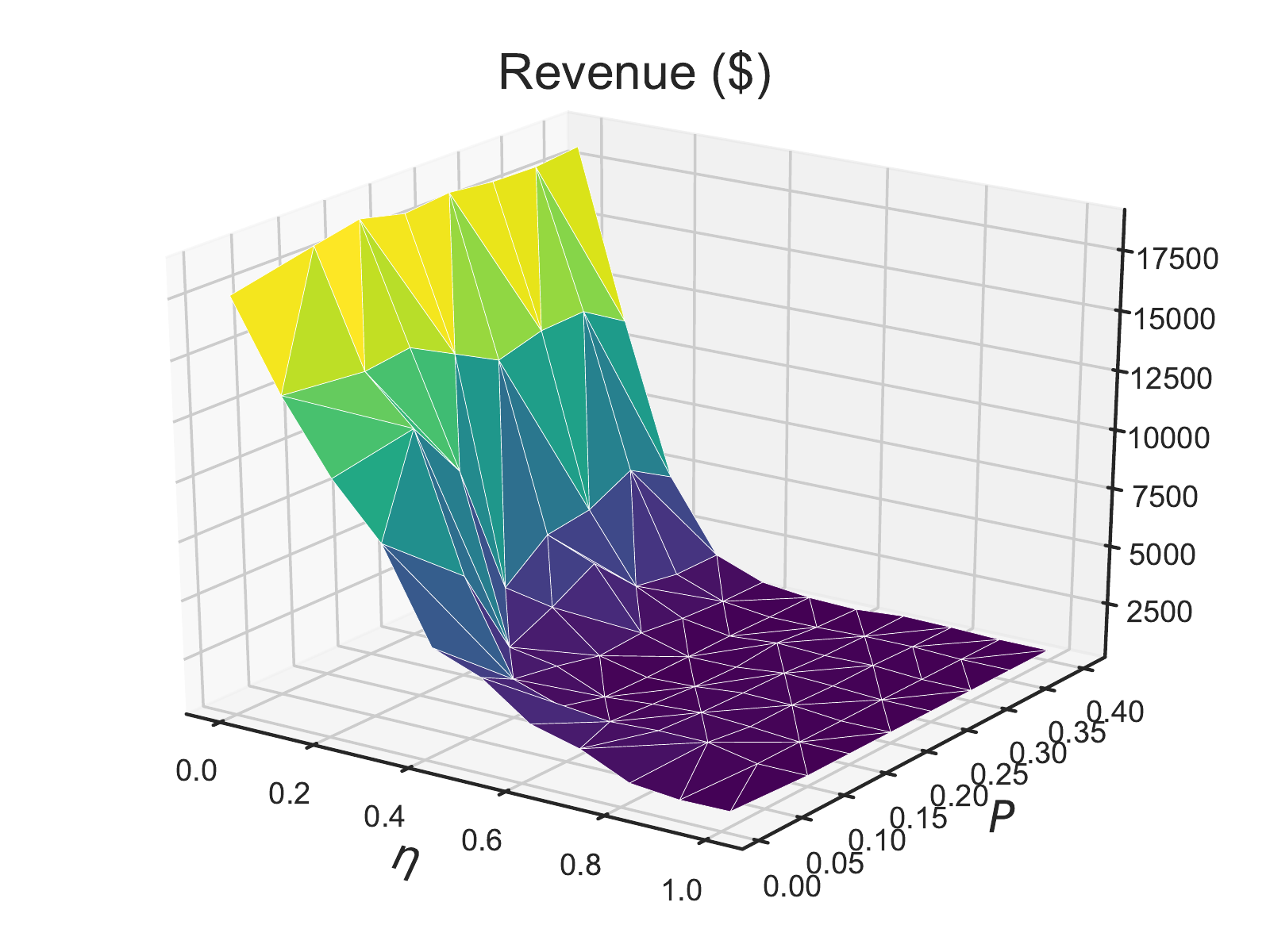}}\hfill
\subfloat[SESE\label{fig:density_tsttmin_sese}] {\includegraphics[width=0.3\textwidth]{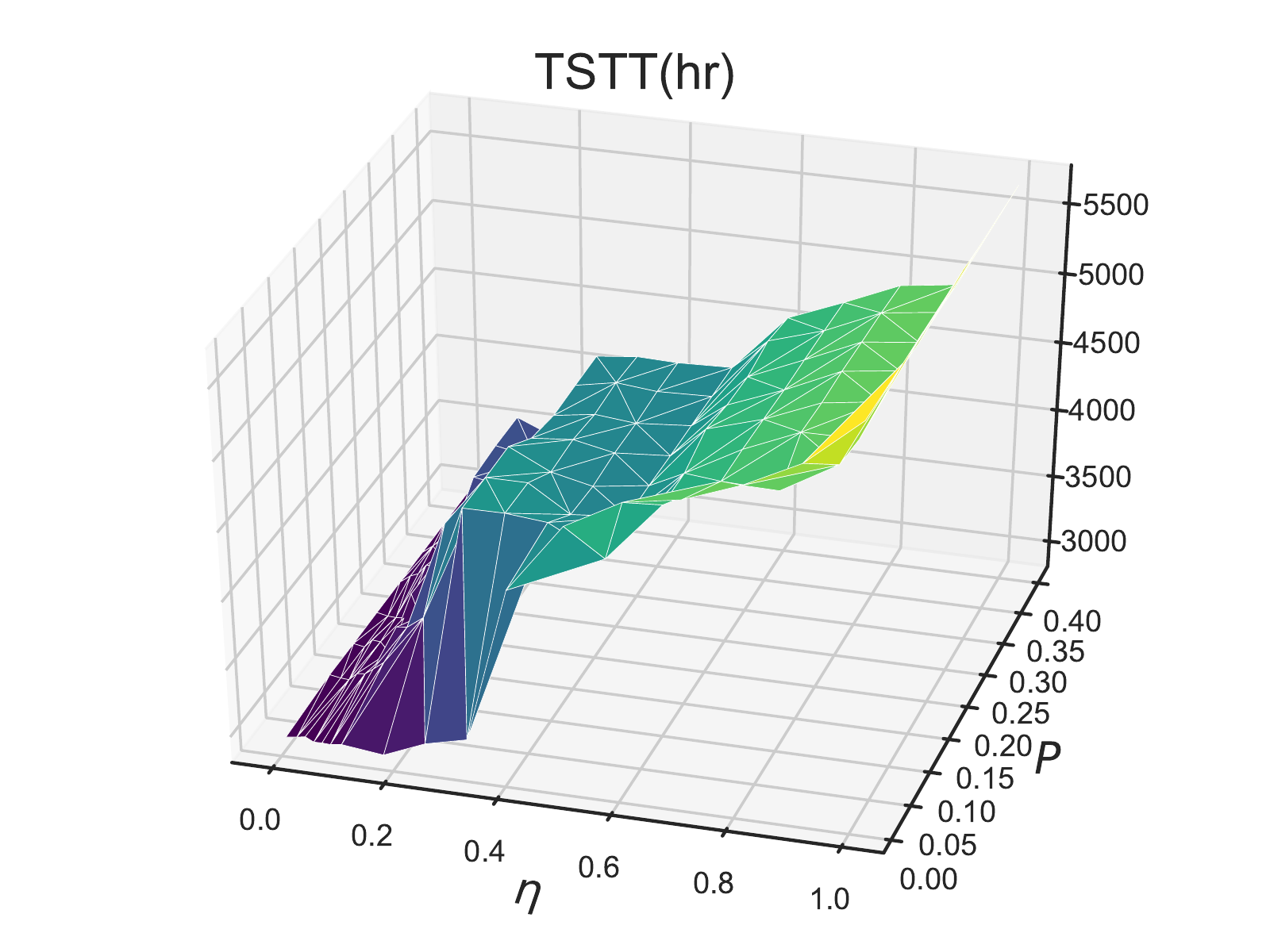}}\hfill
\subfloat[LBJ \label{fig:density_tsttmin_lbj}] {\includegraphics[width=0.3\textwidth]{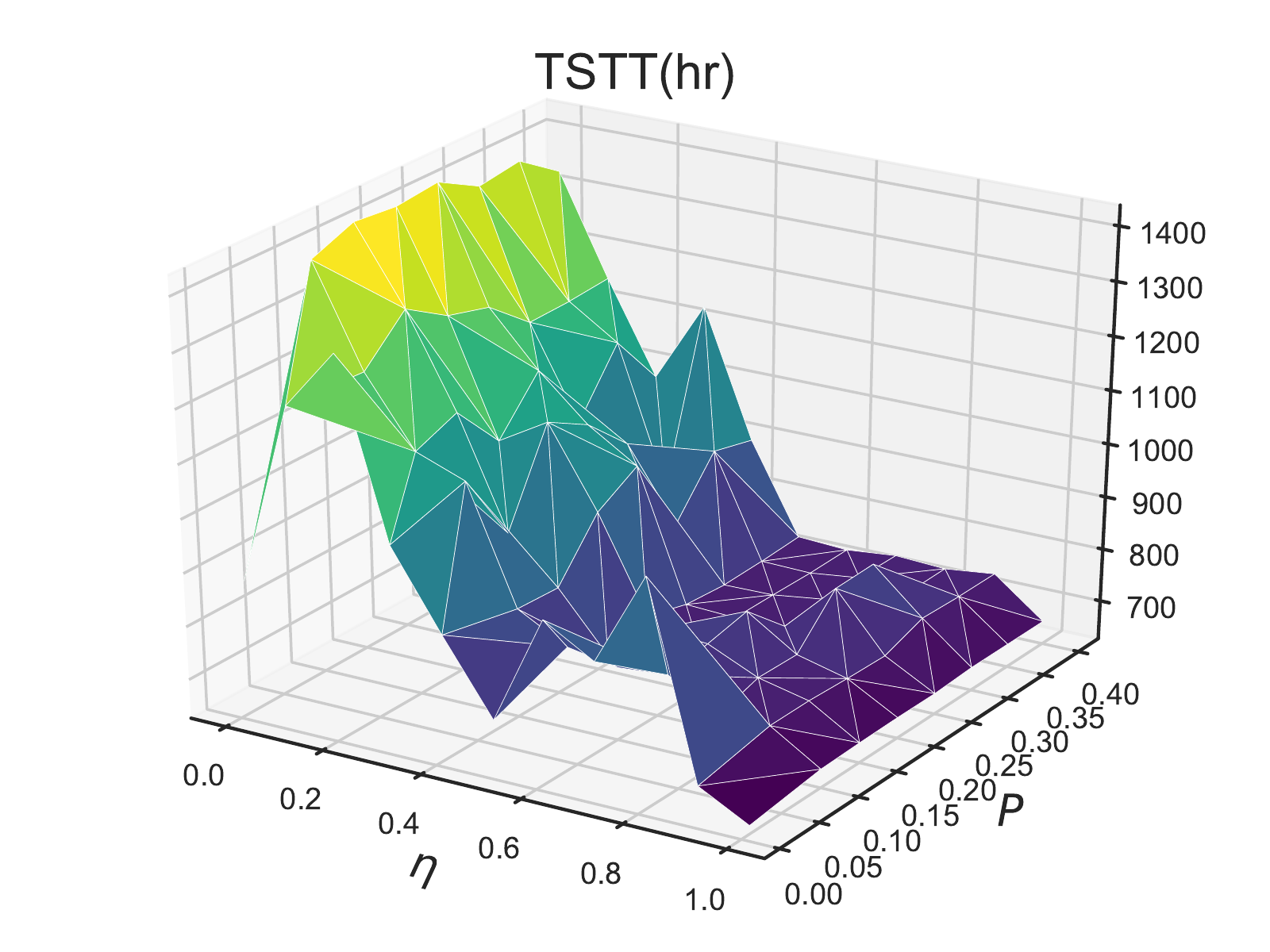}}\hfill
\subfloat[MoPac\label{fig:density_tsttmin_mopac}] {\includegraphics[width=0.3\textwidth]{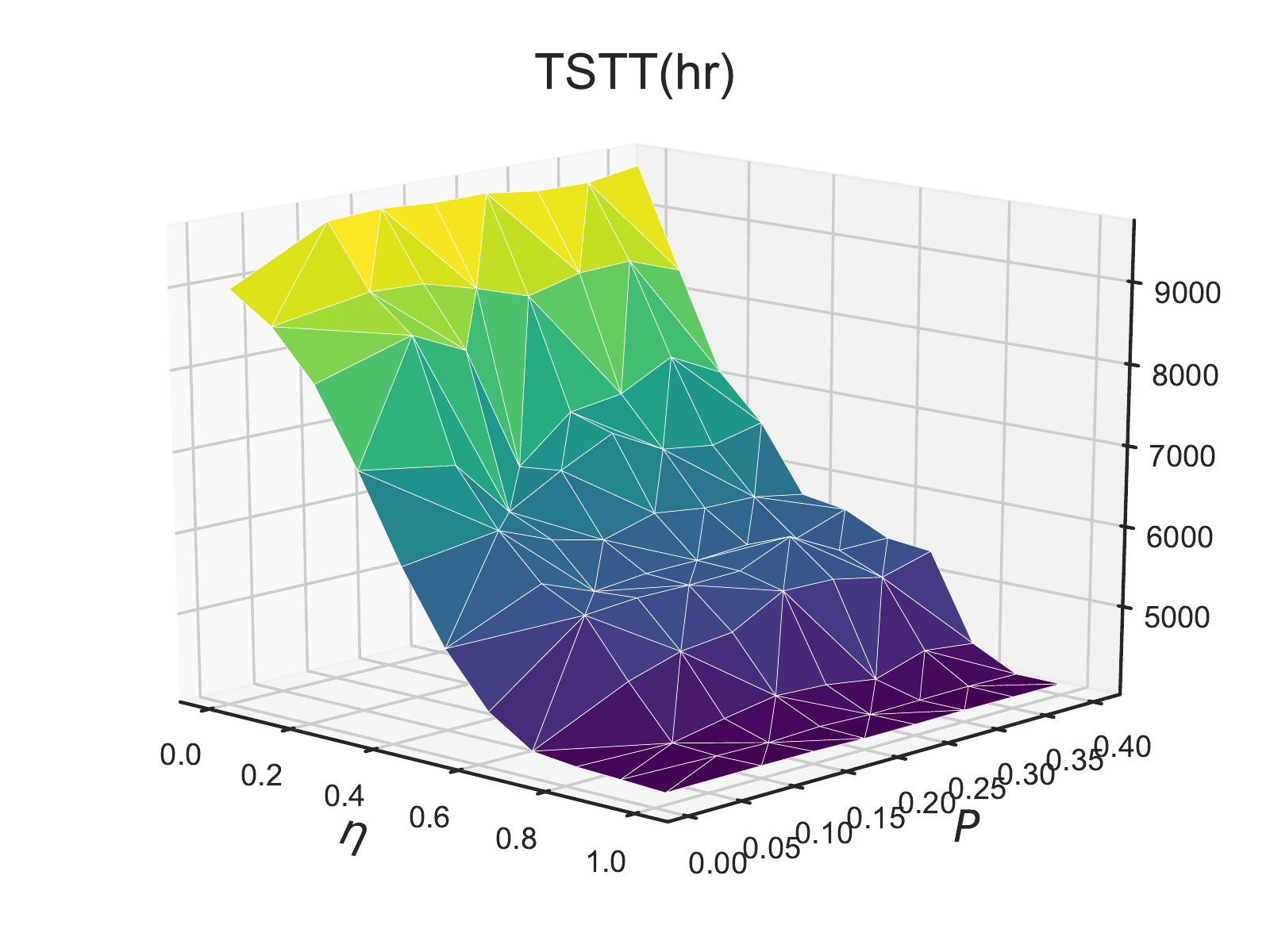}}\hfill

\caption{Variation of revenue ((a),(b),(c)) and TSTT ((d),(e),(f)) for different values of $\eta$ and $P$ parameters for the feedback control heuristic tested on SESE, LBJ, and MoPac networks} 
\label{fig:densityHeuristic}
\end{figure}

In this section, we compare the performance of Deep-RL algorithm against the feedback control heuristic. First, we study the variation of different objectives from the feedback control heuristic for different values of $\eta$ and $P$ values to identify the best performance for benchmarking. Figure \ref{fig:densityHeuristic} shows the variation of revenue and TSTT values for the SESE, LBJ, and MoPac networks. The values for each combination of parameters are reported as an average over 10 random seeds where the initial tolls on all toll links are set randomly between the minimum and maximum values for different seeds.

\vpB{As observed, low values of $\eta$ generate the highest average revenue across all combinations. Lower values of $\eta$ ensure that ML is kept relatively more congestion free than the case when $\eta$ value is high. A low value of $\eta$ charges high toll in the beginning and ensures that GPLs are more jammed  promoting jam-and-harvest nature and generating more revenue.}

\vpB{In contrast to this, low values of TSTT are obtained for high values of $\eta$ for both LBJ and MoPac networks. This is also intuitive: tolls minimizing TSTT operate the managed lane close to its critical density at all times. The  contrary behavior of the SESE network, where low values of $\eta$ also generate low TSTT, is because of the reasons explained in Section 4.4. For a given value of $\eta$, the variation of TSTT and revenue with $P$ is not significant, indicating that the performance of feedback control heuristic is more sensitive to the $\eta$ parameter.}

Next we compare the performance of feedback control heuristic against Deep-RL algorithms. Table \ref{tab:heuristicComparison} shows the values of different statistics reported as five-tuple: (revenue, TSTT, JAH$_1$, JAH$_2$, \texttt{\%-violation}) for both the revenue maximization and the TSTT minimization objectives for Deep-RL algorithms (we report the better objective value between VPG and PPO) and the feedback control heuristic. We highlight the value of the optimization objective in bold. \vpB{We also include the standard deviation in the objective value for both algorithms; the Deep-RL algorithm generates stochastic objective values due to the stochastic nature of the policy, while the feedback control heuristic generates stochastic objective values for different random initializations, given values of $\eta$ and $P$.}

\begin{table}[h]
\centering
\caption{Comparison of Deep-RL against the feedback control heuristic for the two optimization objectives. Results are reported as a five-tuple: (revenue, TSTT, JAH$_1$, JAH$_2$, \texttt{\%-violation})}
\label{tab:heuristicComparison}
\begin{tabular}{|c|c|c|}
\hline
  \multicolumn{3}{|c|}{\textbf{Revenue maximization objective}} \\ \hline
 & \textbf{Deep-RL} & \textbf{Feedback Control} \\ \hline
\textbf{SESE} & \begin{tabular}[c]{@{}c@{}}(\textbf{\$11889.80 $\pm$ 3.77}, 2933.88 hr, \\ 1166.43 veh, 0.34, 0\%)\end{tabular} & {\color[HTML]{333333} \begin{tabular}[c]{@{}c@{}}(\textbf{\$11881.70 $\pm$ 7.92}, 2933.70 hr, \\ 1166.44 veh, 0.34, 0\%)\end{tabular}} \\ \hline
\textbf{DESE} & \begin{tabular}[c]{@{}c@{}}(\textbf{\$497.97 $\pm$ 4.94}, 221.52 hr, \\ 159.47 veh, 0.32, 0\%)\end{tabular} & \begin{tabular}[c]{@{}c@{}}(\textbf{\$489.08 $\pm$ 0}, 223.26 hr, \\ 160.43 veh, 0.32, 0\%)\end{tabular} \\ \hline
\textbf{LBJ} & \begin{tabular}[c]{@{}c@{}}(\textbf{\$4718.43 $\pm$ 255.70}, 1396.15 hr, \\ 986.81 veh, 0.49, 1.62\%))\end{tabular} & \begin{tabular}[c]{@{}c@{}}(\textbf{\$4307.74 $\pm$ 275.59}, 1356.89 hr, \\ 929.57 veh, 0.43, 0.77\%)\end{tabular} \\ \hline
\textbf{MoPac} & \begin{tabular}[c]{@{}c@{}}(\textbf{\$18740.40 $\pm$ 61.64}, 9618.04 hr, \\ 3102.17 veh, 0.32, 1.26\%)\end{tabular} & \begin{tabular}[c]{@{}c@{}}(\textbf{\$18544.77 $\pm$ 133.36}, 9600.08 hr, \\ 3097.71 veh, 0.32, 1.28\%)\end{tabular} \\ \hline
  \multicolumn{3}{|c|}{\textbf{TSTT minimization objective}} \\ \hline
 & \textbf{Deep-RL} & \textbf{Feedback Control} \\ \hline
\textbf{SESE} & \begin{tabular}[c]{@{}c@{}}(\$11705.9, \textbf{2894.27 $\pm$ 16.22 hr}, \\ 1166.38 veh, 0.34, 0\%)\end{tabular} & \begin{tabular}[c]{@{}c@{}}(\$11530.38, \textbf{2897.41 $\pm$ 18.72 hr}, \\ 1166.53 veh, 0.34, 0\%)\end{tabular} \\ \hline
\textbf{DESE} & \begin{tabular}[c]{@{}c@{}}(\$271.46, \textbf{191.40 $\pm$ 7.53 hr}, \\ 128.23 veh, 0.22, 0\%)\end{tabular} & \begin{tabular}[c]{@{}c@{}}(\$275.91, \textbf{213.57 $\pm$ 5.64 hr}, \\ 128.00 veh, 0.25, 0\%)\end{tabular} \\ \hline
\textbf{LBJ} & \begin{tabular}[c]{@{}c@{}}(\$254.43, \textbf{641.72 $\pm$ 15.67 hr}, \\ 541.18 veh, 0.25, 0.24\%)\end{tabular} & \begin{tabular}[c]{@{}c@{}}(\$158.46, \textbf{661.40 $\pm$ 0 hr}, \\ 421.67 veh, 0.21, 0.32\%)\end{tabular} \\ \hline
\textbf{MoPac} & \begin{tabular}[c]{@{}c@{}}(\$655.45, \textbf{4022.45 $\pm$ 4.21 hr}, \\ 1199.22 veh, 0.11, 0.07\%)\end{tabular} & \begin{tabular}[c]{@{}c@{}}(\$606.01, \textbf{4024.83 $\pm$ 11.01 hr}, \\ 1141.37 veh, 0.11, 0.03\%)\end{tabular} \\ \hline
\end{tabular}
\end{table}

\vpB{The Deep-RL algorithms always finds tolls with slightly better objective values compared to the feedback control heuristic. For the revenue maximization objective, the average revenues from Deep-RL are 0.07--9.5\% higher than the ones obtained from the feedback control heuristic. Similarly, for the TSTT minimization objective, the average TSTT values obtained from the Deep-RL algorithm are 0.09--10.38\% lower than the average TSTT from the feedback control heuristic.} Similar to the observations made earlier, the tolls maximizing the revenue also generate a high value of JAH$_2$ statistic and the tolls generating high revenue generate low TSTT (with an exception of SESE network). \vpB{The value of \texttt{\%-violation} on the ML is less than 2\% on an average for all toll profiles, with insignificant differences between the Deep-RL algorithm and the feedback control heuristic.}

\section{Conclusion}

In this article, we developed Deep-RL algorithms for dynamic pricing of express lanes with multiple access points. We showed that the Deep-RL algorithms are able to learn toll profiles for multiple objectives, even capable of generating toll profiles lying on the Pareto frontier. The average objective value converged within 200 iterations for the four networks tests. The number of sensors and sensor locations were found to have little impact on the learning due to the spatial correlation of congestion pattern. We also conducted transferability tests and showed that policies trained using Deep-RL algorithm can be transferred to setting with new demand distribution and VOT distribution without losing performance; however, if the lane choice model is changed the transferred policy performs poorly. \vp{We analyzed the variation of multiple objectives together and found that TSTT-minimizing profiles may be similar to revenue-maximizing profiles for certain network characteristics where the GPL invariably becomes congested early in the simulation.} 
\vpB{We also compared the performance of Deep-RL algorithms against the feedback control heuristic and found that it outperformed the heuristic for the revenue maximization objective generating average revenue up to 9.5\% higher than the heuristic and generating average TSTT up to 10.4\% lower than the heuristic.}

\vp{The Deep-RL model in this article requires training, which is dependent on the input data and the parameters. We make following implementation recommendations. If a toll operator has access to the input data including the demand distribution and driver lane choice behavior, we recommend first calibrating a lane-choice model using the data and then using the calibrated model to train the policy for the desired objective under desired constraints. If the driver lane choice data is very detailed and can exactly identify how many travelers chose the ML at each time, then that data can be directly used in training without calibrating a lane-choice model; however, a calibrated model is still recommended as it can assist in conducting sensitivity analysis to other inputs and/or long-term planning. If the input data is not available or has poor accuracy, we recommend two alternatives. A toll operator can either train the Deep-RL model considering high stochasticity by choosing a large values for the standard deviations ($\sigma_d$ and $\sigma_o$), or train several policies for different combinations of inputs and use the policy based on the expected realization of inputs from field data for real-time implementation. Lastly, we also recommend retraining the toll policy using real-time data. For example, a policy can be trained from the historic data and then improved based on the observations from a specific day and the improved policy can then be applied to the next day. Additionally, though the model in this article trains a stochastic policy, for implementation purposes, we can use a deterministic policy with the tolls set as the mean value predicted by the policy.}

In addition to the future work ideas discussed earlier (marked as $\text{FW}\#$), there are additional topics that should be studied. First, the choice of traffic flow model is critical to the performance of Deep-RL algorithms. The macroscopic multiclass cell transmission model used in our analysis does not capture the impacts of lane changes and the second-order stop-and-go waves. Future work can be devoted to developing efficient Deep-RL algorithm using microscopic simulation models and on testing the transferability of algorithms trained on a macroscopic scale to microscopic scales. Second, we only considered loop detector density measurements in the simulations. Other types of observations like speeds, toll-tag readings, and measurements using Lagrangian sensors like GPS devices on vehicles require redefining the POMDP to handle such measurements and can be looked into as part of the future work. \vp{Third, for real-time implementation of Deep-RL algorithms, the minimum speed limit constraint on ML (constraint 2 defined in Section \ref{subsec:episodicRL}) should be satisfied throughout the learning phase, which requires analysis of constrained policy optimization methods like in Achiam et al.~\cite{achiam2017constrained}. Last, the future work should also analyze the equity impacts of the tolls generated by Deep-RL across multiple vehicle classes and investigate if generating equitable toll policies can be included as part of the Deep-RL problem.} 

\section*{Acknowledgment}
Partial support for this research was provided by the North Central Texas Council of Governments, Data-Supported Transportation Planning and Operations University Transportation Center, and the National Science Foundation (Grants No. $1254921$, $1562291$, and $1826230$.) The authors are grateful for this support. The authors would also like to thank Natalia Ruiz Juri and Tianxin Li at the Center for Transportation Research, The University of Texas at Austin for their help in providing us the data for the MoPac Express lanes, and Josiah Hanna for his comments on the paper draft.

\bibliography{references}

\end{document}